\documentclass[11pt,english]{elsarticle}
\usepackage{mathptmx}
\usepackage[T1]{fontenc}
\usepackage[latin9]{inputenc}
\usepackage[letterpaper]{geometry}
\usepackage{color}
\usepackage{amsmath}
\usepackage{amsthm}
\usepackage{graphicx}

\makeatletter
\numberwithin{equation}{section}
\numberwithin{figure}{section}

\DeclareMathAlphabet{\mathsfsl}{OT1}{cmss}{m}{sl}
\newcommand{\tensor}[1]{\stackrel{\leftrightarrow}{\mathsfsl{#1}}}
\renewcommand{\vec}[1]{\mathbf{#1}}

\newcommand{\nab}{\nabla}

\newcommand{\rb}[1]{\left( #1 \right)}
\renewcommand{\sb}[1]{\left[ #1 \right]}

\makeatother

\usepackage{babel}
\begin{document}

\title{An adaptive, implicit, conservative 1D-2V multi-species Vlasov-Fokker-Planck
multiscale solver in planar geometry}

\author[lanl]{W. T. Taitano\corref{cor1}}

\ead{taitano@lanl.gov}

\author[lanl]{L. Chacón}

\author[lanl2]{A. N. Simakov}

\cortext[cor1]{Corresponding author}

\address[lanl]{Theoretical Division Los Alamos National Laboratory, Los Alamos,
NM 87545}

\address[lanl2]{Theoretical Design Division, Los Alamos National Laboratory, Los
Alamos, NM 87545}

\address{}
\begin{abstract}
We consider a 1D-2V Vlasov-Fokker-Planck multi-species ionic description
coupled to fluid electrons. We address temporal stiffness with implicit
time stepping, suitably preconditioned. To address temperature disparity
in time and space, we extend the conservative \emph{adaptive} velocity-space
discretization scheme proposed in {[}Taitano \emph{et al}., \emph{J.
Comput. Phys.}, 318, 391\textendash 420\textbf{,} (2016){]} to a spatially
inhomogeneous system. In this approach, we normalize the velocity-space
coordinate to a temporally and spatially varying local characteristic
speed per species. We explicitly consider the resulting inertial terms
in the Vlasov equation, and derive a discrete formulation that conserves
mass, momentum, and energy up to a prescribed nonlinear tolerance
upon convergence. Our conservation strategy employs nonlinear constraints
to enforce these properties discretely for both the Vlasov operator
and the Fokker-Planck collision operator. Numerical examples of varying
degrees of complexity, including shock-wave propagation, demonstrate
the favorable efficiency and accuracy properties of the scheme. 
\end{abstract}
\maketitle

\section{Introduction\label{sec:Introduction}}

The Vlasov-Fokker-Planck (VFP) collisional kinetic description, coupled
with Maxwell's equations, is regarded as a first-principles physical
model for describing weakly coupled plasmas in all collisionality
regimes, and accordingly, has a wide range of applications in laboratory
(e.g., magnetic and inertial thermonuclear fusion), space (e.g., Earth's
magnetosphere), and astrophysical (e.g., stellar mass ejections) plasmas.
In the VFP system, collisions are modeled by the Fokker-Planck collision
operator, which describes collisional relaxation of particle distribution
functions in plasmas under the assumption of binary, grazing-angle
collisions \citep{rosenbluth,arsenev_1991conn_boltz_lfpe,desvillettes_1992_asymp_boltz_eqn_graze_col,degond_1992_fp_asymp_bolt_col_op_coul,goudon_1997_jsp,landau_1937_fokker_planck}.
Mathematically, the Fokker-Planck operator is integro-differential,
non-local, and very difficult to invert.

The system of VFP equations for various plasma species supports disparate
length and time scales, as well as arbitrary temperature disparity
in time and space, which makes this system particularly challenging
to solve with grid-based approaches. The challenges of temperature
disparity are evident when one considers the thermal speed, $v_{th}=\sqrt{2T/m}$,
which provides a characteristic width of the plasma species distribution
function and is a function of the plasma temperature, $T$, and particle
species mass, $m$. In many practical applications of interest, $v_{th}$
variation for a given species can span several orders of magnitude
in configuration space. In addition, mass differences result in strong
$v_{th}$ disparities for different species. Since the velocity-space
domain size is determined for a given species by the hottest region
(large $v_{th}$), and the velocity-grid spacing must resolve the
coldest region (small $v_{th}$), velocity-space discretizations with
uniform Cartesian grids in such scenarios may lead to impractical
grid size requirements. 

Several studies recognized and tried to address these challenges by
normalizing the velocity coordinate to the local thermal velocity
\citep{larroche_2007_lsse_jcp,larroche_EPJ_2003_icf_fuel_ion_implosion_sim,jarema_CPC_2015_block_structured_grid_adaptive_vpace,peigney_JCP_2014_fp_kinetic_modeling_of_alpha}.
In this fashion, the grid will expand as the plasma heats, and contract
as it cools. Particularly relevant to this study is the work in Ref.
\citep{larroche_EPJ_2003_icf_fuel_ion_implosion_sim}, where the velocity-space
domain was adapted for multiple ion species based on a single local
\emph{average} $v_{th}$ (over the ion species) and hydrodynamic velocity
of the plasma. This powerful strategy enabled the fully kinetic implosion
simulations of inertial confinement fusion (ICF) capsules \citep{larroche_EPJ_2003_icf_fuel_ion_implosion_sim,larroche_pop_2012_Dhe_3_icf_sim,inglebert_epl_2014_species_separation_kinetic_effect_neutron_diagnostics},
but required \textcolor{black}{intermittent remapping in both the
physical and velocity space}. \textcolor{black}{None of these strategies
conserve mass, momentum, and energy, and some of them \citep{jarema_CPC_2015_block_structured_grid_adaptive_vpace}
break the structured nature of the underlying computational mesh.}

Recently, a novel strategy that deals with strong temperature disparity,
avoids remapping, and works on structured meshes was proposed in Ref.
\citep{Taitano_2016_rfp_0d2v_implicit} for the 0D-2V multispecies
Fokker-Planck equation. The strategy employs a multiple-grid approach
by normalizing each species' velocity to its thermal speed. The Fokker-Planck
equation was transformed analytically, and then discretized on a mesh.
The transformed equations exposed the continuum conservation symmetries,
which were then enforced in the discrete via nonlinear constraints.
This strategy ensures that the species' distribution function is always
well resolved regardless of temperature or mass disparity. 

In this study, we extend the conservative, multiple-dynamic velocity-space
adaptive strategy developed in Ref. \citep{Taitano_2016_rfp_0d2v_implicit}
to a spatially inhomogeneous, 1D Cartesian system. We consider a quasi-neutral
plasma with multiple kinetic ion species and fluid electrons. As before,
ionic species are evolved on a velocity-space grid normalized to a
temporally and spatially varying characteristic speed, $v^{*}$ (a
function of their $v_{th}$), and the VFP equation is analytically
transformed. This transformation introduces additional inertial terms,
which are carefully discretized to ensure simultaneous conservation
of mass, momentum, and energy. 

The rest of the paper is organized as follows. Section \ref{sec:VRFP}
introduces the VFP and fluid-electron equations and discusses their
conservation properties. In Sec. \ref{sec:Numerical-implementation},
we introduce the normalized Vlasov equation, and provide a detailed
discussion on the implementation of the proposed schemes in the following
order: 1) a discretization of the Vlasov-Fokker-Planck equation with
the additional inertial terms, 2) a discretization of the fluid electron
equation, 3) our discrete conservation strategy for fluid electrons
and kinetic ions, 4) a discrete conservation strategy for the Vlasov
component with the added inertial terms, and 5) temporal and spatial
evolution of $v^{*}$. The numerical performance of the scheme is
demonstrated with various multi-species tests of varying degrees of
complexity in Sec. \ref{sec:numerical-results}. Finally, we conclude
in Sec. \ref{sec:conclusions}.

\section{The multi-species Vlasov-Rosenbluth-Fokker-Planck equation with fluid
electrons\label{sec:VRFP}}

A dynamic evolution of weakly-coupled collisional plasmas is described
by the Vlasov-Fokker-Planck equation for the particle distribution
function (PDF), $f\left(\vec{\vec{x},\vec{v},t}\right)$ in configuration
space, $\vec{x}$, velocity space, $\vec{v}$, and time, $t$: 
\begin{equation}
\partial_{t}f_{\alpha}+\nabla_{x}\cdot\left(\vec{v}f_{\alpha}\right)+\frac{q_{\alpha}}{m_{\alpha}}\nabla_{v}\cdot\left[\left(\vec{E}+\vec{v}\times\vec{B}\right)f_{\alpha}\right]=\sum_{\beta=1}^{N_{s}}C_{\alpha\beta},
\end{equation}
where $\vec{E}$ is the electric field, $\vec{B}$ is the magnetic
field, $N_{s}$ is the total number of plasma species in the system,
and $C_{\alpha\beta}$ is the Fokker-Planck collision operator for
species $\alpha$ colliding with species $\beta$: 

\begin{equation}
C_{\alpha\beta}=\Gamma_{\alpha\beta}\nab_{v}\cdot\left[\tensor{D}_{\beta}\cdot\nabla_{v}f_{\alpha}-\frac{m_{\alpha}}{m_{\beta}}\vec{{A}}_{\beta}f_{\alpha}\right].\label{eq:fokker_planck_operator}
\end{equation}
Here, $\Gamma_{\alpha\beta}=\frac{2\pi Z_{\alpha}^{2}Z_{\beta}^{2}e^{4}\Lambda_{\alpha\beta}}{m_{\alpha}^{2}},$
$\tensor D{}_{\beta}$ and $\vec{{A}}_{\beta}$ are the tensor-diffusion
and friction coefficients for species $\beta$, $m_{\alpha}$ and
$m_{\beta}$ are the masses of species $\alpha$ and $\beta$, respectively,
$Z_{\alpha}=q_{\alpha}/e$ is the ionization state of species $\alpha$,
$e$ is the proton charge, and $\Lambda_{\alpha\beta}$ is the Coulomb
logarithm ($\Lambda_{\alpha\beta}=10$ is assumed for simplicity in
this study for all species). 

The Rosenbluth formulation of the Fokker-Planck collision operator
\citep{rosenbluth} computes the velocity-space-transport coefficients
from the so-called Rosenbluth potentials $G_{\beta}$, $H_{\beta}$
as:
\begin{eqnarray}
\tensor D_{\beta} & = & \nab_{v}\nab_{v}G_{\beta},\label{eq:diffusion_tensor}\\
\vec{A}_{\beta} & = & \nab_{v}H_{\beta},\label{eq:friction_vector}
\end{eqnarray}
which, in turn, are computed from the distribution function of species
$\beta$ as:
\begin{equation}
\nabla_{v}^{2}H_{\beta}=-8\pi f_{\beta},\label{eq:h_poisson}
\end{equation}
\begin{equation}
\nabla_{v}^{2}G_{\beta}=H_{\beta}.\label{eq:g_poisson}
\end{equation}
The Rosenbluth form is completely equivalent to the integral Landau
form \citep{landau_1937_fokker_planck}, but more advantageous algorithmically
because it can be inverted with $\mathcal{O}(N)$ complexity (with
$N$ the number of degrees of freedom in velocity space) \citep{Taitano_2015_rfp_0d2v_implicit}.

The collision operator, Eq. (\ref{eq:fokker_planck_operator}), preserves
the positivity of $f_{\alpha}$, and conserves mass, momentum, and
energy. The conservation properties stem from the following symmetries
\citep{braginskii}:
\begin{eqnarray}
\left<1,C_{\alpha\beta}\right>_{v} & = & 0,\label{eq:charge-cons}\\
m_{\alpha}\left<\vec{v},C_{\alpha\beta}\right>_{v} & = & -m_{\beta}\left<\vec{v},C_{\beta\alpha}\right>_{v},\label{eq:momentum-cons}\\
m_{\alpha}\left<\frac{v^{2}}{2},C_{\alpha\beta}\right>_{v} & = & -m_{\beta}\left<\frac{v^{2}}{2},C_{\beta\alpha}\right>_{v},\label{eq:energy-cons}
\end{eqnarray}
where the inner product is defined as $\left<A,B\right>_{v}=2\pi\int_{-\infty}^{\infty}dv_{||}\int_{0}^{\infty}dv_{\perp}v_{\perp}A(\vec{v})\,B(\vec{v})$
(for the cylindrically symmetric coordinate system in the velocity
space employed herein). These conservation symmetries can be enforced
in the discrete following the general procedures discussed in Refs.
\citep{Taitano_2015_rfp_0d2v_implicit,Taitano_2016_rfp_0d2v_implicit}.

In this study, we consider a 1D planar geometry in the configuration
space without a magnetic field. Without loss of generality, we consider
a 2V cylindrically symmetric coordinate system in the velocity space.
We adopt a fluid-electron model with a reduced ion-electron collision
operator. We obtain the following simplified system of equations comprised
of the ion-Vlasov-Fokker-Planck equation (per species $\alpha$),

\begin{equation}
\partial_{t}f_{\alpha}+\partial_{x}\left(v_{||}f_{\alpha}\right)+\frac{q_{\alpha}}{m_{\alpha}}E_{||}\partial_{v_{||}}f_{\alpha}=\sum_{\beta}^{N_{s}}C_{\alpha\beta}+C_{\alpha e}\label{eq:1d_vfp_eqn}
\end{equation}
and the electron-temperature equation,

\begin{equation}
\frac{3}{2}\frac{\partial}{\partial t}\left[n_{e}T_{e}\right]+\frac{5}{2}\partial_{x}\left[u_{||,e}n_{e}T_{e}\right]+\partial_{x}Q_{||,e}-q_{e}n_{e}u_{||,e}E_{||}=\sum_{\alpha}^{N_{s}}W_{e\alpha}.\label{eq:1d_electron_equation}
\end{equation}
Here, $Q_{||,e}$ is the parallel electron-heat flux, $n_{e}$ is
the electron density, $T_{e}$ is the electron temperature, $u_{||,e}$
is the parallel electron fluid velocity, $P_{e}=n_{e}T_{e}$ is the
electron pressure, $W_{e\alpha}$ describes the electron-ion energy
exchange,

\begin{equation}
W_{e\alpha}=F_{||,\alpha e}u_{||,\alpha}+3\nu_{e\alpha}\frac{m_{e}}{m_{\alpha}}n_{e}\left(T_{\alpha}-T_{e}\right),
\end{equation}
$\vec{F}_{\alpha e}^{T}=\left[F_{||,\alpha e},0\right]$ is the friction
force between the $\alpha$-ion species and electrons, and 

\begin{equation}
\nu_{e\alpha}=\frac{4\sqrt{2\pi}n_{\alpha}q_{\alpha}^{2}e^{4}\Lambda_{e\alpha}}{3\sqrt{m_{e}}T_{e}^{3/2}}
\end{equation}
is the electron-ion collision frequency. The frictional force between
the $\alpha$-ion species and electrons is given by,

\begin{equation}
\vec{F}_{\alpha e}=-m_{e}n_{e}\nu_{e\alpha}\left(\vec{u}_{\alpha}-\left\langle \vec{u}_{\alpha}\right\rangle \right)+\alpha_{0}m_{e}n_{e}\nu_{e\alpha}\left(\vec{u}_{e}-\left\langle \vec{u}_{\alpha}\right\rangle \right)+\beta_{0}\frac{n_{e}\nu_{e\alpha}\nabla T_{e}}{\sum_{\alpha}^{N_{s}}\nu_{e\alpha}},\label{eq:ion_electron_frictional_force}
\end{equation}
and the electron heat flux by

\begin{equation}
\vec{Q}_{e}=\beta_{0}n_{e}T_{e}\left(\vec{u}_{e}-\left\langle \vec{u}_{\alpha}\right\rangle \right)-\kappa_{e}\nabla T_{e}.
\end{equation}
Definitions of coefficients $\alpha_{0},$$\beta_{0}$, $\kappa_{e}$,
and the collision-frequency-averaged ion velocity, $\left\langle \vec{u}_{\alpha}\right\rangle $,
can be found in App. \ref{app:sophisticated_fluid_electron_model}
and in Ref. \citep{simakov_PoP_2014_e_transp_wh_multi_ion} .

The parallel (to the x-axis) component of the ambipolar-electric field,
$E_{||}$, in Eq. (\ref{eq:1d_electron_equation}) is found from the
inertialess electron momentum equation ,

\begin{equation}
E_{||}=\frac{\sum_{\alpha}^{N_{s}}F_{||,\alpha e}+\partial_{x}P_{e}}{q_{e}n_{e}}.\label{eq:ambipolar_field}
\end{equation}
Finally, we assume quasi-neutrality,

\begin{equation}
n_{e}=\sum_{\alpha}^{N_{s}}Z_{\alpha}n_{\alpha},\label{eq:quasi-neutrality}
\end{equation}
and ambipolarity,

\begin{equation}
u_{||,e}=\frac{\sum_{\alpha}^{N_{s}}Z_{\alpha}nu_{||,\alpha}}{n_{e}},\label{eq:ambipolarity}
\end{equation}
to close the system.

The electron-ion collision operator $C_{\alpha e}$ in Eq. (\ref{eq:1d_vfp_eqn})
is given by:

\begin{equation}
C_{\alpha e}=\Gamma_{\alpha e}\nabla_{v}\cdot\left[\tensor D_{\alpha e}\cdot\nabla_{v}f_{\alpha}-\frac{m_{\alpha}}{m_{e}}\vec{A}_{\alpha e}f_{\alpha}\right],\label{eq:ion_electron_collision_operator}
\end{equation}
where we adopt the reduced ion-electron potentials \citep{hazeltine_1991_plas_confin}:

\begin{equation}
G_{\alpha e}=\frac{2}{3}\frac{n_{e}}{\sqrt{\pi}}\frac{\left(\vec{v}-\vec{u}_{\alpha}\right)\cdot\left(\vec{v}-\vec{u}_{\alpha}\right)}{v_{th,e}},
\end{equation}

\begin{equation}
H_{\alpha e}=-\frac{8}{3}\frac{n_{e}}{\sqrt{\pi}v_{th,e}^{3}}\left[\frac{\left(\vec{v}-\vec{u}_{\alpha}\right)\cdot\left(\vec{v}-\vec{u}_{\alpha}\right)}{2}-\frac{\vec{F}_{\alpha e}}{m_{e}n_{e}\nu_{e\alpha}}\cdot\left(\vec{v}-\vec{u}_{\alpha}\right)\right].
\end{equation}
Here, $\vec{v}^{T}=\left[v_{||},v_{\perp}\right]$, $\vec{u}_{\alpha}^{T}=\left[u_{||,\alpha},0\right]$,
$v_{th,e}$ is the electron thermal speed, and the transport coefficients
are:

\begin{equation}
\tensor D_{\alpha e}=\nabla_{v}\nabla_{v}G_{\alpha e}=\frac{4}{3}\frac{n_{e}}{\sqrt{\pi}}\frac{\tensor I}{v_{th,e}},\label{eq:D_ion_elecron}
\end{equation}
\begin{equation}
\vec{A}_{\alpha e}=\text{\ensuremath{\vec{A}}}_{\alpha e,u}+\vec{A}_{\alpha e,F},\label{eq:A_ion_electron}
\end{equation}
where $\tensor I$ is the unit dyad,

\begin{equation}
\vec{A}_{\alpha e,u}=\nabla_{v}H_{\alpha e,u}=-\frac{8}{3}\frac{n_{e}}{\sqrt{\pi}v_{th,e}^{3}}\left(\vec{v}-\vec{u}_{\alpha}\right)
\end{equation}
and

\begin{equation}
\vec{A}_{\alpha e,F}=\nabla_{v}H_{\alpha e,F}=\frac{8}{3}\frac{n_{e}}{\sqrt{\pi}v_{th,e}^{3}}\frac{\vec{F}_{\alpha e}}{m_{e}n_{e}\nu_{e\alpha}}.\label{eq:A_ion_electron_F}
\end{equation}

\subsection{Conservation properties of the kinetic-ion/fluid-electron system\label{subsec:conservation_properties_for_ion_fluid_electron_system}}

The coupled kinetic-ion and fluid-electron system possesses continuum
conservation properties. We remark that, although these properties
are well known and have been discussed by others in the past \citep{hazeltine},
we reproduce them here to explicitly expose the continuum symmetries
that are required to ensure the conservation properties. The goal
is to develop a strategy that can ensure these properties discretely
(Sec. \ref{subsec:simul_disc_mass_mom_ener_cons_for_kin_ion_fe_sys}).

Mass conservation follows trivially from the ion mass conservation
and quasi-neutrality. Momentum conservation follows from taking the
$m_{\alpha}v_{||}$ moment of the ion Vlasov equation summed over
all ions:
\begin{equation}
\sum_{\alpha}^{N_{s}}\left[m_{\alpha}\partial_{t}I_{||,\alpha}+m_{\alpha}\partial_{x}S_{2,||,\alpha}-q_{\alpha}n_{\alpha}E_{||}\right]=\sum_{\alpha}^{N_{s}}m_{\alpha}\left[\sum_{\beta}^{N_{s}}\left\langle v_{||},C\left(f_{\alpha},f_{\beta}\right)\right\rangle _{\vec{v}}+\left\langle v_{||},C_{\alpha e}\right\rangle _{\vec{v}}\right],\label{eq:vfp_continuum_momentum_moment}
\end{equation}
where $I_{||,\alpha}=\left\langle v_{||},f_{\alpha}\right\rangle _{\vec{v}}$
is the parallel specific momentum density flux, $S_{2,||,\alpha}=\left\langle v_{||},v_{||}f_{\alpha}\right\rangle _{\vec{v}}$
is the parallel-parallel component of the pressure tensor, and $q_{\alpha}n_{\alpha}E_{||}=-q_{\alpha}E_{||}\left\langle v_{||},\partial_{v_{||}}f_{\alpha}\right\rangle _{v}$
is the parallel electrostatic force. The first term on the right-hand
side (RHS) of Eq. (\ref{eq:vfp_continuum_momentum_moment}) vanishes
due to momentum conservation across ion species \citep{Taitano_2015_rfp_0d2v_implicit}.
The second term on the RHS uses the reduced-collision operator between
ion species and electrons, and it can be expanded as follows:
\begin{eqnarray}
 &  & m_{\alpha}\left\langle v_{||},C_{\alpha e}\right\rangle _{\vec{v}}=m_{\alpha}\Gamma_{\alpha e}\left\langle v_{||},\nabla_{v}\cdot\left[\tensor D_{\alpha e}\cdot\nabla f_{\alpha}-\frac{m_{\alpha}}{m_{e}}\vec{A}_{\alpha e}f_{\alpha}\right]\right\rangle _{\vec{v}}\nonumber \\
 &  & =m_{\alpha}\Gamma_{\alpha e}\left[\underbrace{\left\langle v_{||},\nabla_{v}\cdot\left[\tensor D_{\alpha e}\cdot\nabla_{v}f_{\alpha}\right]\right\rangle _{\vec{v}}}_{\textcircled a}-\frac{m_{\alpha}}{m_{e}}\left\langle v_{||},\nabla_{v}\cdot\left[\left(\underbrace{\vec{A}_{\alpha e,u}}_{\textcircled b}+\underbrace{\vec{A}_{\alpha e,F}}_{\textcircled c}\right)f_{\alpha}\right]\right\rangle _{\vec{v}}\right].
\end{eqnarray}
Here, the terms $\textcircled a$ and $\textcircled b$ vanish independently:

\begin{equation}
m_{\alpha}\left\langle v_{||},\Gamma_{\alpha e}\nabla_{v}\cdot\left[\tensor D_{\alpha e}\cdot\nabla_{v}f_{\alpha}\right]\right\rangle _{v}=m_{\alpha}\Gamma_{\alpha e}\frac{4}{3}\frac{n_{e}}{v_{th,e}\sqrt{\pi}}\left\langle v_{||},\nabla_{v}\cdot\left[\tensor I\cdot\nabla_{v}f_{\alpha}\right]\right\rangle _{v}=0\label{eq:ie_col_mom_cons_diff}
\end{equation}
and
\begin{equation}
m_{\alpha}\Gamma_{\alpha e}\left\langle v_{||},\nabla_{v}\cdot\left[\frac{m_{\alpha}}{m_{e}}\vec{A}_{\alpha e,u}f_{\alpha}\right]\right\rangle _{v}=-m_{\alpha}\Gamma_{\alpha e}\frac{m_{\alpha}}{m_{e}}\frac{8}{3}\frac{n_{e}}{v_{th,e}^{3}\sqrt{\pi}}\left\langle v_{||},\nabla_{v}\cdot\left[\left(\vec{v}-\vec{u}_{\alpha}\right)f_{\alpha}\right]\right\rangle _{v}=0.\label{eq:ie_col_mom_cons_fric}
\end{equation}
The term $\textcircled c$ becomes {[}by Eq. (\ref{eq:A_ion_electron_F}){]}:

\begin{equation}
m_{\alpha}\Gamma_{\alpha e}\left\langle v_{||},\nabla_{v}\cdot\left[\frac{m_{\alpha}}{m_{e}}\vec{A}_{\alpha e,F}f_{\alpha}\right]\right\rangle _{\vec{v}}=-F_{||,\alpha e}.\label{eq:ie_col_mom_cons_fric_F}
\end{equation}
When combined with quasi-neutrality, Eq. (\ref{eq:quasi-neutrality}),
and the ion electrostatic acceleration term:

\begin{equation}
\sum_{\alpha}^{N_{s}}q_{\alpha}n_{\alpha}E_{||}=-\left(\partial_{x}P_{e}+\sum_{\alpha}^{N_{s}}F_{||,\alpha e}\right),\label{eq:ion_electrostatic_acceleration}
\end{equation}
 Eq. \ref{eq:vfp_continuum_momentum_moment} yields the total plasma
momentum equation

\begin{equation}
\sum_{\alpha}^{N_{s}}m_{\alpha}\left[\partial_{t}I_{\alpha}+\partial_{x}S_{2,||,\alpha}\right]+\partial_{x}P_{e}=0,
\end{equation}
which is in a conservative form. 

To show energy conservation, we take the second moment ($m_{\alpha}\frac{v^{2}}{2}$)
of the ion Vlasov equation as:
\begin{equation}
\sum_{\alpha}^{N_{s}}\left[m_{\alpha}\partial_{t}U_{\alpha}+m_{\alpha}\partial_{x}S_{3,||,\alpha}-q_{\alpha}n_{\alpha}u_{||,\alpha}E_{||}\right]=\sum_{\alpha}^{N_{s}}m_{\alpha}\left[\sum_{\beta}^{N_{s}}\left\langle \frac{v^{2}}{2},C\left(f_{\alpha},f_{\beta}\right)\right\rangle _{\vec{v}}+\left\langle \frac{v^{2}}{2},C_{\alpha e}\right\rangle _{\vec{v}}\right].
\end{equation}
Here, $U_{\alpha}=\left\langle \frac{v^{2}}{2},f_{\alpha}\right\rangle _{\vec{v}}$
is the specific-energy density, $S_{3,\alpha,||}=\left\langle v_{||}\frac{v^{2}}{2},f_{\alpha}\right\rangle _{\vec{v}}$
is the parallel component of the specific-energy density flux, and
$q_{\alpha}n_{\alpha}u_{||,\alpha}E_{||}=-q_{\alpha}E_{||}\left\langle \frac{v^{2}}{2},\partial_{v_{||}}f_{\alpha}\right\rangle $.
The first term on the RHS vanishes owing to energy conservation across
ion species. The second term on the RHS can be expanded as follows:
\begin{equation}
m_{\alpha}\left\langle \frac{v^{2}}{2},C_{\alpha e}\right\rangle _{\vec{v}}=m_{\alpha}\Gamma_{\alpha e}\left[\underbrace{\left\langle \frac{v^{2}}{2},\nabla_{v}\cdot\left[\tensor D_{\alpha e}\cdot\nabla f_{\alpha}\right]\right\rangle _{\vec{v}}}_{\textcircled a}-\frac{m_{\alpha}}{m_{e}}\left\langle \frac{v^{2}}{2},\nabla_{v}\cdot\left[\left(\underbrace{\vec{A}_{\alpha e,u}}_{\textcircled b}+\underbrace{\vec{A}_{\alpha e,F}}_{\textcircled c}\right)f_{\alpha}\right]\right\rangle _{\vec{v}}\right].
\end{equation}
Here, terms $\textcircled a$, $\textcircled b$, and $\textcircled c$
independently yield {[}using Eq. (\ref{eq:D_ion_elecron}) and (\ref{eq:A_ion_electron}){]}: 

\begin{equation}
\Gamma_{\alpha e}m_{\alpha}\left\langle \frac{v^{2}}{2},\nabla_{v}\cdot\left[\tensor D_{\alpha e}\cdot\nabla_{v}f_{\alpha}\right]\right\rangle _{v}=m_{\alpha}\Gamma_{\alpha e}\frac{4}{3}\frac{n_{e}}{v_{th,e}\sqrt{\pi}}\left\langle \frac{v^{2}}{2},\nabla_{v}\cdot\left[\tensor I\cdot\nabla_{v}f_{\alpha}\right]\right\rangle _{v}=3\nu_{e\alpha}\frac{m_{e}}{m_{\alpha}}n_{e}T_{e},\label{eq:ie_col_ener_cons_diff}
\end{equation}

\begin{eqnarray}
m_{\alpha}\Gamma_{\alpha e}\left\langle \frac{v^{2}}{2},\nabla_{v}\cdot\left[\frac{m_{\alpha}}{m_{e}}\vec{A}_{\alpha e,u}f_{\alpha}\right]\right\rangle _{v}=3\nu_{e\alpha}\frac{m_{e}}{m_{\alpha}}n_{e}T_{\alpha},\label{eq:ie_col_ener_cons_fric}
\end{eqnarray}
and
\begin{eqnarray}
m_{\alpha}\Gamma_{\alpha e}\left\langle \frac{v^{2}}{2},\nabla_{v}\cdot\left[\frac{m_{\alpha}}{m_{e}}\vec{A}_{\alpha e,F}f_{\alpha}\right]\right\rangle _{v}=-u_{||,\alpha}F_{||,\alpha e}.\label{eq:ie_col_ener_cons_fric-F}
\end{eqnarray}
Gathering terms, the energy moment of the ion-electron collision operator
yields:

\begin{equation}
m_{\alpha}\left\langle \frac{v^{2}}{2},C_{\alpha e}\right\rangle _{v}=3\nu_{e\alpha}\frac{m_{e}}{m_{\alpha}}n_{e}\left(T_{e}-T_{\alpha}\right)+u_{||,\alpha}F_{||,\alpha e}=-W_{e\alpha}.
\end{equation}
Using the fluid electron temperature equation and ambipolarity, we
finally obtain:

\begin{equation}
\partial_{t}\left(\frac{3}{2}n_{e}T_{e}+\sum_{\alpha}^{N_{s}}m_{\alpha}U_{\alpha}\right)+\partial_{x}\left(\frac{5}{2}u_{||,e}n_{e}T_{e}+\sum_{\alpha}^{N_{s}}m_{\alpha}S_{3,\alpha,||}+Q_{||,e}\right)=0,
\end{equation}
which is a conservative form of the evolution equation for the total
plasma energy density $\frac{3}{2}n_{e}T_{e}+\sum_{\alpha}^{N_{s}}m_{\alpha}U_{\alpha}$.

\section{Numerical implementation\label{sec:Numerical-implementation}}

\subsection{VFP equation in normalized velocity variables \textcolor{red}{\label{sec:normalization_of_the_vlasov_equation_to_vstar-1}}}

\begin{figure}[h]
\begin{centering}
\includegraphics[height=6cm]{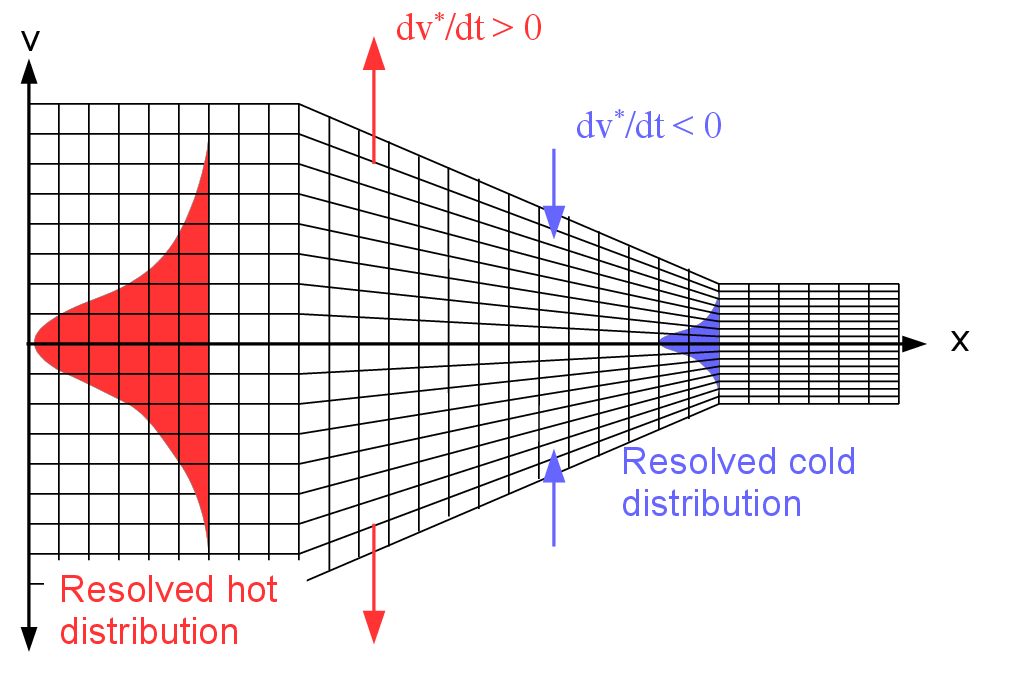}
\par\end{centering}
\caption{Illustration of the velocity space adaptivity.\label{fig:vth-disparity-in-space-wh-adaptivity-1}}
\end{figure}

We consider the normalization of all velocity-space quantities for
a species $\alpha$ to a reference speed, $v_{\alpha}^{*}\left(x,t\right)$,
related to their thermal speed, as follows:

\[
\vec{\widehat{v}}_{\alpha}=\frac{\vec{v}}{v_{\alpha}^{*}},\;\partial_{\widehat{v}_{||}}=v_{\alpha}^{*}\partial_{\widehat{v}_{||}},\;\widehat{f}_{\alpha}=\left(v_{\alpha}^{*}\right)^{3}f_{\alpha}.
\]
Here, the hat denotes quantities normalized to $v_{\alpha}^{*}$.
As an example, the density, drift, and temperature moments are defined
as:

\[
n_{\alpha}=\left\langle 1,\widehat{f}_{\alpha}\right\rangle _{\widehat{v}},\;u_{||,\alpha}=v_{\alpha}^{*}\widehat{u}_{||,\alpha}=v_{\alpha}^{*}\left\langle \widehat{v}_{||},\widehat{f}_{\alpha}\right\rangle _{\widehat{v}},\;T_{\alpha}=\left(v_{\alpha}^{*}\right)^{2}\widehat{T}_{\alpha}=\frac{m_{\alpha}\left(v_{\alpha}^{*}\right)^{2}\left\langle \left(\widehat{\vec{v}}-\widehat{\vec{u}}_{\alpha}\right)^{2},\widehat{f}_{\alpha}\right\rangle _{\widehat{v}}}{3\left\langle 1,\widehat{f}_{\alpha}\right\rangle _{\widehat{v}}}
\]
where $\left\langle \left(\cdot\right),\widehat{f}_{\alpha}\right\rangle _{\widehat{v}}=2\pi\int_{-\infty}^{\infty}d\widehat{v}_{||}\int_{0}^{\infty}\left(\cdot\right)\widehat{f}_{\alpha}\widehat{v}_{\perp}d\widehat{v}_{\perp}$.The
normalization of other relevant quantities and the collision operator
are discussed in Ref. \citep{Taitano_2016_rfp_0d2v_implicit}. We
note that, as $v^{*}$ is a function of local $v_{th}$ for a given
plasma species (elaborated in Sec. \ref{subsec:Evolution-equation-of-vstar}),
the grid will expand as the plasma heats, and contract as it cools;
refer to Fig. \ref{fig:vth-disparity-in-space-wh-adaptivity-1}.

The temporal and spatial dependence of $v_{\alpha}^{*}$ introduces
inertial terms in the Vlasov equation, Eq. (\ref{eq:1d_vfp_eqn}),
which after the transformation reads (App. \ref{app:details_on_vstar_transformation}):
\begin{eqnarray}
\partial_{t}\widehat{f}_{\alpha}+\partial_{x}\left(v_{\alpha}^{*}\widehat{v}_{||}\widehat{f}_{\alpha}\right) & + & \partial_{\widehat{v}_{||}}\left[\left(\frac{q_{\alpha}}{m_{\alpha}}\frac{E_{||}}{v_{\alpha}^{*}}\right)\widehat{f}_{\alpha}\right]-\widehat{\nabla}_{\widehat{v}}\cdot\left[\left(\frac{\partial_{t}v_{\alpha}^{*}}{v_{\alpha}^{*}}\right)\vec{\widehat{v}}\widehat{f}_{\alpha}\right]-\widehat{\nabla}_{\widehat{v}}\cdot\left[\left(\partial_{x}v_{\alpha}^{*}\right)\vec{\widehat{v}}\widehat{v}_{||}\widehat{f}_{\alpha}\right]\nonumber \\
 & = & \left(v_{\alpha}^{*}\right)^{3}\left(\sum_{\beta}^{N_{s}}C_{\alpha\beta}+C_{\alpha e}\right)=\sum_{\beta}^{N_{s}}\widehat{C}_{\alpha\beta}+\widehat{C}_{\alpha e}.\label{eq:vfp_transformed_continuum}
\end{eqnarray}

\subsection{Discretization of the VFP equation with inertial terms\label{subsec:discretization_of_the_vfp_equation_wh_adaptivity}}

We discretize the VFP equation using finite volumes in a 1D planar
configuration space ($x$) and 2V cylindrical-velocity space ($\widehat{v}_{||}$
and $\widehat{v}_{\perp}$) with azimuthal symmetry. We compute the
discrete volume for cell (\emph{i,j,k}) as:
\[
\Delta V_{i,j,k}=2\pi\Delta x\Delta\widehat{v}_{||}\widehat{v}_{\perp,k}\Delta\widehat{v}_{\perp},
\]
where $\Delta x$, $\Delta\widehat{v}_{||}$, and $\Delta\widehat{v}_{\perp}$
are mesh spacings in the configuration space and the parallel- and
perpendicular-velocity space, respectively. For a uniform mesh (assumed
henceforth), we have:

\[
\Delta x=\frac{L_{x}}{N_{x}},\,\,\Delta\widehat{v}_{||}=\frac{L_{||}}{N_{||}},\,\,\Delta\widehat{v}_{\perp}=\frac{L_{\perp}}{N_{\perp}},
\]
where $L_{x}$, $L_{||}$, and $L_{\perp}$ are the configuration
space, and parallel and perpendicular velocity-space domain sizes,
respectively, and $N_{x}$, $N_{||}$, and $N_{\perp}$ are the corresponding
number of cells. The mesh is arranged such that cell faces map to
the domain boundary (and therefore outermost cell centers are half
a mesh-spacing away from the boundary). We define the distribution
function $f$ and the Rosenbluth potentials $H$, $G$ at cell centers. 

Velocity-space inner products are approximated via a mid-point quadrature
rule as
\begin{equation}
\left<A,B\right>_{\widehat{v}}\approx2\pi\sum_{j=1}^{N_{||}}\sum_{k=1}^{N_{\perp}}\widehat{v}_{\perp,k}\Delta\widehat{v}_{||}\Delta\widehat{v}_{\perp}A_{j,k}B{}_{j,k}\label{eq:fv_quadrature}
\end{equation}
for scalars and
\[
\left<\vec{A},\vec{B}\right>_{\widehat{v}}\approx2\pi\left[\sum_{j=0}^{N_{||}}\sum_{k=1}^{N_{\perp}}\widehat{v}_{\perp,k}\Delta\widehat{v}_{||}\Delta\widehat{v}_{\perp}A_{||,j+1/2,k}B{}_{||,j+1/2,k}+\sum_{j=1}^{N_{||}}\sum_{k=0}^{N_{\perp}}\widehat{v}_{\perp,k+1/2}\Delta\widehat{v}_{||}\Delta\widehat{v}_{\perp}A_{\perp,j,k+1/2}B{}_{\perp,j,k+1/2}\right]
\]
for vectors (with components defined at cell faces as denoted by the
half-integer indices $j+1/2$, $k+1/2$).

We discretize Eq. (\ref{eq:vfp_transformed_continuum}) in a conservative
form as

\begin{eqnarray}
\frac{c^{p+1}\widehat{f}_{\alpha,i,j,k}^{p+1}+c^{p}\widehat{f}_{\alpha,i,j,k}^{p}+c^{p-1}\widehat{f}_{\alpha,i,j,k}^{p-1}}{\Delta t}+\underbrace{\frac{F_{\alpha,i+1/2,j,k}^{p+1}-F_{\alpha,i-1/2,j,k}^{p+1}}{\Delta x}}_{\textcircled a}+\underbrace{\frac{J_{\alpha,acc,i,j+1/2,k}^{p+1}-J_{\alpha,acc,i,j-1/2,k}^{p+1}}{\Delta\widehat{v}_{||}}}_{\textcircled b}\nonumber \\
+\underbrace{\left[\frac{J_{||,\alpha,t,i,j+1/2,k}^{p+1}-J_{||,\alpha,t,i,j-1/2,k}^{p+1}}{\Delta\widehat{v}_{||}}+\frac{\widehat{v}_{\perp,k+1/2}J_{\perp,\alpha,t,i,j,k+1/2}^{p+1}-\widehat{v}_{\perp,k-1/2}J_{\perp,\alpha,t,i,j,k-1/2}^{p+1}}{\widehat{v}_{\perp,k}\Delta\widehat{v}_{\perp}}\right]}_{\textcircled c}\nonumber \\
+\underbrace{\left[\frac{J_{||,\alpha,x,i,j+1/2,k}^{p+1}-J_{||,\alpha,x,i,j-1/2,k}^{p+1}}{\Delta\widehat{v}_{||}}+\frac{\widehat{v}_{\perp,k+1/2}J_{\perp,\alpha,x,i,j,k+1/2}^{p+1}-\widehat{v}_{\perp,k-1/2}J_{\perp,\alpha,x,i,j,k-1/2}^{p+1}}{\widehat{v}_{\perp,k}\Delta\widehat{v}_{\perp}}\right]}_{\textcircled d}\nonumber \\
=\sum_{\beta}^{N_{s}}\left.C_{\alpha\beta}^{p+1}\right|_{i,j,k}+\left.C_{\alpha e}^{p+1}\right|_{i,j,k}.\label{eq:vfp_discrete_eqn}
\end{eqnarray}
Here, $c^{p+1}$, $c^{p}$, and $c^{p-1}$ are the coefficients for
the second-order backwards difference formula (BDF2) \citep{Taitano_2015_rfp_0d2v_implicit}
and $p$ is the discrete time index. 

The term $\textcircled a$ corresponds to the discretization of the
spatial streaming term, with

\[
F_{\alpha,i+1/2,j,k}^{p+1}=v_{\alpha,i+1/2}^{*,p}\widehat{v}_{||,j}\textnormal{Interp}\left(\widehat{v}_{||,j},\widehat{f}_{\alpha}^{p+1}\right)_{i+1/2,j,k},
\]

\[
v_{\alpha,i+1/2}^{*}=\frac{v_{\alpha,i+1}^{*}+v_{\alpha,i}^{*}}{2},
\]
where $\textnormal{Interp}\left(a,\phi\right)_{face}$ is an advection
interpolation operator of a scalar $\phi$ at a cell face based on
a given velocity $a$, which can be written in general as
\begin{equation}
\textnormal{Interp}\left(a,\phi\right)_{face}=\sum_{i'=1}^{N}\omega_{face,i'}\left(a,\phi\right)\phi_{i'}.\label{eq:face_interpolation_rule}
\end{equation}
The coefficients $\omega_{face,i'}$ are the interpolation weights
for the spatial cells $i'$ surrounding the cell face of interest
(in this study, they are determined by the SMART discretization \citep{smart}). 

The term $\textcircled b$ corresponds to the electrostatic-acceleration
term with

\begin{equation}
J_{\alpha,acc,i,j+1/2,k}^{p+1}={\cal A}_{\alpha,i}^{p+1}\textnormal{Interp}\left({\cal A}_{\alpha,i}^{p+1},\widehat{f}_{\alpha}^{p+1}\right)_{i,j+1/2,k},
\end{equation}
where
\[
{\cal A}_{\alpha,i}^{p+1}=\frac{q_{\alpha}}{m_{\alpha}}\frac{E_{||,i}^{p+1}}{v_{\alpha,i}^{*,p}}.
\]

The term $\textcircled c$ corresponds to the inertial term due to
temporal variation of the normalization velocity, $v_{\alpha}^{*},$
with

\[
J_{||,\alpha,t,i,j+1/2,k}^{p+1}={\cal I}_{\alpha,t,i}^{p}\widehat{v}_{||,j+1/2}\textnormal{Interp}\left({\cal I}_{\alpha,t,i}^{p}\widehat{v}_{||,j+1/2},\widehat{f}_{\alpha}^{p+1}\right)_{i,j+1/2,k}
\]
and

\[
J_{\perp,\alpha,t,i,j,k+1/2}^{p+1}={\cal I}_{\alpha,t,i}^{p}\widehat{v}_{\perp,k+1/2}\textnormal{Interp}\left({\cal I}_{\alpha,t,i}^{p}\widehat{v}_{\perp,k+1/2},\widehat{f}_{\alpha}^{p+1}\right)_{i,j,k+1/2},
\]
where

\begin{equation}
{\cal I}_{\alpha,t,i}^{p}=-\frac{\left(\partial_{t}v_{\alpha}^{*,p}\right)_{i}}{v_{\alpha,i}^{*,p}}\approx-\frac{c^{p+1}v_{\alpha,i}^{*,p}+c^{p}v_{\alpha,i}^{*,p-1}+c^{p-1}v_{\alpha,i}^{*,p-2}}{v_{\alpha,i}^{*,p}\Delta t}.
\end{equation}
We lag the time level between the BDF2 coefficients and the normalization
velocity for well-posedness of the velocity-space grid motion \citep{Taitano_2016_rfp_0d2v_implicit}. 

The term $\textcircled d$ corresponds to the inertial term due to
the spatial variation of the normalization velocity, $v_{\alpha}^{*}$,
with

\begin{equation}
J_{||,\alpha,x,i,j+1/2,k}^{p+1}={\cal I}_{\alpha,x,i}^{p}\widehat{v}_{||,j+1/2}^{2}\textnormal{Interp}\left({\cal I}_{\alpha,x,i}^{p},\widehat{f}_{\alpha}^{p+1}\right)_{i,j+1/2,k},
\end{equation}

\begin{equation}
J_{\perp,\alpha,x,i,j,k+1/2}^{p+1}={\cal I}_{\alpha,x,i}^{p}\widehat{v}_{||,j}\widehat{v}_{\perp,k+1/2}\textnormal{Interp}\left({\cal I}_{\alpha,x,i}^{p}\widehat{v}_{||,j},\widehat{f}_{\alpha}^{p+1}\right)_{i,j,k+1/2},
\end{equation}
where

\begin{equation}
{\cal I}_{\alpha,x,i}^{p}=-\left(\partial_{x}v_{\alpha}^{*,p}\right)_{i}\approx-\frac{v_{\alpha,i+1/2}^{*,p}-v_{\alpha,i-1/2}^{*,p}}{\Delta x}.
\end{equation}

Finally, the right-hand-side of Eq. (\ref{eq:vfp_discrete_eqn}) corresponds
to the Fokker-Planck-collision operator and its treatment is discussed
in detail in Refs. \citep{Taitano_2015_rfp_0d2v_implicit,Taitano_2016_rfp_0d2v_implicit}.
In this study, we use the mimetic differencing approach for the tensor
diffusion operator in the collision term proposed in Ref. \citep{lipnikov_rjnamm_2012}.

\subsection{Discretization of the electron temperature equation}

The electron temperature equation, Eq. (\ref{eq:1d_electron_equation}),
is also discretized using a finite-volume scheme in space and BDF2
in time:
\begin{eqnarray}
\frac{3}{2}\frac{c^{p+1}n_{e,i}^{p+1}T_{e,i}^{p+1}+c^{p}n_{e,i}^{p}T_{e,i}^{p}+c^{p-1}n_{e,i}^{p-1}T_{e,i}^{p-1}}{\Delta t^{k}}+\frac{5}{2}\frac{u_{||,e,i+1/2}^{p+1}\left(\widetilde{n_{e}T_{e}}\right)_{i+1/2}^{p+1}-u_{||,e,i-1/2}^{p+1}\left(\widetilde{n_{e}T_{e}}\right)_{i-1/2}^{p+1}}{\Delta x}+\nonumber \\
\frac{Q_{||,e,i+1/2}^{p+1}-Q_{||,e,i-1/2}^{p+1}}{\Delta x}-q_{e}\left[\left(n_{e}u_{||,e}\right)_{i}^{E}E_{||,i}\right]^{p+1}=3\nu_{e\alpha,i}^{p+1}\frac{m_{e}}{m_{\alpha}}n_{e,i}^{p+1}\left(T_{\alpha,i}^{p+1}-T_{e,i}^{p+1}\right)+F_{||,\alpha e,i}^{p+1}u_{||,\alpha,i}^{p+1}.\label{eq:discrete_electron_eqn}
\end{eqnarray}
Here the tilde denotes a cell-face discretization for the advection
quantities (SMART in this study). The quantity with a superscript
$E$ in Eq. (\ref{eq:discrete_electron_eqn}) are defined so as to
enforce conservation properties, and will be discussed shortly. Other
terms in Eq. (\ref{eq:discrete_electron_eqn}) are defined as:

\[
u_{||,e,i+1/2}^{p+1}=0.5\left(u_{||,e,i+1}^{p+1}+u_{||,e,i}^{p+1}\right),
\]

\begin{equation}
Q_{||,e,i+1/2}^{p+1}=\left[\beta_{0}n_{e}\left(u_{||,e}-\left\langle u_{||,\alpha}\right\rangle \right)\right]_{i+1/2}^{p+1}\widetilde{T}_{e,i+1/2}^{p+1}-\kappa_{||,e,i+1/2}^{p+1}\frac{T_{e,i+1}^{p+1}-T_{e,i}^{p+1}}{\Delta x},
\end{equation}
where
\begin{equation}
\left[\beta_{0}n_{e}\left(u_{||,e}-\left\langle u_{||,\alpha}\right\rangle \right)\right]_{i+1/2}^{p+1}=0.5\left\{ \left[\beta_{0}n_{e}\left(u_{||,e}-\left\langle u_{||,\alpha}\right\rangle \right)\right]_{i+1}^{p+1}+\left[\beta_{0}n_{e}\left(u_{||,e}-\left\langle u_{||,\alpha}\right\rangle \right)\right]_{i}^{p+1}\right\} ,
\end{equation}

\begin{equation}
\kappa_{||,e,i+1/2}^{p+1}=0.5\left[\kappa_{||,e,i+1}^{p+1}+\kappa_{||,e,i}^{p+1}\right],
\end{equation}

\begin{eqnarray}
F_{||,\alpha e,i}^{p+1} & = & -m_{e}n_{e,i}^{p+1}\nu_{e\alpha,i}^{p+1}\left(u_{||,\alpha,i}^{p+1}-\left\langle u_{||,\alpha,i}\right\rangle ^{p+1}\right)+\alpha_{0,i}^{p+1}m_{e}n_{e,i}^{p+1}\nu_{e\alpha,i}^{p+1}\left(u_{||,e,i}^{p+1}-\left\langle u_{||,\alpha,i}\right\rangle ^{p+1}\right)\nonumber \\
 &  & +\beta_{0,i}^{p+1}\frac{n_{e,i}^{p+1}\nu_{e\alpha,i}^{p+1}}{\sum_{\alpha}^{N_{s}}\nu_{e\alpha,i}^{p+1}}\frac{T_{e,i+1}^{p+1}-T_{e,i-1}^{p+1}}{2\Delta x}.
\end{eqnarray}

\subsection{Definitions for electron quantities to ensure simultaneous discrete
conservation of mass, momentum, and energy in the kinetic-ion/fluid-electron
system\label{subsec:simul_disc_mass_mom_ener_cons_for_kin_ion_fe_sys}}

In this section, we will obtain discrete expressions for the electron
density, the electron drift velocity, and the electric field that
ensure conservation of mass, momentum, and energy within the kinetic-ion/fluid-electron
system. 

In Sec. \ref{subsec:conservation_properties_for_ion_fluid_electron_system},
we proved that the kinetic-ion-fluid-electron system possesses continuum
conservation properties for mass, momentum, and energy. Here, we take
an approach similar to that discussed in Refs. \citep{Taitano_2015_rfp_0d2v_implicit}
and \citep{Taitano_2016_rfp_0d2v_implicit}, and introduce discrete
nonlinear constraints to enforce these properties in the ion-electron
collision operator, the electric field, and the Joule-heating term
(in the electron temperature equation). The ion-electron collision
operator is modified to become

\begin{equation}
\widehat{C}_{\alpha e}=\Gamma_{\alpha e}\widehat{\nabla}_{\widehat{v}}\cdot\left[\gamma_{G,\alpha e}\vec{J}_{G,\alpha e}-\frac{m_{\alpha}}{m_{e}}\left[\gamma_{H,\alpha e,u}\vec{J}_{H,\alpha e,u}+\gamma_{H,\alpha e,F}\vec{J}_{H,\alpha e,F}\right]\right].
\end{equation}
Here, 

\begin{equation}
\vec{J}_{G,\alpha e}=\tensor D_{\alpha e}\cdot\nabla_{v}f_{\alpha},
\end{equation}
\begin{equation}
\vec{J}_{H,\alpha e,u}=\vec{A}_{\alpha e,u}f_{\alpha},
\end{equation}
and

\begin{equation}
\vec{J}_{H,\alpha e,F}=\vec{A}_{\alpha e,F}f_{\alpha}
\end{equation}
are the collisional-velocity-space fluxes and $\gamma_{G,\alpha e}$,
$\gamma_{H,\alpha e,u}$, and $\gamma_{H,\alpha e,F}$ are the nonlinear-constraint
functions that ensure discrete conservation of momentum and energy
for collisions between kinetic ions and fluid electrons (App. \ref{app:details_conservation_strat_col_btw_kin_ion_fluid_e}).

In order to ensure discrete global momentum conservation between
kinetic ions and fluid electrons, we require the following relationship
in the continuum (Sec. \ref{subsec:conservation_properties_for_ion_fluid_electron_system}):

\[
\sum_{\alpha}\left[q_{\alpha}E_{||}\left\langle v_{||},\partial_{v_{||}}f\right\rangle _{v}-m_{\alpha}\left\langle v_{||},C_{\alpha e}\right\rangle _{v}\right]=\partial_{x}P_{e}.
\]
We specialize this expression at cell centers $i$ as:
\[
\sum_{\alpha}\left[q_{\alpha}E_{||}\left\langle v_{||},\partial_{v_{||}}f\right\rangle _{v}-m_{\alpha}\left\langle v_{||},C_{\alpha e}\right\rangle _{v}\right]_{i}=\left.\partial_{x}P_{e}\right|_{i}.
\]
The first term in the left-hand side gives:
\[
\left\langle v_{||},\partial_{v_{||}}f\right\rangle _{v,i}=-n_{\alpha,i}^{E},
\]
where $n_{\alpha,i}^{E}$ is a density computed by integration by
parts, but which accounts for the sign of the electric field in the
original advection operator as:
\begin{equation}
n_{\alpha,i}^{E}=\sum_{j,k}\Delta\widetilde{V}_{j+1/2,k}\textnormal{Interp}\left(E_{\parallel,i}^{p+1},\widehat{f}_{\alpha}^{p+1}\right)_{i,j+1/2,k}.\label{eq:nstar_momentum_cons}
\end{equation}
Here, $\sum_{j,k}\equiv\sum_{j}\sum_{k}$, and $\Delta\widetilde{V}_{j+1/2,k}=2\pi v_{\perp,k}\Delta v_{\perp}\Delta v_{||}$.
We define a corresponding electron density by quasineutrality as:
\begin{equation}
n_{e,i}^{E}=\sum_{\alpha}^{N_{s}}Z_{\alpha}n_{\alpha,i}^{E}.
\end{equation}
The second term in the left-hand side gives:
\[
\sum_{\alpha}\left[m_{\alpha}\left\langle v_{||},C_{\alpha e}\right\rangle _{v}\right]_{i}=\left.\sum_{\alpha}F_{\alpha e,||}\right|_{i}
\]
when the associated ion-electron collision operator symmetries are
satisfied (App. \ref{app:details_conservation_strat_col_btw_kin_ion_fluid_e}).
There results the following definition of the discrete electric field
at spatial cell index $i$:

\begin{equation}
E_{||,i}=\frac{\left(\partial_{x}P_{e}+\sum_{\alpha}F_{\alpha e}\right)_{i}}{q_{e}n_{e,i}^{E}}\label{eq:E_parallel_star}
\end{equation}
This result ensures momentum conservation for the kinetic-ion/fluid-electron
system.

To ensure energy conservation for the kinetic-ion/fluid-electron system,
we require the following relationship in the continuum (which we specialize
at the cell $i$):

\begin{equation}
(n_{e}u_{e,||})_{i}^{E}=-\sum_{\alpha}\left[Z_{\alpha}\left\langle \frac{v^{2}}{2},\partial_{v_{||}}f\right\rangle _{v,i}\right].
\end{equation}
We achieve this discretely as before by computing an electric-field-aware
momentum moment as:

\begin{equation}
-\left\langle \frac{v^{2}}{2},\left(\partial_{v_{||}}f\right)\right\rangle _{v,i}=\sum_{j,k}\Delta\widetilde{V}_{j+1/2,k}v_{||,j+1/2}\textnormal{Interp}\left(E_{\parallel,i}^{p+1},\widehat{f}_{\alpha}^{p+1}\right)_{i,j+1/2,k},\label{eq:nu_star_energy_conservation}
\end{equation}
and compute the fluid-electron Joule-heating term in Eq. (\ref{eq:1d_electron_equation})
as:

\begin{equation}
\left(n_{e}u_{e}\right)_{i}^{E}=\sum_{\alpha}Z_{\alpha}\left(n_{\alpha}u_{||,\alpha}\right)_{i}^{E}.\label{eq:modified_ion_acceleration_energy_moment}
\end{equation}
To summarize, we have defined $n_{e,i}^{E}$, $\left(n_{e}u_{e}\right)_{i}^{E}$,
and $E_{||,i}$ as given by Eqs. (\ref{eq:nstar_momentum_cons}),
(\ref{eq:modified_ion_acceleration_energy_moment}), (\ref{eq:E_parallel_star})
to ensure conservation of momentum and energy within the kinetic-ions/fluid-electron
system. We point out that Eqs. (\ref{eq:E_parallel_star}) and (\ref{eq:modified_ion_acceleration_energy_moment})
are the key innovations in this section.\textcolor{red}{{} }

\subsection{Discretization of ion Vlasov component: exact conservation properties\label{subsec:disc_of_ion_vlasov_comp:exact_cons_properties}}

This section describes the procedure to ensure the set of exact conservation
symmetries of the Vlasov piece in the ion kinetic equation in the
presence of velocity-space grid adaptivity. We begin by developing
separate discretizations for mass, momentum, and energy conservation
in a periodic spatial domain \textit{without} any background field.
In this, we follow a procedure almost identical to the 0D2V case \citep{Taitano_2016_rfp_0d2v_implicit}.
We continue by developing a simultaneous mass and momentum conserving
discretization, and a simultaneous mass and energy conserving discretization.
Finally, we combine all the conservation properties. We remark, that
the detailed derivations of conservation symmetries for the temporal
terms in the Vlasov equation and for the collision operator have been
considered elsewhere \citep{Taitano_2015_rfp_0d2v_implicit,Taitano_2016_rfp_0d2v_implicit},
with a more numerically robust generalization based on a constrained-minimization
approach discussed in App. \ref{app:robust_generalizaiton_of_discrete_cons_scheme_for_col_op},
and, therefore, only the spatial gradient terms are considered here.

\subsubsection{Mass conservation\label{subsubsec:Exact-mass-conservation}}

Consider the spatial gradient terms in the Vlasov equation, (\ref{eq:vfp_transformed_continuum})
:

\begin{equation}
\partial_{x}\left(v^{*}\widehat{v}_{||}\widehat{f}\right)-\partial_{x}v^{*}\widehat{\nabla}_{\widehat{v}}\cdot\left(\vec{\widehat{v}}\widehat{v}_{||}\widehat{f}\right).\label{eq:continuum_spatial_inertial_vlasov}
\end{equation}
Mass conservation is revealed by taking the $m\widehat{v}{}^{0}$
moment to find
\begin{equation}
\partial_{x}\left(v^{*}mn\hat{u}_{||}\right),
\end{equation}
which is in a conservative form. Here, $n\hat{u}_{||}=\left\langle 1,\widehat{v}_{||}\widehat{f}\right\rangle _{\vec{\widehat{v}}}$.
Note that the second term in the expression (\ref{eq:continuum_spatial_inertial_vlasov})
is in a divergence form in velocity space, and therefore its $m\widehat{v}^{0}$
moment trivially vanishes both continuously and discretely.

\subsubsection{Momentum conservation\label{subsubsec:Exact-momentum-conservation}}

Similarly to Ref. \citep{Taitano_2016_rfp_0d2v_implicit}, we re-write
the expression in (\ref{eq:continuum_spatial_inertial_vlasov}) by
multiplying by $v^{*}$ and using the chain rule to obtain

\begin{equation}
\partial_{x}\left(\left(v^{*}\right)^{2}\widehat{v}_{||}f\right)-\partial_{x}v^{*}\left[v^{*}\widehat{v}_{||}f+\widehat{\nabla}_{\widehat{v}}\cdot\left(\vec{\widehat{v}}v^{*}\widehat{v}_{||}f\right)\right].\label{eq:spatial_vlasov_momentum_conserving_form}
\end{equation}
By taking the $m\widehat{v}_{||}$ moment, and noting that $\left\langle v_{||},v^{*}\widehat{v}_{||}\widehat{f}+\widehat{\nabla}_{\widehat{v}}\cdot\left(\vec{\widehat{v}}v^{*}\widehat{v}_{||}\widehat{f}\right)\right\rangle _{\vec{\widehat{v}}}=0$,
we obtain,

\begin{equation}
m\left\langle \widehat{v}_{||},\partial_{x}\left(\left(v^{*}\right)^{2}\widehat{v}_{||}\widehat{f}\right)-\partial_{x}v^{*}\left[v^{*}\widehat{v}_{||}\widehat{f}+\widehat{\nabla}_{\widehat{v}}\cdot\left(\vec{\widehat{v}}v^{*}\widehat{v}_{||}\widehat{f}\right)\right]\right\rangle _{\widehat{v}}=\partial_{x}\left[m\left(v^{*}\right)^{2}\widehat{S}_{2,||}\right],
\end{equation}
which again is in a conservative form. Here, $\widehat{S}_{2,||}=\left\langle \widehat{v}_{||}^{2},\widehat{f}\right\rangle _{\widehat{v}}$. 

It follows that the key requirement for the momentum conservation
is to have the $m\widehat{v}_{||}$ moment of $v^{*}\widehat{v}_{||}\widehat{f}+\widehat{\nabla}_{\widehat{v}}\cdot\left(\vec{\widehat{v}}v^{*}\widehat{v}_{||}\widehat{f}\right)$
be zero, which is generally not true discretely. In order to enforce
this property, we modify expression (\ref{eq:spatial_vlasov_momentum_conserving_form})
by introducing a constraint function, $\Upsilon_{x}$, as follows:

\begin{equation}
\partial_{x}\left(\left(v^{*}\right)^{2}\widehat{v}_{||}\widehat{f}\right)-\partial_{x}v^{*}\left[v^{*}\widehat{v}_{||}\widehat{f}+\widehat{\nabla}_{\widehat{v}}\cdot\left(\Upsilon_{x}\vec{\widehat{v}}v^{*}\widehat{v}_{||}\widehat{f}\right)\right],\label{eq:spatial_vlasov_discrete_mom_cons}
\end{equation}
where

\begin{equation}
\Upsilon_{x}=1+\sum_{l=0}^{N}C_{l}^{\Upsilon}P_{l}^{\Upsilon}\left(\widehat{v}_{||},\widehat{v}_{\perp}\right).\label{eq:upsilon_momentum_conservation}
\end{equation}
Here, $C_{l}^{\Upsilon}$ is the constraint coefficient for the basis
function $P_{l}^{\Upsilon}$, which is obtained by solving a constrained-minimization
problem for the following objective function:
\begin{equation}
F\left(\vec{C}^{\Upsilon},\lambda\right)=\frac{1}{2}\sum_{l=0}^{N}\left(C_{l}^{\Upsilon}\right)^{2}-\lambda\left(\left\langle \widehat{v}_{||},v^{*}\widehat{v}_{||}\widehat{f}\right\rangle _{\vec{\widehat{v}}}+\left\langle \widehat{v}_{||},\widehat{\nabla}_{\widehat{v}}\cdot\left(\left[1+\sum_{l=0}^{N}C_{l}^{\Upsilon}P_{l}^{\Upsilon}\right]\vec{\widehat{v}}v^{*}\widehat{v}_{||}\widehat{f}\right)\right\rangle _{\vec{\widehat{v}}}\right),
\end{equation}
where $\lambda$ is a Lagrange multiplier, $\vec{C}^{\Upsilon}$ is
a vector of the constraint coefficients, and $P^{\Upsilon}$ is the
constraint basis. The particular choice of basis functions in this
study is described in App. \ref{app:details_conservation_strat_col_btw_kin_ion_fluid_e}.
We remark that this minimization procedure is a generalization of
the conservation strategy in Ref. \citep{Taitano_2016_rfp_0d2v_implicit}.

\subsubsection{Energy conservation\label{subsubsec:Exact-energy-conservation}}

As before, we re-write the conservation equation by multiplying Eq.
(\ref{eq:continuum_spatial_inertial_vlasov}) by $\left(v^{*}\right)^{2}$
and using the chain rule to cast it into the energy-conserving form: 

\begin{equation}
\partial_{x}\left(\left(v^{*}\right)^{3}\widehat{v}_{||}\widehat{f}\right)-\partial_{x}\left(v^{*}\right)^{2}\left[v^{*}\widehat{v}_{||}\widehat{f}+\frac{\widehat{\nabla}_{\widehat{v}}}{2}\cdot\left(\vec{\widehat{v}}v^{*}\widehat{v}_{||}\widehat{f}\right)\right].\label{eq:spatial_vlasov_energy_conserving_form}
\end{equation}
Taking the $m\widehat{v}^{2}/2$ moment of this expression and noting
that $\left\langle \frac{\widehat{v}^{2}}{2},v^{*}\widehat{v}_{||}\widehat{f}+\frac{\widehat{\nabla}_{\widehat{v}}}{2}\cdot\left(\widehat{\vec{v}}v^{*}\widehat{v}_{||}\widehat{f}\right)\right\rangle _{\vec{v}}=0$,
we find: 

\begin{equation}
\frac{m}{2}\left\langle \widehat{v}^{2},\partial_{x}\left(\left(v^{*}\right)^{3}\widehat{v}_{||}\widehat{f}\right)-\partial_{x}\left(v^{*}\right)^{2}\left[v^{*}\widehat{v}_{||}\widehat{f}+\frac{\widehat{\nabla}_{\widehat{v}}}{2}\cdot\left(\vec{\widehat{v}}v^{*}\widehat{v}_{||}\widehat{f}\right)\right]\right\rangle _{\vec{\widehat{v}}}=\partial_{x}\left(m\left(v^{*}\right)^{3}\widehat{S}_{3,||}\right),
\end{equation}
which is again in a conservative form. Here, $\widehat{S}_{3,||}=\frac{1}{2}\left\langle \widehat{v}^{2},\widehat{v}_{||}\widehat{f}\right\rangle _{\vec{\widehat{v}}}$.

The key requirement for a discrete energy conservation is to have
the $m\widehat{v}^{2}/2$ moment of the quantity $v^{*}\widehat{v}_{||}\widehat{f}+\frac{\widehat{\nabla}_{\widehat{}v}}{2}\cdot\left(\vec{\widehat{v}}v^{*}\widehat{v}_{||}\widehat{f}\right)$
cancel discretely. As before, in order to enforce this constraint,
we modify expression (\ref{eq:spatial_vlasov_energy_conserving_form})
by introducing a constraint function, $\gamma_{x}$, as follows:

\begin{equation}
\partial_{x}\left[\left(v^{*}\right)^{3}\widehat{v}_{||}\widehat{f}\right]-\partial_{x}\left(v^{*}\right)^{2}\left[v^{*}\widehat{v}_{||}\widehat{f}+\frac{\widehat{\nabla}_{\widehat{v}}}{2}\cdot\left(\gamma_{x}\vec{\widehat{v}}v^{*}\widehat{v}_{||}\widehat{f}\right)\right],\label{eq:vlasov_spatial_inertial_energy_cons_constraint}
\end{equation}
where:

\begin{equation}
\gamma_{x}=1+\sum_{l=0}^{N}C_{l}^{\gamma}P_{l}^{\gamma}\left(\widehat{v}_{||},\widehat{v}_{\perp}\right)\label{eq:gamma_energy_conservation}
\end{equation}
with similar definitions for $C_{l}^{\Upsilon}$ and $P_{l}^{\Upsilon}$
as in the momentum-only conserving formulation. The coefficients $C_{l}^{\gamma}$
are obtained in an manner identical to $C_{l}^{\Upsilon}$, but with
a different constraint, Eq. (\ref{eq:vlasov_spatial_inertial_energy_cons_constraint}). 

\subsubsection{Simultaneous conservation of mass and momentum\label{subsubsec:Exact-mass-momentum-conservation1}}

Next, we obtain a discretization that simultaneously enforces mass
and momentum conservation. The conservation scheme is developed by
recursively applying the chain rule discussed earlier {[}expressions
(\ref{eq:spatial_vlasov_momentum_conserving_form}) and (\ref{eq:continuum_spatial_inertial_vlasov}){]}.
The recursive application follows a procedure similar to that outlined
in Ref. \citep{Taitano_2016_rfp_0d2v_implicit} for the temporal terms. 

For the spatial terms, we employ the following transformation to derive
the momentum-conserving form (\ref{eq:spatial_vlasov_momentum_conserving_form})
from the mass-conserving form (\ref{eq:continuum_spatial_inertial_vlasov})
of the spatial-gradient terms in the Vlasov equation: 
\begin{equation}
\left(v^{*}\right)^{2}\partial_{x}\left(v^{*}\widehat{v}_{||}\widehat{f}\right)-\frac{\partial_{x}\left(v^{*}\right)^{2}}{2}\widehat{\nabla}_{\widehat{v}}\cdot\left(v^{*}\widehat{v}_{||}\vec{v}\widehat{f}\right)=v^{*}\left\{ \partial_{x}\left(\left(v^{*}\right)^{2}\widehat{v}_{||}\widehat{f}\right)-\partial_{x}v^{*}\left[v^{*}\widehat{v}_{||}\widehat{f}+\widehat{\nabla}_{\widehat{v}}\cdot\left(v^{*}\vec{\widehat{v}}\widehat{v}_{||}\widehat{f}\right)\right]\right\} .\label{eq:spatial_inertial_momentum_cons_chain_rule_1}
\end{equation}
This exact relationship is not enforced discretely due to truncation
errors, leading to a momentum-conservation error when using the mass-conserving
form and vise-versa, i.e.,
\begin{equation}
\eta_{x}=\left(v^{*}\right)^{2}\partial_{x}\left(v^{*}\widehat{v}_{||}\widehat{f}\right)-\frac{\partial_{x}\left(v^{*}\right)^{2}}{2}\widehat{\nabla}_{\widehat{v}}\cdot\left(v^{*}\widehat{v}_{||}\vec{v}\widehat{f}\right)-v^{*}\left\{ \partial_{x}\left(\left(v^{*}\right)^{2}\widehat{v}_{||}\widehat{f}\right)-\partial_{x}v^{*}\left[v^{*}\widehat{v}_{||}\widehat{f}+\widehat{\nabla}_{\widehat{v}}\cdot\left(v^{*}\vec{\widehat{v}}\widehat{v}_{||}\widehat{f}\right)\right]\right\} \neq0.\label{eqn:eta_x_alpha-1}
\end{equation}

In order to account for truncation errors in the chain rule, we modify
the momentum-conserving form of the spatial-gradient terms in the
Vlasov equation (\ref{eq:spatial_vlasov_discrete_mom_cons}) to become,

\begin{equation}
v^{*}\left\{ \partial_{x}\left(\left(v^{*}\right)^{2}\widehat{v}_{||}\widehat{f}\right)-\partial_{x}v^{*}\left[v^{*}\widehat{v}_{||}\widehat{f}+\widehat{\nabla}_{\widehat{v}}\cdot\left(\Upsilon_{x}\vec{\widehat{v}}v^{*}\widehat{v}_{||}\widehat{f}\right)\right]\right\} +\eta_{x}.\label{eq:spatial_inertial_mass_momentum_cons_form}
\end{equation}
and has the role of enforcing the discrete chain rule on the spatial
quantities ($\eta_{x}=0$ in the continuum). In particular, with $\Upsilon_{x}$
defined as in Eq. (\ref{eq:upsilon_momentum_conservation}), momentum
conservation requires $\int_{L_{x,min}}^{L_{x,max}}\left\langle \widehat{v}_{||},\eta_{x}\right\rangle _{\vec{\widehat{v}}}dx=0$
discretely, where $L_{x,min}$ and $L_{x,max}$ are the bounds on
the domain. We achieve this by a careful discretization of the various
spatial gradients in Eq. (\ref{eqn:eta_x_alpha-1}), as we discuss
next. We begin by rewriting
\begin{eqnarray}
\left\langle \widehat{v}_{||},\eta_{x}\right\rangle _{\vec{\widehat{v}}}=\left\{ \left(v^{*}\right)^{2}\partial_{x}\left(v^{*}\widehat{S}_{2,||}\right)\right\} -v^{*}\left\{ \partial_{x}\left[\left(v^{*}\right)^{2}\widehat{S}_{2,||}\right]-\partial_{x}v^{*}\left[v^{*}\widehat{S}_{2,||}\right]\right\} .
\end{eqnarray}
Dividing by $v^{*}\neq0$, we obtain for the right-hand side

\begin{equation}
\underbrace{v^{*}\partial_{x}\left(v^{*}\widehat{S}_{2,||}\right)}_{\textcircled a}=\underbrace{\partial_{x}\left[\left(v^{*}\right)^{2}\widehat{S}_{2,||}\right]}_{\textcircled b}-\underbrace{\partial_{x}v^{*}\left[v^{*}\widehat{S}_{2,||}\right]}_{\textcircled c}.
\end{equation}
We discretize the individual terms as follows:

\begin{equation}
\textcircled a\approx v_{i}^{*}\frac{v_{i+1/2}^{*}\widehat{S}_{2,||,i+1/2}-v_{i-1/2}^{*}\widehat{S}_{2,||,i-1/2}}{\Delta x},\label{eq:discrete_mass_mom_cons_term_a}
\end{equation}

\begin{equation}
\textcircled b\approx\frac{\left(v_{i+1/2}^{*}\right)^{2}\widehat{S}_{2,||,i+1/2}-\left(v_{i-1/2}^{*}\right)^{2}\widehat{S}_{2,||,i-1/2}}{\Delta x},\label{eq:discrete_mass_mom_cons_term_b}
\end{equation}
and

\begin{equation}
\textcircled c\approx\frac{1}{2}\left\{ \frac{v_{i+1}^{*}-v_{i}^{*}}{\Delta x}v_{i+1/2}^{*}\widehat{S}_{2,||,i+1/2}+\frac{v_{i}^{*}-v_{i-1}^{*}}{\Delta x}v_{i-1/2}^{*}\widehat{S}_{2,||,i-1/2}\right\} .\label{eq:discrete_mass_mom_cons_term_c}
\end{equation}
Here, $v_{i+1/2}^{*}=\frac{1}{2}\left(v_{i+1}^{*}+v_{i}^{*}\right)$
and $v_{i+1/2}^{*}\widehat{S}_{2,||,i+1/2}=\left\langle \widehat{v}_{||},F_{i+1/2}\right\rangle _{\vec{\widehat{v}}}$
with $F_{i+1/2}$ the configuration-space cell-face discretization
of the streaming operator {[}term $\textcircled a$ in Eq. (\ref{eq:vfp_discrete_eqn}){]}. 

Local mass conservation can be shown by substituting $\eta_{x}$,
Eq. (\ref{eqn:eta_x_alpha-1}), into expression (\ref{eq:spatial_inertial_mass_momentum_cons_form})
and dividing by $\left(v^{*}\right)^{2}$ to find: 
\begin{equation}
\partial_{x}\left(v^{*}\widehat{v}_{||}\widehat{f}\right)-\widehat{\nabla}_{\widehat{v}}\cdot\left(\left[\frac{1}{2}\left(v^{*}\right)^{-1}\partial_{x}\left(v^{*}\right)^{2}+\partial_{x}v^{*}\left(\Upsilon_{x}-1\right)\right]\vec{\widehat{v}}\widehat{v}_{||}\widehat{f}\right).
\end{equation}
The equation is in a conservative form, guaranteeing that mass is
locally conserved when evaluating the zeroth velocity moment. We show
in App. \ref{app:Discrete-mass-and-momentum-conservation-proof} that
this formulation also leads to the discrete momentum conservation.
We note, that we introduced $\eta_{x}$ simply to \emph{expose} analytically
the conservation symmetry for $\Upsilon_{x}$. In practice, $\eta_{x}$
is explicitly replaced in expression (\ref{eq:spatial_inertial_mass_momentum_cons_form})
and the resulting expression is simplified as much as possible. 

\subsubsection{Simultaneous conservation of mass and energy\label{subsubsec:Exact-mass-energy-conservation}}

Next, we obtain a discretization to simultaneously conserve mass and
energy. Similarly to the simultaneous mass- and momentum-conserving
scheme, we use the chain rule for spatial terms to derive the energy-conserving
form (\ref{eq:spatial_vlasov_energy_conserving_form}) starting from
the mass-conserving form (\ref{eq:continuum_spatial_inertial_vlasov})
of the spatial-gradient terms in the Vlasov equation. We obtain 
\begin{eqnarray}
\xi_{x} & = & \left\{ \left(v^{*}\right)^{2}\partial_{x}\left(v^{*}\widehat{v}_{||}\widehat{f}\right)-\partial_{x}\left(v^{*}\right)^{2}\frac{\widehat{\nabla}_{\widehat{v}}}{2}\cdot\left(\vec{\widehat{v}}v^{*}\widehat{v}_{||}\widehat{f}\right)\right\} \nonumber \\
 & - & \left\{ \partial_{x}\left(\left(v^{*}\right)^{3}\widehat{v}_{||}\widehat{f}\right)-\partial_{x}\left(v^{*}\right)^{2}\left[v^{*}\widehat{v}_{||}\widehat{f}+\frac{\widehat{\nabla}_{\widehat{v}}}{2}\cdot\left(\vec{\widehat{v}}v^{*}\widehat{v}_{||}\widehat{f}\right)\right]\right\} .\label{eqn:xi_x_alpha-1}
\end{eqnarray}
As before, this relationship is not enforced discretely due to a truncation
error, leading to an energy conservation error when using the mass-conserving
form and vise-versa. In order to simultaneously remove these truncation
errors, we modify the energy-conserving form (\ref{eq:vlasov_spatial_inertial_energy_cons_constraint})
of the relevant terms in the Vlasov equation to become
\begin{equation}
\left\{ \partial_{x}\left(\left(v^{*}\right)^{3}\widehat{v}_{||}\widehat{f}\right)-\partial_{x}\left(v^{*}\right)^{2}\left[v^{*}\widehat{v}_{||}\widehat{f}+\frac{\widehat{\nabla}_{\widehat{v}}}{2}\cdot\left(\gamma_{x}\vec{\widehat{v}}v^{*}\widehat{v}_{||}\widehat{f}\right)\right]\right\} +\xi_{x}.\label{eq:spatial_inertial_mass_energy_cons_form}
\end{equation}
With $\gamma_{x}$ defined as in Eq. (\ref{eq:gamma_energy_conservation}),
energy conservation requires $\int_{L_{x,min}}^{L_{x,max}}\left\langle \frac{\widehat{v}^{2}}{2},\xi_{x}\right\rangle _{\vec{\widehat{v}}}dx=0$
discretely. As before, we achieve this by careful discretization of
spatial gradient terms, as we show next. We begin by rewriting
\begin{equation}
\left\langle \frac{\widehat{v}^{2}}{2},\xi_{x}\right\rangle _{\vec{\widehat{v}}}=\underbrace{\left(v^{*}\right)^{2}\partial_{x}\left(v^{*}\widehat{S}_{3,||}\right)}_{\textcircled a}-\underbrace{\partial_{x}\left(\left(v^{*}\right)^{3}\widehat{S}_{3,||}\right)}_{\textcircled b}+\underbrace{\partial_{x}\left(v^{*}\right)^{2}\left[v^{*}\widehat{S}_{3,||}\right]}_{\textcircled c}.
\end{equation}
We discretize the individual terms as follows:

\begin{equation}
\textcircled a\approx\left(v_{i}^{*}\right)^{2}\frac{v_{i+1/2}^{*}\widehat{S}_{3,||,i+1/2}-v_{i-1/2}^{*}\widehat{S}_{3,||,i-1/2}}{\Delta x},\label{eq:discrete_mass_ener_cons_term_a}
\end{equation}

\begin{equation}
\textcircled b\approx\frac{\left(v_{i+1/2}^{*}\right)^{3}\widehat{S}_{3,||,i+1/2}-\left(v_{i-1/2}^{*}\right)^{3}\widehat{S}_{3,||,i-1/2}}{\Delta x},\label{eq:discrete_mass_ener_cons_term_b}
\end{equation}
and

\begin{equation}
\textcircled c\approx\frac{v_{i+1/2}^{*}}{2}\left\{ \frac{\left(v_{i+1}^{*}\right)^{2}-\left(v_{i}^{*}\right)^{2}}{\Delta x}\widehat{S}_{3,||,i+1/2}+\frac{\left(v_{i}^{*}\right)^{2}-\left(v_{i-1}^{*}\right)^{2}}{\Delta x}\widehat{S}_{3,||,i-1/2}\right\} ,\label{eq:discrete_mass_ener_cons_term_c}
\end{equation}
where $v_{i+1/2}^{*}\widehat{S}_{3,||,i+1/2}=\left\langle \frac{\widehat{v}^{2}}{2},F_{i+1/2}\right\rangle _{\vec{\widehat{v}}}$. 

As before, local-mass conservation can be shown by substituting $\xi_{x}$,
Eq. (\ref{eqn:xi_x_alpha-1}), into expression (\ref{eq:spatial_inertial_mass_energy_cons_form})
and dividing by $\left(v^{*}\right)^{2}\neq0$ to find: 
\begin{equation}
\partial_{x}\left(v^{*}\widehat{v}_{||}\widehat{f}\right)-\widehat{\nabla}_{\widehat{v}}\cdot\left(\left[\left(v^{*}\right)^{-1}\partial_{x}\left(v^{*}\right)^{2}\left(\gamma_{x}-\frac{1}{2}\right)\right]\vec{\widehat{v}}\widehat{v}_{||}\widehat{f}\right).
\end{equation}
The equation is in a purely conservative form, guaranteeing that mass
is locally conserved when taking the zeroth velocity moment. We give
in App. \ref{app:Discrete-mass-and-energy-conservation-proof} a proof
of energy conservation for the above discretizations. We note that,
similarly to $\eta_{x}$, we introduced $\xi_{x}$ simply to \emph{expose
analytically} the conservation symmetry for $\gamma_{x}$. In practice,
it is substituted into expression (\ref{eq:spatial_inertial_mass_energy_cons_form}),
and is not explicitly computed. 

\subsubsection{Simultaneous conservation of mass, momentum, and energy\label{subsubsec:Exact-mass-momentum-energy-conservation}}

Finally, we combine the previous ideas to develop a simultaneously
mass-, momentum-, and energy-conserving discretization scheme. As
before, the idea is to correct for chain-rule discretization errors.
We begin by modifying the energy-conserving form of the relevant terms
in the Vlasov equation (\ref{eq:spatial_vlasov_energy_conserving_form})
to become 
\begin{eqnarray}
\partial_{x}\left(\left(v^{*}\right)^{3}\widehat{v}_{||}\widehat{f}\right)-\partial_{x}\left(v^{*}\right)^{2}\left[v^{*}\widehat{v}_{||}\widehat{f}+\frac{\widehat{\nabla}_{\widehat{v}}}{2}\cdot\left(\gamma_{x}\vec{\widehat{v}}v^{*}\widehat{v}_{||}\widehat{f}\right)\right]+\xi_{x}^{*}.\label{eq:vlasov_spatial_inertial_full_cons_1st_eq}
\end{eqnarray}
Here, $\xi_{x}^{*}$ enforces the discrete-chain rule on the spatial
quantities: 
\begin{eqnarray}
\xi_{x}^{*}=\left\{ v^{*}\partial_{x}\left(\left(v^{*}\right)^{2}\widehat{v}_{||}\widehat{f}\right)-v^{*}\partial_{x}v^{*}\left[v^{*}\widehat{v}_{||}\widehat{f}+\widehat{\nabla}_{\widehat{v}}\cdot\left(\Upsilon_{x}\vec{\widehat{v}}v^{*}\widehat{v}_{||}\widehat{f}\right)\right]\right\} -\nonumber \\
\left\{ \partial_{x}\left(\left(v^{*}\right)^{3}\widehat{v}_{||}\widehat{f}\right)-\partial_{x}\left(v^{*}\right)^{2}\left[v^{*}\widehat{v}_{||}\widehat{f}+\frac{\widehat{\nabla}_{v}}{2}\cdot\left(\zeta_{x}\vec{\widehat{v}}v^{*}\widehat{v}_{||}\widehat{f}\right)\right]\right\} +\eta_{x}
\end{eqnarray}
where $\eta_{x}$ is defined in Eq. (\ref{eqn:eta_x_alpha-1}), and
$\gamma_{x}$ and $\Upsilon_{x}$ are defined in Eqs. (\ref{eq:upsilon_momentum_conservation}),
(\ref{eq:gamma_energy_conservation}). A new conservation constraint
coefficient, $\zeta_{x}$, has been introduced in the definition of
$\xi_{x}^{*}$ to ensure simultaneous conservation of mass, momentum,
and energy, and is defined as:

\begin{equation}
\zeta_{x}=1+\sum_{l}C_{l}^{\zeta}P_{l}^{\zeta}\left(\widehat{v}_{||},\widehat{v}_{\perp}\right).
\end{equation}
Here, $P_{l}^{\zeta}$ is the constraint function and $C_{l}^{\zeta}$
is the corresponding coefficient, which is obtained by solving a constrained-minimization
problem for the following objective function:

\begin{equation}
F\left(\vec{C}^{\zeta},\vec{\lambda}\right)=\frac{1}{2}\sum_{l=0}^{N}\left(C_{l}^{\zeta}\right)^{2}-\vec{\lambda}\cdot\left(\vec{S}-\vec{M}\right)
\end{equation}
where

\begin{equation}
\vec{S}=\begin{bmatrix}\left\langle \widehat{v}_{||},\widehat{\nabla}_{\widehat{v}}\cdot\left(\gamma_{x}\vec{\widehat{v}}v^{*}\widehat{v}_{||}\widehat{f}\right)\right\rangle _{\vec{\widehat{v}}}\\
v^{*}\partial_{x}v^{*}\left\langle \frac{\widehat{v}^{2}}{2},\widehat{\nabla}_{\widehat{v}}\cdot\left(\Upsilon_{x}\vec{\widehat{v}}v^{*}\widehat{v}_{||}\widehat{f}\right)\right\rangle _{\vec{\widehat{v}}}
\end{bmatrix},\;\vec{M}=\begin{bmatrix}\left\langle \widehat{v}_{||},\widehat{\nabla}_{\widehat{v}}\cdot\left(\zeta_{x}\vec{\widehat{v}}v^{*}\widehat{v}_{||}\widehat{f}\right)\right\rangle _{\vec{\widehat{v}}}\\
\frac{\partial_{x}\left(v^{*}\right)^{2}}{2}\left\langle \frac{\widehat{v}^{2}}{2},\widehat{\nabla}_{\widehat{v}}\cdot\left(\zeta_{x}\vec{\widehat{v}}v^{*}\widehat{v}_{||}\widehat{f}\right)\right\rangle _{\vec{\widehat{v}}}
\end{bmatrix}.\label{eq:vlasov_spatial_inertial_full_cons_last_eq}
\end{equation}
We stress that $\eta_{x}$ and $\xi_{x}^{*}$ are used only to expose
analytically the conservation symmetries and are not explicitly computed.

\subsection{Evolution strategy of $v^{*}$\label{subsec:Evolution-equation-of-vstar}}

We discuss next the temporal and spatial update strategy for the normalization
velocity, $v^{*}$, used to transform the Vlasov-Fokker-Planck equation.
In Ref. \citep{Taitano_2016_rfp_0d2v_implicit}, the Fokker-Planck
equation for each ion species was normalized to its $v_{th}$ for
homogeneous plasmas. In spatially inhomogeneous plasmas, this strategy
lacks robustness.

\begin{figure}
\begin{centering}
\includegraphics[height=5.5cm]{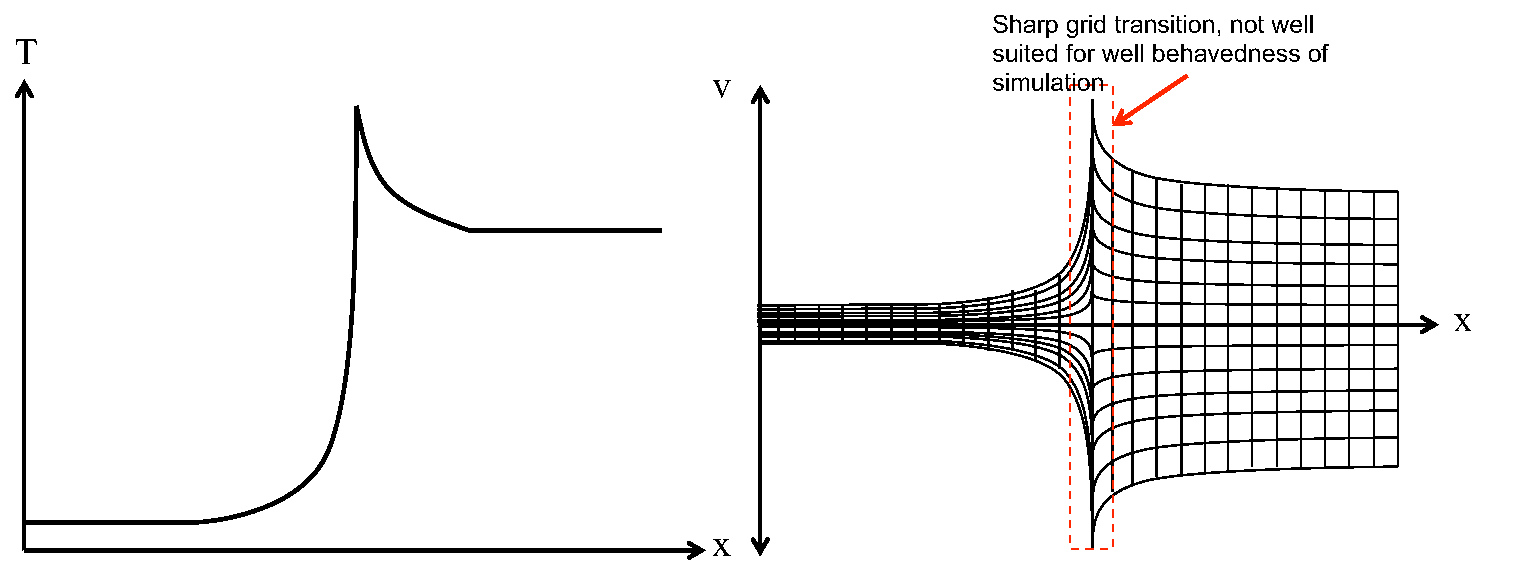}
\par\end{centering}
\caption{Illustration of a plasma shock temperature profile and the corresponding
grid quality near the temperature jump\label{fig:illustration_1d_plasma_shock_T_and_mesh}}
\end{figure}

Consider a scenario where a planar plasma shock propagates through
a medium {[}Fig. (\ref{fig:illustration_1d_plasma_shock_T_and_mesh}){]}\citep{casanova_1991_prl_kin_sim_of_col_shock_plasma}.
 For strong shocks, a sharp temperature variation exists near the
shock front. This large variation in $v_{th}$ will cause the velocity
space grid to be expanded too rapidly (both in space and time), resulting
in numerical brittleness. In this study, we address this issue by
combining: 1) an empirical temporal limiter, and 2) a spatial smoothing
operation. We note that neither of these strategies results in loss
of numerical accuracy in principle, as the transformed equations are
correct for an arbitrary $v^{*}$. We will demonstrate this numerically
later in this paper. We elaborate on these strategies next.

In order to limit the velocity grid expansion/contraction rate in
time, we limit the change of update of $v^{*}$ by $10\%$ from time
step to time step, i.e.:

\begin{equation}
\left(v_{\alpha}^{*}\right)^{p+1}=\begin{cases}
\left(v_{\alpha}^{*}\right)^{p}+\Delta t\left(\dot{v}_{\alpha}^{*}\right)^{p} & \textnormal{if}\;\Delta t\frac{\left|\left(\dot{v}_{\alpha}^{*}\right)^{p}\right|}{\left(v_{\alpha}^{*}\right)^{p}}\le0.1\\
\left(v_{\alpha}^{*}\right)^{p}\left[1+0.1\textnormal{sign}\left(\left(\dot{v}_{\alpha}^{*}\right)^{p}\right)\right] & \textnormal{otherwise}
\end{cases},
\end{equation}
where

\[
\left(\dot{v}_{\alpha}^{*}\right)^{p}=\frac{\left(\widetilde{v}_{\alpha}^{*}\right)^{p+1}-\left(v_{\alpha}^{*}\right)^{p}}{\Delta t}
\]
and

\[
\left(\widetilde{v}_{\alpha}^{*}\right)^{p+1}=\sqrt{\frac{2T_{\alpha}^{p+1}}{m_{\alpha}}}.
\]
To ensure that the profile of $v_{\alpha}^{*}$ is smooth \emph{in
space}, we perform a binomial filtering operation,

\begin{equation}
\left(v_{\alpha,i}^{*}\right)^{p+1}\leftarrow\textnormal{SM}\left(\left(v_{\alpha,i}^{*}\right)^{p+1}\right),
\end{equation}
where

\begin{equation}
\textnormal{SM}\left(a_{i}\right)=\frac{a_{i+1}+2a_{i}+a_{i-1}}{4}.
\end{equation}
The number of smoothing operations, $N_{sm}$, can be varied depending
on the size of expected temperature gradients in the problem. We define
$N_{sm}$ passes of binomial smoothing operation as,

\[
a_{i}\leftarrow\textnormal{SM}^{N_{sm}}\left(a_{i}\right)=\textnormal{SM}\left(\textnormal{SM}^{N_{sm}-1}\left(a_{i}\right)\right).
\]

\begin{figure}
\begin{centering}
\includegraphics[height=6cm]{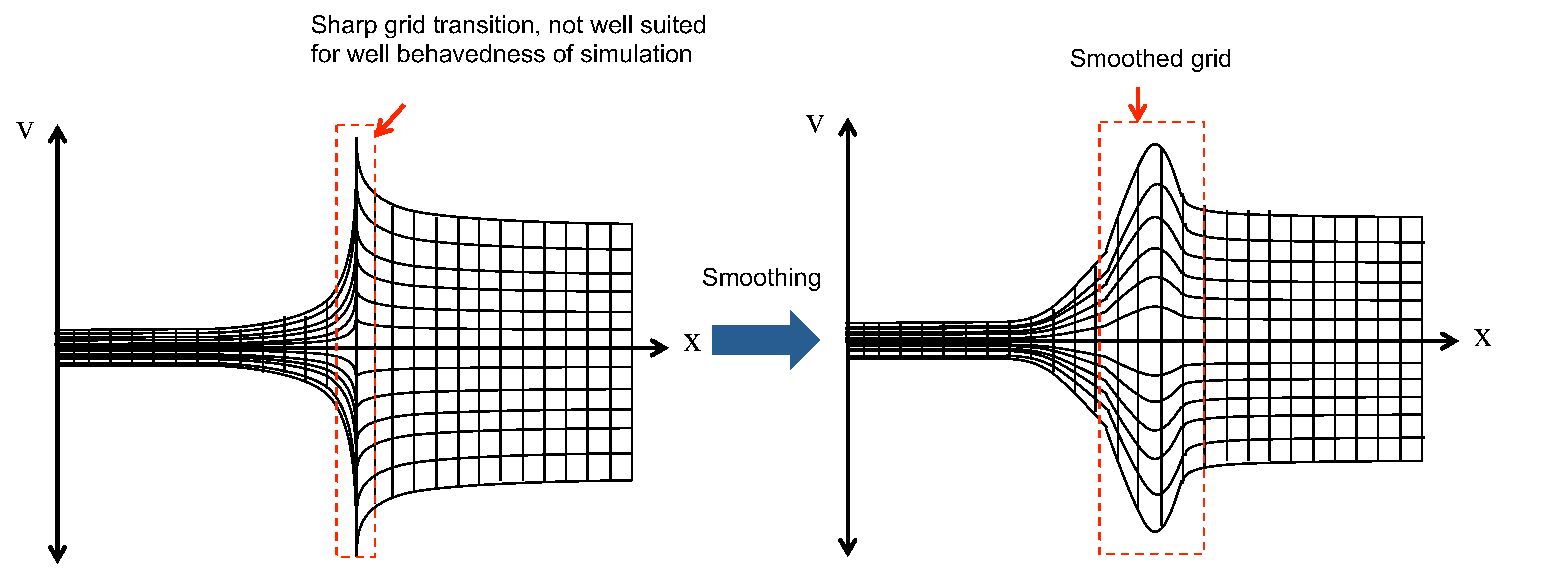}
\par\end{centering}
\caption{Illustration of unsmoothed (left) and smoothed phase-space grid on
a plasma shock problem. \label{fig:illustration_grid_smoothing}}
\end{figure}

The sensitivity of the solution with respect to the number of smoothing
passes is discussed in Sec. \ref{fig:periodic_hyperbolic_angent_temp_equil_sol_quality_wrt_vth_smoothing}.
Refer to Fig. \ref{fig:illustration_grid_smoothing} for an illustration
of the effects of post-grid-smoothing operation.

\section{Numerical results\label{sec:numerical-results}}

In this section, we demonstrate the properties of our numerical implementation,
both in terms of conservation and order of accuracy, with various
examples of varying degrees of complexity. For all problems, we normalize
the mass, charge, temperature, density, velocity, and time to the
proton mass, $m_{0}$, proton charge, $e$, reference temperature,
$T_{0}$, density, $n_{0}$, characteristic speed, $v_{0}=\sqrt{T_{0}/m_{0}}$,
and time-scale, $\tau_{0}=\frac{3\sqrt{m_{0}}T_{0}^{3/2}}{4\sqrt{\pi}n_{0}\Lambda e^{4}}$,
respectively. A fixed Coulomb logarithm, $\Lambda=10$, is used throughout
this study. All normalized distribution functions are initialized
as Maxwellians, with prescribed moments in $n$, $u$, and $T$ as:

\begin{equation}
\widehat{f}_{M}=\frac{nv^{*}}{\left(\pi v_{th}^{3}\right)^{3/2}}\textnormal{exp}\left[-\frac{\left(\widehat{v}_{||}v^{*}-u_{,||}\right)^{2}+\left(v^{*}\widehat{v}_{\perp}\right)^{2}}{v_{th}^{2}}\right].
\end{equation}
The initial normalization velocity profile, $v^{*}(x)$, is found
by applying a few binomial smoothing passes, $v^{*}=\textnormal{SM}^{N_{sm}}\left(v_{th}\right)$
(unless otherwise stated, $N_{sm}=5$), such that high wavenumber
components of the initial temperature profile (if present) are smoothed
out to prevent large numerical errors stemming from the computation
of spatial gradients of $v^{*}$ in the inertial term to pollute the
accuracy of the solutions. We note that in this study, we use a discrete
quadrature error accounting technique to ensure discrete Maxwellian
moments agree with prescribed ones \citep{Taitano_JCP_2017_FP_equil_disc}. 

For the solver, we employ an Anderson acceleration scheme \citep{anderson_aa_JACM_1965}
with nonlinear elimination strategies for the Rosenbluth potential
and fluid electrons (similar to Ref. \citep{Taitano_2015_rfp_0d2v_implicit})
and similar preconditioning strategies (multigrid and operator splitting)
as discussed in Refs. \citep{Taitano_2015_rfp_0d2v_implicit,Gasteiger_JOPP_2016_ADI_type_PC_for_SS_Vlasov}.
Finally, unless otherwise stated, we employ a nonlinear convergence
tolerance of $\epsilon_{r}=10^{-3}$. 

\subsection{Periodic sinusoidal ion-electron temperature equilibration\label{subsec:periodic_sin_ion_electron_temp_equil}}

\begin{figure}
\begin{centering}
\includegraphics[width=0.45\columnwidth]{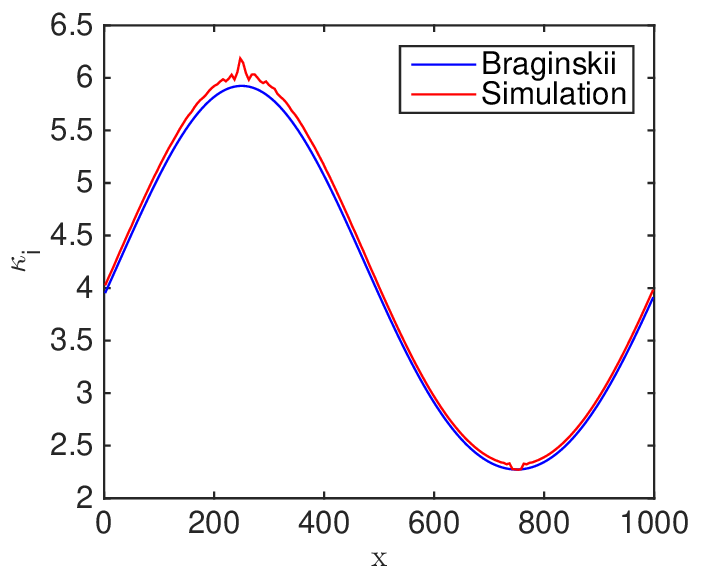}
\includegraphics[width=0.45\columnwidth]{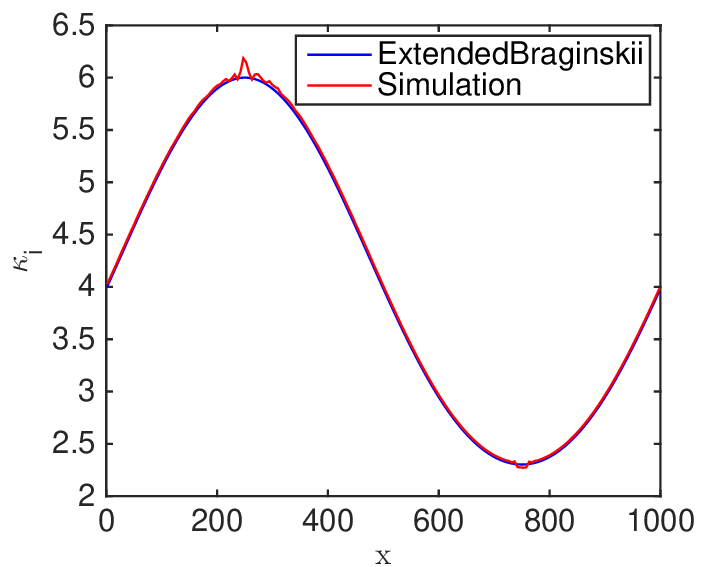}
\par\end{centering}
\caption{Periodic sinusoidal ion-electron temperature equilibration. Braginskii
thermal conductivity for protons versus simulation (left) and the
same but for a a higher-order truncation in the Laguerre polynomial
expansion of Braginskii's thermal conductivity coefficient (right).\label{fig:braginskii_kappa_i}}
\end{figure}

We begin by demonstrating that our proposed grid adaptivity and discretization
strategy recovers Braginskii's fluid solution in a short mean-free
path plasma \citep{braginskii}. Consider an initially stationary
proton-electron plasma in hydrodynamic equilibrium with the total
pressure, $P=nT=1$, and a sinusoidal temperature profile of $T_{i}=T_{e}=1+0.2\textnormal{sin}\left(k_{x}x\right)$,
where $k_{x}=\frac{2\pi}{L_{x}}$ with $L_{x}=1000$ the system size.
We consider a domain of $x\in\left[0,L_{x}\right]$ and $\widehat{L}_{||}=\left[-6,6\right]$,
$\widehat{L}_{\perp}=[0,6]$ with grids $N_{x}=192$, and $N_{v}=N_{||}\times N_{\perp}=64\times32$.
To test our simulation against theory, we focus on the ion collisional
heat flux. The numerical ion thermal conductivity is computed from
Fick's law as
\begin{equation}
\kappa_{i,sim}=-\frac{Q_{||,i}}{\partial T_{i}/\partial x},\label{eq:numerical_ion_conductivity}
\end{equation}
where

\[
Q_{||,i}=\frac{m_{i}}{2}\left\langle \left(v_{||}-u_{||}\right)\left(\vec{v}-\vec{u}_{i}\right)^{2},f_{i}\right\rangle _{v}.
\]
Here, the subscript $i$ denotes ions. This is to be compared with
Braginskii's theoretical result \citep{braginskii}

\begin{equation}
\kappa_{i}=3.9\frac{n_{i}T_{i}\tau_{i}}{m_{i}},\label{eq:brag_ion_conductivity}
\end{equation}
where $\tau_{i}=\frac{3\sqrt{m_{i}}T_{i}^{3/2}}{4\sqrt{\pi}n_{i}\Lambda e^{4}}$
is the ion collision time. In Fig. \ref{fig:braginskii_kappa_i}-left,
the Braginskii ion thermal conductivity is plotted for both simulation
and theory, and an excellent agreement is found.
 We point out that the computation of $\kappa_{i}$ is a bit noisy
at the extrema owing to the vanishing temperature gradient in the
denominator of Eq. (\ref{eq:numerical_ion_conductivity}). We note
that, for the chosen domain size, gradient-scale length, and mean
free path, the maximum Knudsen number, $Kn=\lambda_{i,mfp}/L_{T}$,
with $L_{T}=T_{i}/\partial_{x}T_{i}$, is $\sim{\cal O}\left(10^{-3}\right)$,
making the Braginskii approximation, Eq. (\ref{eq:brag_ion_conductivity}),
appropriate. We point out that there is a roughly 2\% uniform discrepancy
between theory and simulation, which is caused by only retaining two
terms in the truncation of the Laguerre polynomial expansion of the
distribution function in Braginskii's result \citep{braginskii}.
Fig. \ref{fig:braginskii_kappa_i}-right depicts a comparison with
the analytical result when three terms in the expansion are retained,
removing the discrepancy.

\begin{figure}
\begin{centering}
\includegraphics[height=4.5cm]{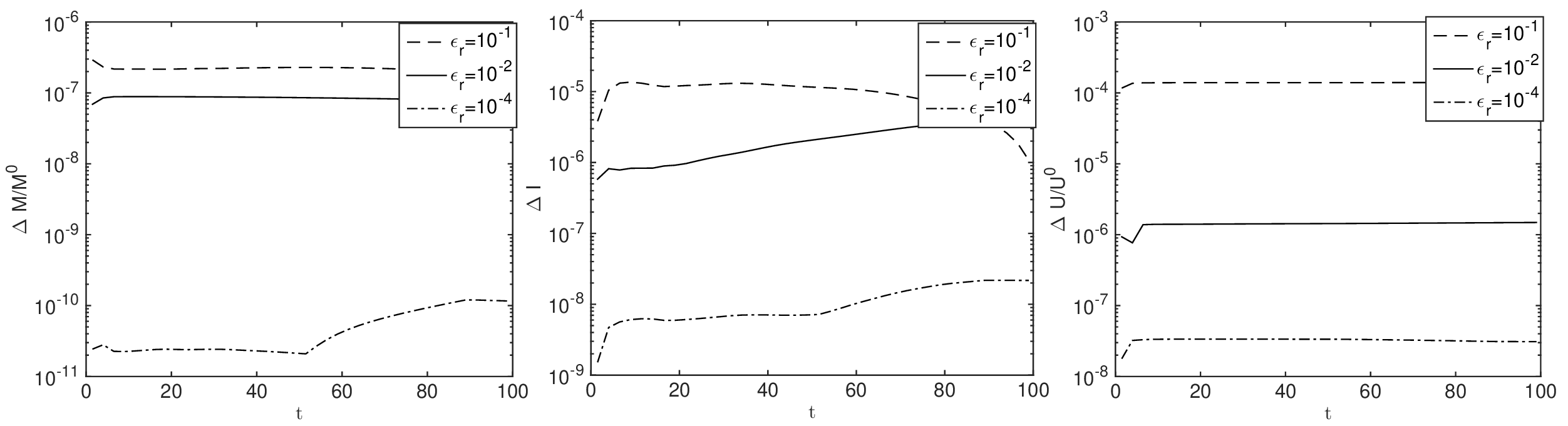}
\par\end{centering}
\caption{Periodic sinusoidal ion-electron temperature equilibration. Conservation
of mass (left), momentum (middle), and energy (right) versus time
for varying nonlinear convergence tolerance, $\epsilon_{r}$. \label{fig:periodic_sinusoidal_dTdx_conservation_plots}}
\end{figure}

We examine next the quality of the conservation properties with varying
nonlinear convergence tolerance; refer to Fig. \ref{fig:periodic_sinusoidal_dTdx_conservation_plots}.
Here, 

\[
\frac{\Delta M}{M^{0}}=\left|\frac{\int_{L_{min}}^{L_{max}}dx\left[\left\langle 1,\widehat{f}\right\rangle _{\vec{\widehat{v}}}-\left\langle 1,\widehat{f}^{0}\right\rangle _{\vec{\widehat{v}}}\right]}{\int_{L_{min}}^{L_{max}}dx\left\langle 1,\widehat{f}^{0}\right\rangle _{\vec{\widehat{v}}}}\right|,
\]

\[
\Delta I=\left|\int_{L_{min}}^{L_{max}}dx\left[\left\langle \widehat{v}_{||}v^{*},\widehat{f}\right\rangle _{\vec{\widehat{v}}}-\left\langle \widehat{v}_{||}v^{*,0},\widehat{f}^{0}\right\rangle _{\vec{\widehat{v}}}\right]\right|,
\]
and
\[
\frac{\Delta U}{U_{0}}=\left|\frac{\int_{L_{min}}^{L_{max}}dx\left\{ \left[m\left\langle \frac{\widehat{v}^{2}\left(v^{*}\right)^{2}}{2},\widehat{f}\right\rangle _{\vec{\widehat{v}}}+\frac{3}{2}n_{e}T_{e}\right]-\left[m\left\langle \frac{\widehat{v}^{2}\left(v^{*,0}\right)^{2}}{2},\widehat{f}^{0}\right\rangle _{\vec{\widehat{v}}}+\frac{3}{2}n_{e}^{0}T_{e}^{0}\right]\right\} }{\int_{L_{min}}^{L_{max}}dx\left[m\left\langle \frac{\widehat{v}^{2}\left(v^{*,0}\right)^{2}}{2},\widehat{f}^{0}\right\rangle _{\vec{\widehat{v}}}+\frac{3}{2}n_{e}^{0}T_{e}^{0}\right]}\right|
\]
are the measures of discrete conservation error in mass, momentum,
and energy, respectively. As can be seen, the conservation quality
improves with tighter nonlinear convergence tolerances, as expected.

Using this test example, we demonstrate next that our proposed scheme
is second-order accurate in configuration space, velocity space, and
time. We remark that, owing to the velocity-space adaptivity, it is
unsuitable to use the $L_{2}$-norm of the distribution function,

\begin{equation}
L_{2}=\sum_{\alpha=1}^{N_{s}}\sqrt{\text{\ensuremath{\int}dx}\int dv\left(f_{\alpha}-f_{\alpha}^{ref}\right)^{2}},
\end{equation}
to quantify the error, because $f_{\alpha}^{\Delta t}=\left(v_{\alpha}^{*,\Delta t}\right)^{-3}\widehat{f}\left(v_{\alpha}^{*,\Delta t}\widehat{v}\right)$
and $f_{\alpha}^{\Delta t,ref}=\left(v_{\alpha}^{*,\Delta t,ref}\right)^{-3}\widehat{f}\left(v_{\alpha}^{*,\Delta t,ref}\widehat{v}\right)$
live on \emph{different }spatial meshes (where the superscripts $\text{\ensuremath{\Delta}t }$
and $\Delta t,ref$ correspond to a prescribed time-step size and
a time-step size for the reference solution, respectively). We recall,
that this difference in the mesh stems from the fact that the velocity
space is normalized by $v^{*}$, which is lagged by a time-step and
undergoes a smoothing operation. As a proxy measure of the numerical
error, which is much simpler to compute and is independent of a normalization
choice, we consider the temperature. 

\begin{figure}[t]
\begin{centering}
\includegraphics[width=0.3\columnwidth]{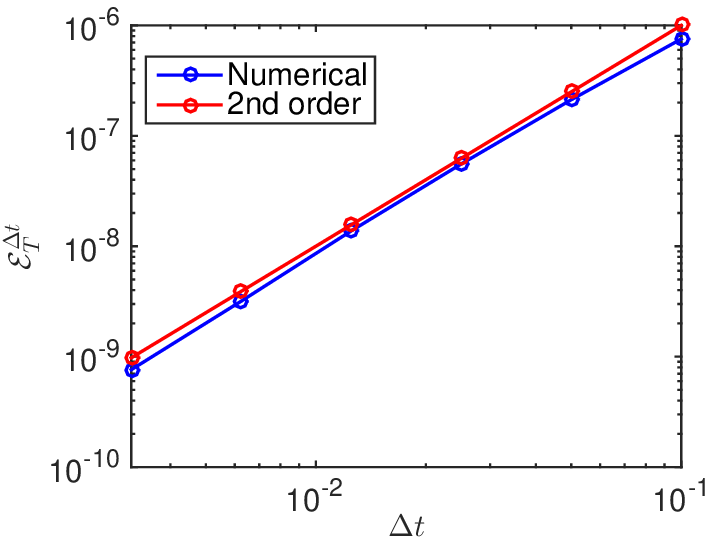}
\includegraphics[width=0.3\columnwidth]{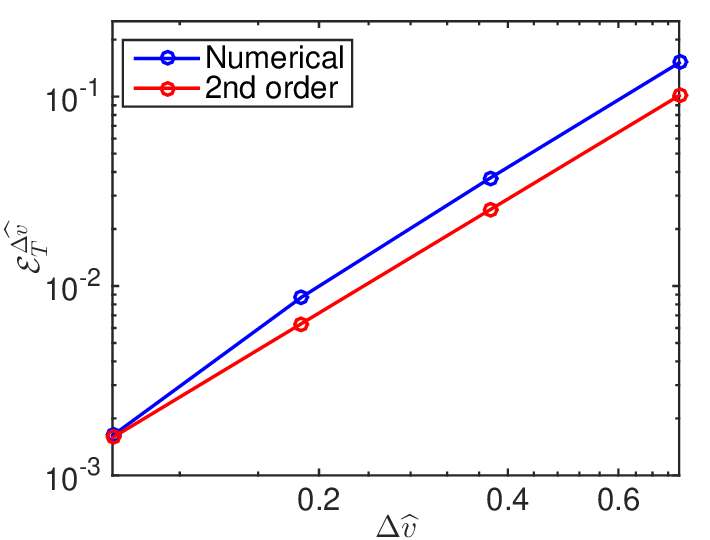}
\includegraphics[width=0.3\columnwidth]{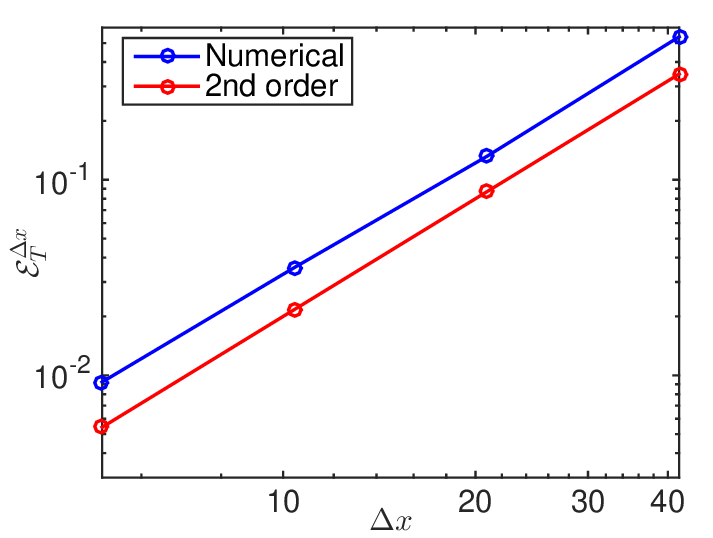}
\par\end{centering}
\centering{}\caption{Periodic sinusoidal ion-electron temperature equilibration. On the
left, we demonstrate a second-order convergence with time-step refinement.
At the center, we demonstrate a second-order convergence with velocity-space
refinement. On the right, we demonstrate a second-order convergence
with physical-space refinement.\label{fig:convergence-tests}}
\end{figure}

To demonstrate the second-order temporal convergence of the BDF2 scheme,
we compute a relative difference of the temperature with respect to
a reference temperature,

\begin{equation}
{\cal E}_{T}^{\Delta t}=\sum_{i=1}^{N_{x}}\Delta x\sum_{\alpha=1}^{N_{s}}\frac{\left|T_{\alpha,i}^{\Delta t,ref}-T_{\alpha,i}^{\Delta t}\right|}{T_{\alpha,i}^{\Delta t,ref}}.\label{eq:temporal-error-measure}
\end{equation}
Here, $T^{\Delta t,ref}$ is the reference temperature solution obtained
using a reference time-step size ($\Delta t_{ref}=10^{-4})$ at the
final time $t_{max}=1$. For all cases, we use a grid size of $N_{x}=24$
and $N_{v}=64\times32$ and a nonlinear convergence tolerance of $\epsilon_{r}=10^{-8}$
(to adequately capture small signals for small $\Delta t$). Fig.
\ref{fig:convergence-tests}-left shows that the expected order of
accuracy with $\Delta t$ refinement is confirmed.

Second-order accuracy in velocity-space is demonstrated similarly
by computing:

\begin{equation}
{\cal E}_{T}^{\Delta\widehat{v}}=\sum_{i=1}^{N_{x}}\Delta x\sum_{\alpha=1}^{N_{s}}\frac{\left|T_{\alpha,i}^{\Delta\widehat{v},ref}-T_{\alpha,i}^{\Delta\widehat{v}}\right|}{T_{\alpha,i}^{\Delta\widehat{v},ref}}.\label{eq:spatial-error-measure}
\end{equation}
Here, $T^{\Delta\widehat{v},ref}$ is the reference temperature solution
obtained using a reference grid resolution of $N_{v}^{ref}=512\times256$.
A uniform grid refinement is performed in both velocity-space directions.
For all cases, we use $\Delta t=1$ and a final time $t_{max}=10$
with $N_{x}=96$. Fig. \ref{fig:convergence-tests}-center confirms
second-order convergence with $\Delta\widehat{v}$ refinement. 

Finally, to demonstrate second-order accuracy of the spatial discretization,
we use a similar approach and compute

\begin{equation}
{\cal E}_{T}^{\Delta x}=\sum_{i=1}^{N_{x,ref}}\Delta x_{ref}\sum_{\alpha=1}^{N_{s}}\frac{\left|T_{\alpha,i}^{\Delta x,ref}-T_{\alpha,i}^{\Delta x}\right|}{T_{\alpha,i}^{\Delta x,ref}}.\label{eq:physical-space-error-measure}
\end{equation}
Here, $T^{\Delta x,ref}$ is the reference-temperature solution obtained
using a reference-grid resolution ($N_{x,ref}=768)$ with a final
time $t_{max}=1$. To compute the norm in Eq. (\ref{eq:physical-space-error-measure}),
we interpolate the coarse solution onto the fine grid via a $4^{th}$
order spline. For all cases, we use a velocity space grid size of
$N_{v}=32\times16$. Fig. \ref{fig:convergence-tests}-right confirms
the expected order of accuracy of our spatial discretization. 

\subsection{Ion temperature relaxation with an initial periodic hyperbolic tangent
profile\label{subsec:periodic_tanh_ion_temp_equil}}

This example highlights the importance of enforcing discrete conservation
in the Vlasov equation when gradients (both in physical and velocity
space) are marginally resolved. We consider a single ion species with
$m=1$, $q=1$, without electrons, on a periodic spatial domain of
$L_{x}\in\left[-50,50\right]$, and a velocity domain $\widehat{L}_{||}\in\left[-6,6\right]$,
$\widehat{L}_{\perp}\in\left[0,5\right]$. We consider a mesh of $N_{x}=96$
and $N_{v}=64\times32$. We assume an initially stationary distribution,
$u=0$, with a homogeneous density, $n=1$, and a hyperbolic tangent
temperature profile,

\[
T=0.495\begin{cases}
1+\textnormal{tanh}\left(\chi\left[x+25\right]\right)+b_{T0} & \textnormal{if}\;-50\le x\le0\\
1-\textnormal{tanh}\left(\chi\left[x-25\right]\right)+b_{T1} & \textnormal{otherwise}
\end{cases}.
\]
 Here, $\chi$ is a parameter that controls the gradient scale length
of the hyperbolic tangent; values for $\chi$ are provided later.
We remark, that for these parameters, $v_{th,max}/v_{th,min}=\text{\ensuremath{\sqrt{T_{max}/T_{min}}}}=10$
and a static uniform grid will require on order of $(v_{th,max}/v_{th,min})^{2}=100$
times more grid points than our velocity adaptivity strategy to resolve
the cold distribution function adequately.

\begin{figure}[t]
\begin{centering}
\includegraphics[height=4.5cm]{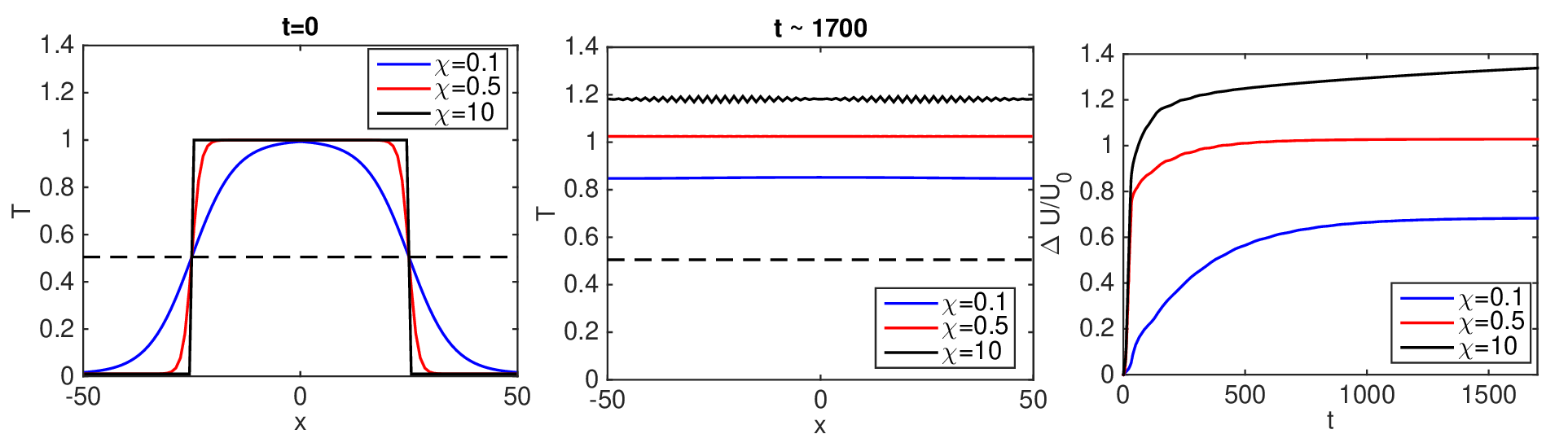}
\par\end{centering}
\caption{Ion temperature relaxation with an initial periodic hyperbolic tangent
profile. The quality of solution for different $\chi$ and $N_{sm}=0$
at $t=0$ (left) and $t\approx1700$ (center). The dashed black line
is the analytical equilibrium solution. The energy conservation error
is shown on the right \label{fig:periodic_hyperbolic_tangent_temp_equil_sol_quality_wrt_chi_T}}
\end{figure}

We investigate the impact of a lack of conservation on long-term accuracy
with respect to various parameters. We turn off the conservation scheme
for the inertial term arising from the spatial dependence of $v^{*}$.
We demonstrate first that the \emph{quality of conservation} depends
on grid resolution. We choose $\chi$ equal to $0.5$, $1$, and $10$
without ensuring either momentum nor energy conservation symmetries
for this inertial term. In Fig. \ref{fig:periodic_hyperbolic_tangent_temp_equil_sol_quality_wrt_chi_T},
we show the solution profile for all $\chi$ and $N_{sm}=0$ at $t\approx1700$.
 As can be seen, in all cases a large energy conservation error is
accumulated over time, leading to significant numerical heating. For
the $\chi=10$ case, the initial numerical heating coming from the
sharp gradients is strong enough that a grid-scale mode is excited. 

\begin{figure}[t]
\begin{centering}
\includegraphics[height=4.5cm]{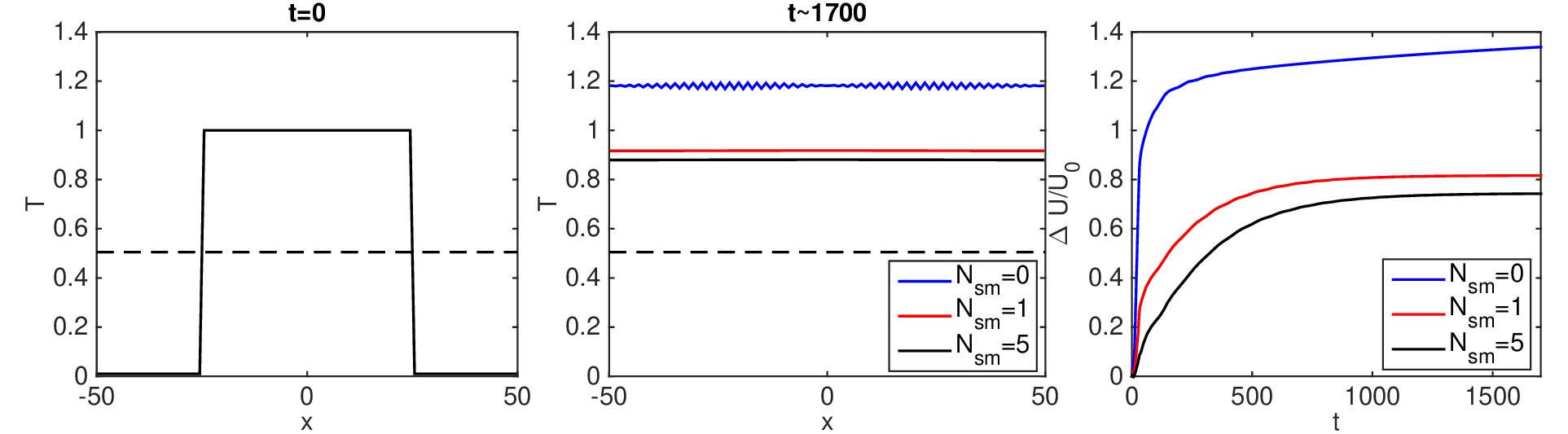}
\par\end{centering}
\caption{Ion temperature relaxation with an initial periodic hyperbolic tangent
profile. The quality of solution for a varying number of binomial
smoothing iterations, $N_{sm}$, on $v^{*}$ at $t=0$ (left) and
$t\approx1700$ (center). The dashed black line is the analytical
equilibrium solution. The energy conservation error is shown on the
right \label{fig:periodic_hyperbolic_angent_temp_equil_sol_quality_wrt_vth_smoothing}}
\end{figure}

Numerical accuracy is improved by either increasing velocity space
resolution or by smoothing $v^{*}$ (as the spatial inertial term
vanishes in the limit of $\partial_{x}v^{*}=0$). We recall that the
introduction of $v^{*}$ is simply a numerical trick and that the
smoothing of $v^{*}$ does not change the \emph{physics} of the problem.
In Fig. \ref{fig:periodic_hyperbolic_angent_temp_equil_sol_quality_wrt_vth_smoothing},
we show the impact of varying $N_{sm}$ on the quality of energy conservation,
with the quality improving for enhanced smoothing of $v^{*}$, as
expected.

\begin{figure}[t]
\begin{centering}
\includegraphics[height=4.5cm]{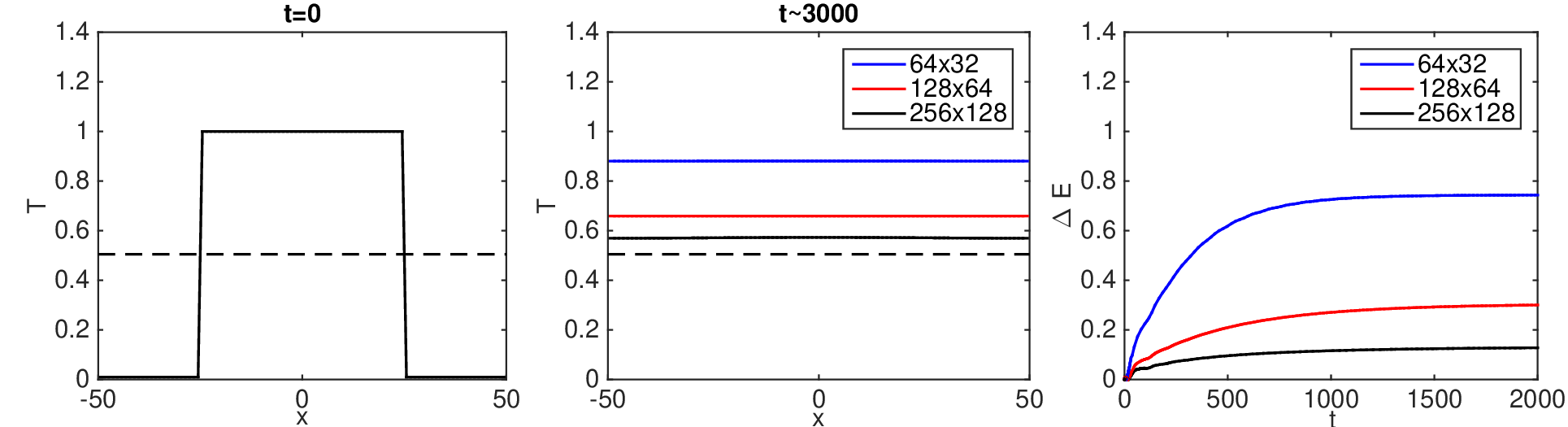}
\par\end{centering}
\caption{Ion temperature relaxation with an initial periodic hyperbolic tangent
profile. The quality of solution for varying velocity-space grid resolution
at $t=0$ (left) and $t\approx3000$ (center). The dashed black line
is the analytical equilibrium solution. The energy conservation error
is shown on the right \label{fig:periodic_hyperbolic_angent_temp_equil_sol_quality_wrt_Nv}}
\end{figure}

Next, we investigate the impact of increasing velocity-space resolution
with $N_{x}=96$, $\chi=10$, and $N_{sm}=5$. In Fig. \ref{fig:periodic_hyperbolic_angent_temp_equil_sol_quality_wrt_Nv},
we show the solution profile for different velocity-space grids. 
 As can be seen, a grid of $N_{v}=256\times128$ is required to reduce
the energy conservation error to within $10\%$. At this point, the
error in conservation is mostly dominated by configuration space discretization
errors (due to the spatial interpolation procedure embedded in the
definition of the conservation symmetries, Eq. (\ref{eq:vlasov_spatial_inertial_energy_cons_constraint})
and (\ref{eq:vlasov_spatial_inertial_full_cons_last_eq})), and further
improvement in conservation via refinement in velocity space will
require refinement in configuration space. 

\begin{figure}[t]
\begin{centering}
\includegraphics[height=4.5cm]{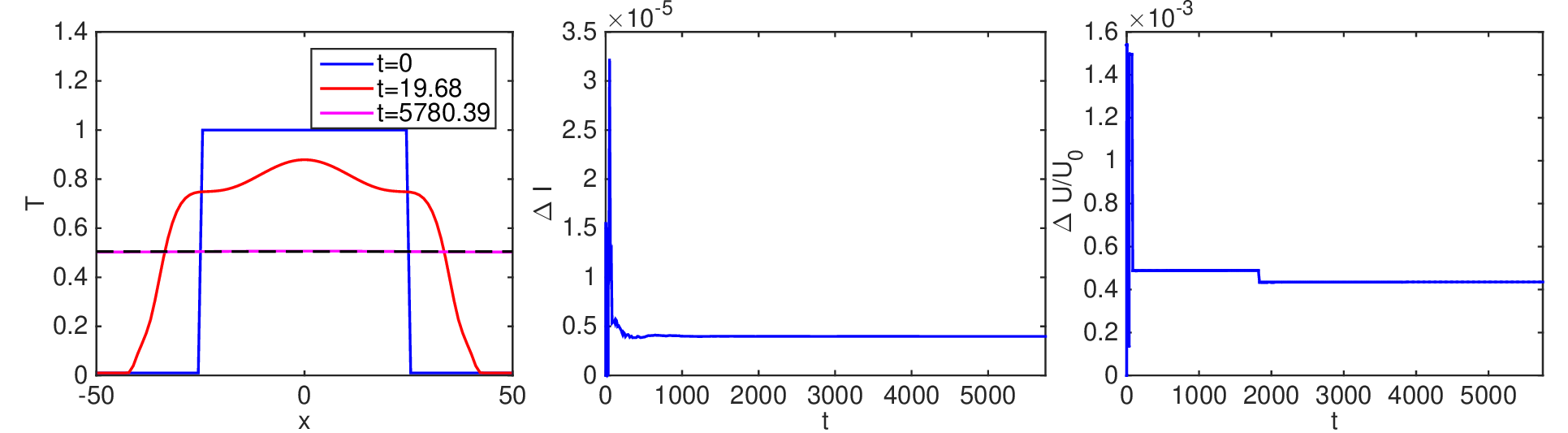}
\par\end{centering}
\caption{Ion temperature relaxation with an initial periodic hyperbolic tangent
profile. The solution is obtained by enforcing the discrete Vlasov
spatial inertial conservation symmetries in Eqs. (\ref{eq:vlasov_spatial_inertial_full_cons_1st_eq})
to (\ref{eq:vlasov_spatial_inertial_full_cons_last_eq}) (left). The
dashed black line is the analytical equilibrium solution. Total momentum
and energy conservation error is shown in the center and right, respectively.
\label{fig:periodic_hyperbolic_angent_temp_equil_sol_quality_wh_conservation}}
\end{figure}

Finally, we show that by ensuring the conservation symmetries in the
inertial term, the numerical heating effect can be suppressed to nonlinear
convergence tolerance even with coarser grids. We employ a grid of
$N_{x}=96$ and $N_{v}=64\times32$, $\chi=10$ and $N_{sm}=5$; refer
to Fig. \ref{fig:periodic_hyperbolic_angent_temp_equil_sol_quality_wh_conservation}.
The correct asymptotic solution is obtained. Conservation errors are
kept small throughout the simulation and commensurate with the default
nonlinear relative convergence tolerance ($10^{-3}$). 

\subsection{Mach 5 steady-state shock\label{subsec:M=00003D5_ssss}}

We simulate a Mach 5 shock in a proton-electron plasma in the frame
of the shock. The purpose of this test problem is to demonstrate that
the correct steady-state solution is obtained for a non-trivial problem.
We obtain the hydrodynamic jump conditions from the Hugoniot relationship:

\begin{equation}
\frac{P_{1}}{P_{0}}=\frac{2\gamma M^{2}-(\gamma-1)}{\gamma+1},
\end{equation}

\begin{equation}
\frac{\rho_{1}}{\rho_{0}}=\frac{u_{0}}{u_{1}}=\frac{M^{2}\left(\gamma+1\right)}{M^{2}\left(\gamma-1\right)+2}.
\end{equation}
Here, the subscript $0$ denotes the upstream (un-shocked) region
and $1$ denotes the downstream (shocked) region. Combining these
equations gives:

\begin{equation}
\frac{u_{0}}{u_{1}}=\frac{\left(\gamma-1\right)P_{0}+\left(\gamma+1\right)P_{1}}{\left(\gamma+1\right)P_{0}+\left(\gamma-1\right)P_{1}}.
\end{equation}
Here, $\gamma$ is the specific heat ratio ($\gamma=5/3$ for fully
ionized plasmas), $P$ is the total static pressure (i.e., $P=P_{i}+P_{e}$),
$\rho=\sum_{\alpha=1}^{N_{s}}m_{\alpha}n_{\alpha}$ is the total mass
density, and $u=\sum_{\alpha=1}^{N_{s}}m_{\alpha}n_{\alpha}u_{\alpha}/\sum_{\alpha=1}^{N_{s}}m_{\alpha}n_{\alpha}$
is the mass averaged drift velocity of the respective regions. The
upstream velocity can be expressed as

\begin{equation}
u_{0}=Mc_{0},
\end{equation}
where $c_{0}$ is the upstream sound speed,

\begin{equation}
c_{0}=\sqrt{\gamma\frac{P_{0}}{\rho_{0}}}.
\end{equation}
Employing the downstream condition of $\rho_{1}=mn_{1}=1,$ $n_{1}=1$,
$m_{1}=1$, $P_{1}=P_{1,i}+P_{1,e}=2P_{1,i}=2n_{1,i}T_{1,i}=2$, $T_{1,i}=1$,
and $M=5$ gives for the upstream conditions $\rho_{0}=mn_{0}=0.28$,
$P_{0}=0.1290$, and $T_{0}=0.1152$ . Then, $c_{0}=0.619697$, $u_{0}=Mc_{0}=3.0984$,
and $u_{1}=0.8676$. 

We consider a computational domain of $L_{x}\in\left[0,160\right]$,
$\widehat{L}_{||}\in\left[-4,13\right]$, $\widehat{L}_{\perp}\in\left[0,6\right]$
with a grid of $N_{x}=96$, $N_{v}=128\times64$. The solution is
initialized with 

\begin{equation}
n=\begin{cases}
0.28 & \textnormal{if}\;x\le80\\
1 & \textnormal{otherwise}
\end{cases},\;u=\begin{cases}
3.0984 & \textnormal{if}\;x\le80\\
0.8676 & \textnormal{otherwise}
\end{cases},\;T=\begin{cases}
0.1152 & \textnormal{if}\;x\le80\\
1 & \textnormal{otherwise}
\end{cases}.
\end{equation}
 In the configuration space, we consider in-flow and out-flow boundary
conditions for the ion distribution functions:

\begin{equation}
f_{B}\left(v_{||},v_{\perp}\right)=\begin{cases}
f_{M}\left(n_{B},u_{B},T_{B}\right) & \textnormal{if}\;\widehat{l}_{B}v_{||}\le0\\
f_{C} & \textnormal{otherwise}
\end{cases}.
\end{equation}
Here, $n_{B}$, $u_{B}$, and $T_{B}$ are the moments defined by
the Hugoniot conditions at the boundary, $\widehat{l}_{B}$ is the
$x$-component of the boundary surface normal vector ($\pm1$ in 1D),
and $f_{C}$ is the distribution function defined in the computational
cell adjacent to the boundary. For the fluid electron temperature,
we use the Dirichlet boundary conditions to impose the Hugoniot asymptotic
jump.

The simulation is run for $t_{max}=250$ until transient structures
have equilibrated. 
\begin{figure}[t]
\begin{centering}
\includegraphics[height=5cm]{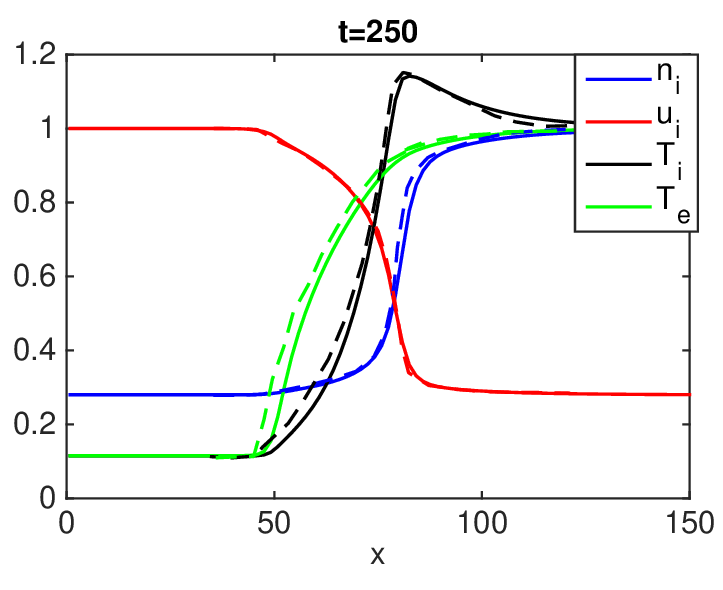}
\par\end{centering}
\caption{Mach 5 steady-state shock. The solid lines are from our simulation,
while the dashed lines are from Ref. \citep{Vidal_PoP_1993_ion_kin_sim_of_plan_col_shock}
for a similar setup (data points were manually extracted from Fig
2.a in Ref. \citep{Vidal_PoP_1993_ion_kin_sim_of_plan_col_shock}
using the WebPlotDigitizer software \citep{webplotdigitizer}). We
note, that in the plot, the drift velocity, $u$, is normalized to
the upstream value. Additionally, the \emph{x}-axis was scaled by
a factor of $\frac{\sqrt{2}}{3\pi}$ to convert the data from \citep{Vidal_PoP_1993_ion_kin_sim_of_plan_col_shock}
to our normalization. \label{fig:M=00003D5_standing_shock_plot}}
\end{figure}
A very good agreement is found with respect to the reference solution
\citep{Vidal_PoP_1993_ion_kin_sim_of_plan_col_shock}; refer to Fig.
\ref{fig:M=00003D5_standing_shock_plot}. 

\subsection{Shock interaction with a density jump\label{subsec:shock_breakout_across_density_increase}}

In this example, we simulate a shock propagating through a mass-density
discontinuity. Unlike the standing shock case, where a steady-state
solution can be found, this problem is inherently dynamic and tests
the robustness of the overall approach. The analytical solution is
well-iknown and given in App. \ref{app:physics_of_shock_traveling_through_a_density_jump}
for reference. We test our approach against this solution.

\begin{figure}[t]
\begin{centering}
\includegraphics[height=5.5cm]{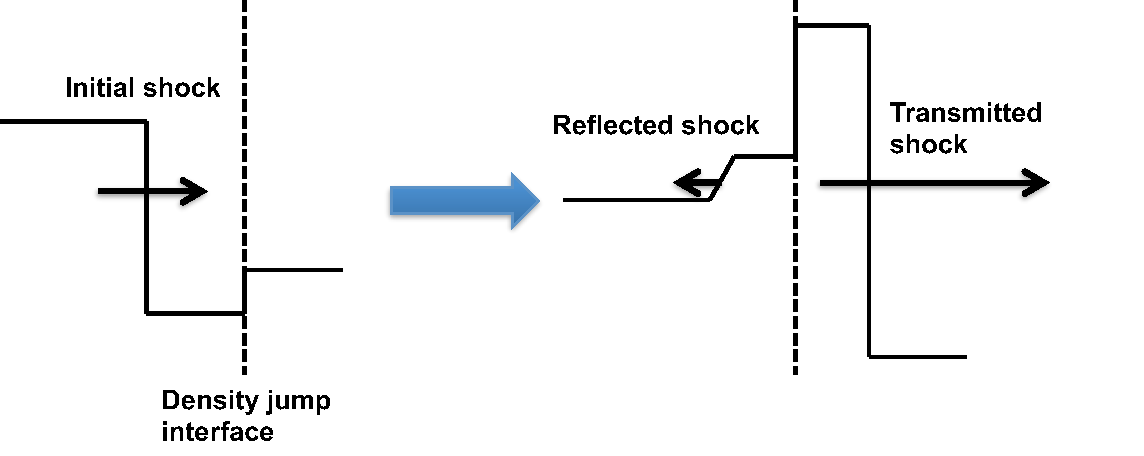}
\par\end{centering}
\caption{Shock interaction with a density jump. Illustration of the problem
setup before (left) and after (right) shock breakout through the interface.\label{fig:illustration_shock_breakout_across_density_increase}}
\end{figure}

We consider an $M=5$ shock propagating from left to right through
a plasma comprised of protons on the left and deuterons on the right.
The ions are initially in pressure equilibrium at the mass-density-jump
interface; refer to Fig \ref{fig:illustration_shock_breakout_across_density_increase}.
The problem is simulated in a domain of $L_{x}\in\left[0,2500\right]$,
$\widehat{L}_{||}\in\left[-7,7\right]$, and $\widehat{L}_{\perp}\in\left[0,7\right]$,
on a grid of $N_{x}=192$ and $N_{v}=128\times64$, with in-/out-flow
boundary conditions in the configuration space for the ion-distribution
functions and Dirichlet boundary conditions for fluid-electron temperature.
The initial conditions are, for protons,

\begin{equation}
n_{P}=\begin{cases}
n_{P0} & \textnormal{if}\;0\le x<x_{0}\\
n_{P1} & \textnormal{if}\;x_{0}\le x<x_{1}\\
n_{P2} & \textnormal{otherwise}\;
\end{cases},\;u_{P}=\begin{cases}
u_{P0} & \textnormal{if}\;0\le x<x_{0}\\
0 & \textnormal{otherwise}
\end{cases},\;T_{P}=\begin{cases}
T_{P0} & \textnormal{if}\;0\le x<x_{0}\\
T_{P1} & \textnormal{otherwise}
\end{cases},
\end{equation}
for deuterons,

\begin{equation}
n_{D}=\begin{cases}
n_{D0} & \textnormal{if}\;0\le x<x_{0}\\
n_{D1} & \textnormal{otherwise}
\end{cases},\;u_{D}=\begin{cases}
u_{D0} & \textnormal{if}\;0\le x<x_{0}\\
0 & \textnormal{otherwise}
\end{cases},\;T_{D}=\begin{cases}
T_{D0} & \textnormal{if}\;0\le x<x_{0}\\
T_{D1} & \textnormal{otherwise}
\end{cases},
\end{equation}
and electron temperature,

\begin{equation}
T_{e}=\begin{cases}
T_{e0} & \textnormal{if}\;0\le x<x_{0}\\
T_{e1} & \textnormal{otherwise}
\end{cases}.
\end{equation}
Here, $x_{0}=150$ and $x_{1}=250$, $n_{P0}=1$, $n_{P1}=0.28$,
$n_{P2}=0.005$, $n_{D0}=0.0028$, $n_{D1}=0.28$, $u_{P0}=u_{D0}=2.2308$,
$T_{P0}=T_{D0}=T_{e0}=1$, and $T_{P1}=T_{D1}=T_{e1}=0.1152$. We
use the initial conditions at the boundary to provide the in-flow
conditions for ions and Dirichlet conditions for electrons.
\begin{figure}[t]
\begin{centering}
\includegraphics[height=4cm]{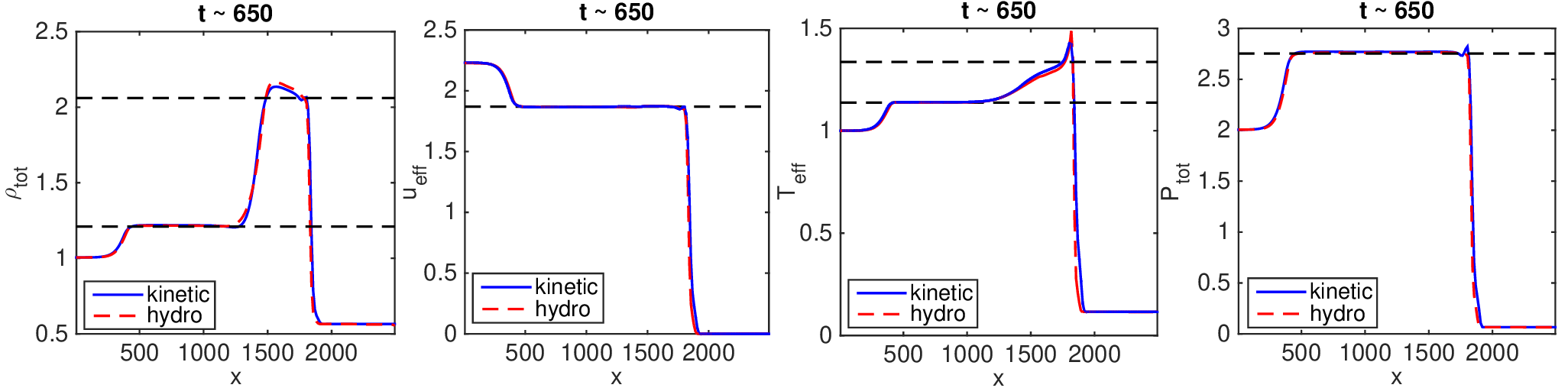}
\par\end{centering}
\begin{centering}
\includegraphics[height=4cm]{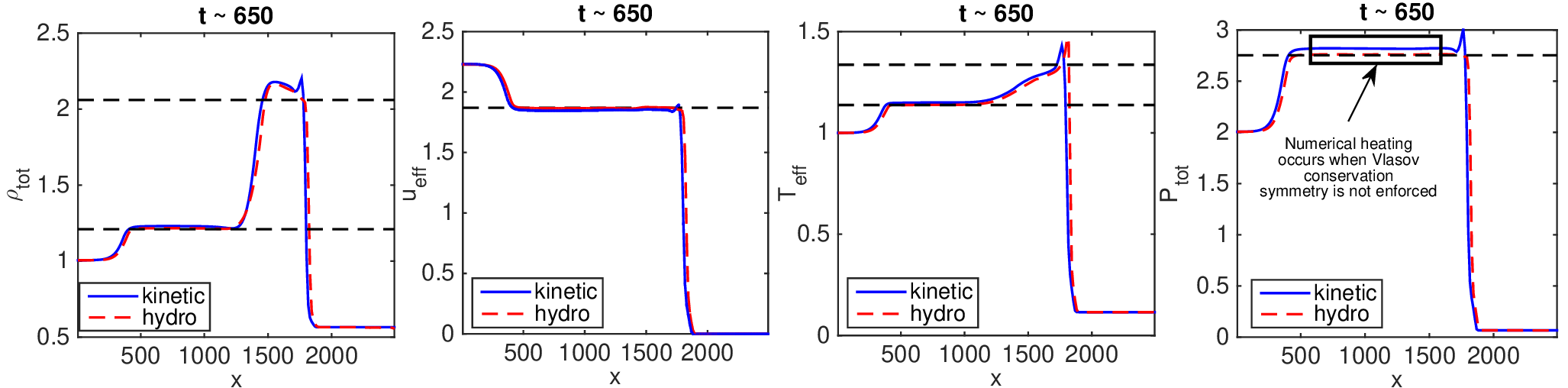}
\par\end{centering}
\caption{Shock interaction with a density jump. From left to right, we show
the total hydrodynamic mass density, $\rho_{tot}=\sum_{\alpha}^{N_{s}}m_{\alpha}n_{\alpha}$,
the center of mass velocity, $u_{eff}=\sum_{\alpha}^{N_{s}}m_{\alpha}n_{\alpha}u_{\alpha}/\rho_{tot}$,
the effective ion temperature, $T_{eff}=\sum_{\alpha}^{N_{s}}n_{\alpha}T_{\alpha}/\sum_{\alpha}^{N_{s}}n_{\alpha}$,
and the total pressure, $P_{tot}=\left(\sum_{\alpha}^{N_{s}}n_{\alpha}T_{\alpha}+n_{e}T_{e}\right)$.
The top row represents the case with all conservation symmetries enforced
and the bottom without enforcing the Vlasov conservation symmetries.
Here, the blue lines represent the solution obtained from our kinetic
code, while the red lines show the solution obtained from a hydro
code. As can be seen, our implementation with grid adaptivity and
discrete conservation strategy correctly recovers the shock jump conditions
(black dashed lines) across the density jump, while numerical heating
is observed when the Vlasov conservation symmetries are not enforced.\label{fig:M=00003D5_shock_across_density_increase}}
\end{figure}
As can be seen in Fig. \ref{fig:M=00003D5_shock_across_density_increase},
the long term kinetic solution agrees very well with a hydro simulation
(obtained from an in-house multi-fluid Euler code), demonstrating
the capability of the proposed approach to capture the hydrodynamic
limit. We stress that this limit is rigorously obtained by ensuring
strict conservation. The bottom row of Fig. \ref{fig:M=00003D5_shock_across_density_increase}
shows that numerical heating \textendash result of \emph{not} \emph{enforcing}
the conservation symmetry for the Vlasov operator\textendash{} results
in the pressure in the solution departing from the correct asymptotics
(\textasciitilde{}5\% error for the current grid). While this error
may seem small (a consequence of this problem setup being much more
constrained than the periodic case), its impact in highly nonlinear
applications could be very large (for instance, in inertial confinement
fusion, up to 4 shocks are used to compress and heat the fuel to fusion
conditions, and this level of numerical heating can result in a \textasciitilde{}20\%
error in the final fuel temperature, significantly altering implosion
dynamics).

\section{Conclusion\label{sec:conclusions}}

In this study, we have demonstrated, for the first time, an approach
that is fully conservative and optimally adaptive for the multi-species,
1D2V VFP ion plasma equations with fluid electrons. The approach features
exact (in practice, up to a nonlinear tolerance) mass, momentum, and
energy conservation and allows for a large temperature variation in
time and in space. Our approach analytically adapts the velocity-space
mesh for each species by normalizing the velocity space to each species
reference velocity, $v^{*}$ (i.e., we consider multiple velocity-space
grids). The analytical formulation allows us to expose the continuum-conservation
symmetries in the inertial terms arising from the normalization, which
are then enforced discretely via the use of nonlinear constraints,
as proposed in earlier studies \citep{Taitano_2015_jcp_cmec_va,Taitano_2015_rfp_0d2v_implicit}.
We have demonstrated the ability of the scheme to capture transport
and hydrodynamic asymptotic solutions correctly, which is exceedingly
challenging for VFP codes.

\textcolor{black}{We remark that the present approach cannot handle
well situations where the bulk velocity is much larger than the thermal
velocity of the plasma. In these situations, one must }\textcolor{black}{\emph{shift}}\textcolor{black}{{}
the velocity space by the bulk velocity, as was done in Ref. \citep{larroche_EPJ_2003_icf_fuel_ion_implosion_sim}.
This will give rise to an additional inertial term, which will be
considered in future work. We close by noting that the methodology
developed in this study has been extended to a spherical geometry
with grid adaptivity in configuration space. This work will be documented
in a follow-on manuscript. }

\section*{Acknowledgments}

This work was sponsored by the Metropolis Postdoctoral Fellowship
for W.T.T., the LDRD office, the Institutional Computing, and the
Thermonuclear Burn Initiative of the Advanced Simulation and Computing
Program at the Los Alamos National Laboratory. This work was performed
under the auspices of the National Nuclear Security Administration
of the U.S. Department of Energy at Los Alamos National Laboratory,
managed by LANS, LLC under contract DE-AC52-06NA25396.

\bibliographystyle{ieeetr}
\bibliography{./mybib,./kinetic,./fokker-planck,./transport,./numerics,./general,./icf_physics}

\appendix

\section{Details on the fluid-electron model\label{app:sophisticated_fluid_electron_model}}

The frictional force between the $\alpha$-ion species and electrons
is given as,

\begin{equation}
\vec{F}_{\alpha e}=-m_{e}n_{e}\nu_{e\alpha}\left(\vec{u}_{\alpha}-\left\langle \vec{u}_{\alpha}\right\rangle \right)+\alpha_{0}m_{e}n_{e}\nu_{e\alpha}\left(\vec{u}_{e}-\left\langle \vec{u}_{\alpha}\right\rangle \right)+\beta_{0}\frac{n_{e}\nu_{e\alpha}\nabla T_{e}}{\sum_{\alpha}^{N_{s}}\nu_{e\alpha}},
\end{equation}
where

\begin{equation}
\left\langle \vec{u}_{\alpha}\right\rangle \equiv\frac{\sum_{\alpha}^{N_{s}}\nu_{e\alpha}\vec{u}_{\alpha}}{\sum_{\alpha}^{N_{s}}\nu_{e\alpha}},
\end{equation}

\begin{equation}
\nu_{e\alpha}=\frac{4\sqrt{2\pi}n_{\alpha}q_{\alpha}^{2}e^{4}\Lambda_{e\alpha}}{3\sqrt{m_{e}}T_{e}^{3/2}},
\end{equation}

\begin{equation}
\alpha_{0}=\frac{4\left(16Z_{eff}^{2}+61\sqrt{2}Z_{eff}+72\right)}{217Z_{eff}^{2}+604\sqrt{2}Z_{eff}+288},
\end{equation}

\begin{equation}
\beta_{0}=\frac{30Z_{eff}\left(11Z_{eff}+15\sqrt{2}\right)}{217Z_{eff}^{2}+604\sqrt{2}Z_{eff}+288},
\end{equation}
and the effective charge is defined as

\begin{equation}
Z_{eff}=-\frac{\sum_{\alpha}^{N_{s}}q_{\alpha}^{2}n_{\alpha}}{q_{e}n_{e}}.
\end{equation}
The electron heat flux is given as

\begin{equation}
\vec{Q}_{e}=\beta_{0}n_{e}T_{e}\left(\vec{u}_{e}-\left\langle \vec{u}_{\alpha}\right\rangle \right)-\kappa_{e}\nabla T_{e},
\end{equation}
where the electron-thermal conductivity is given as

\begin{equation}
\kappa_{e}=\frac{\gamma_{0}n_{e}T_{e}}{m_{e}\sum_{\alpha}^{N_{s}}\nu_{e\alpha}},
\end{equation}
with

\begin{equation}
\gamma_{0}=\frac{25Z_{eff}\left(433Z_{eff}+180\sqrt{2}\right)}{4\left(217Z_{eff}^{2}+604\sqrt{2}Z_{eff}+288\right)}.
\end{equation}
See Ref. \citep{simakov_PoP_2014_e_transp_wh_multi_ion} for a complete
derivation and discussion of the coefficients $\alpha_{0}$, $\beta_{0}$,
and $\gamma_{0}$. 

\section{Vlasov-Fokker-Planck equation expressed in normalized velocity variables
\label{app:details_on_vstar_transformation}}

We consider the Vlasov equation under the velocity coordinate transformation
$\vec{v}=v_{\alpha}^{*}\left(x,t\right)\widehat{\vec{v}}$. The total
derivative of $\widehat{f}_{\alpha}\left(x,\widehat{v},t\right)$
keeping $x$ and $\vec{v}$ constant is given by

\begin{equation}
\left.\partial_{t}\widehat{f}_{\alpha}\right|_{x,\vec{v}}=\left.\partial_{t}\widehat{f}_{\alpha}\right|_{x,\vec{\widehat{v}}}+\left.\partial_{t}\vec{\widehat{v}}\right|_{x,\vec{v}}\cdot\left.\frac{\partial\widehat{f}_{\alpha}}{\partial\vec{\widehat{v}}}\right|_{x,\vec{v}},
\end{equation}
where $\left.\partial_{t}\vec{\widehat{v}}\right|_{x,\vec{v}}$ can
be expressed as

\begin{equation}
\left.\partial_{t}\vec{\widehat{v}}\right|_{x,\vec{v}}=-\frac{\vec{\widehat{v}}}{v_{\alpha}^{*}}\partial_{t}v_{\alpha}^{*}.
\end{equation}
There results

\begin{equation}
\left.\partial_{t}\widehat{f}_{\alpha}\right|_{x,\vec{v}}=\left.\partial_{t}\widehat{f}_{\alpha}\right|_{x,\vec{\widehat{v}}}-\frac{\partial_{t}v_{\alpha}^{*}}{v_{\alpha}^{*}}\vec{\widehat{v}}\cdot\widehat{\nabla}_{v}\widehat{f}_{\alpha}.
\end{equation}
\textcolor{black}{Similarly, we have
\[
\left.\partial_{x}\widehat{f}_{\alpha}\right|_{t,\vec{v}}=\left.\partial_{x}\widehat{f}_{\alpha}\right|_{t,\vec{\widehat{v}}}+\left.\partial_{x}\vec{\widehat{v}}\right|_{t,\vec{v}}\cdot\widehat{\nabla}_{v}\widehat{f}_{\alpha},
\]
where
\[
\left.\partial_{x}\vec{\widehat{v}}\right|_{t,\vec{v}}=-\frac{\vec{\widehat{v}}}{v_{\alpha}^{*}}\partial_{x}v_{\alpha}^{*}.
\]
Therefore
\[
\left.\partial_{x}\widehat{f}_{\alpha}\right|_{t,\vec{v}}=\left.\partial_{x}\widehat{f}_{\alpha}\right|_{t,\vec{\widehat{v}}}-\frac{\partial_{x}v_{\alpha}^{*}}{v_{\alpha}^{*}}\vec{\widehat{v}}\cdot\widehat{\nabla}_{v}\widehat{f}_{\alpha}.
\]
}

\textcolor{black}{From the Vlasov equation we have
\begin{eqnarray*}
\left.\partial_{t}f_{\alpha}\right|_{x,\vec{v}}+\partial_{x}\left.\left(v_{||}f_{\alpha}\right)\right|_{\vec{v},t}=\left.\partial_{t}\left[\frac{\hat{f}_{\alpha}}{\left(v_{\alpha}^{*}\right)^{3}}\right]\right|_{x,\vec{v}}+v_{\parallel}\partial_{x}\left.\left[\frac{\widehat{f}_{\alpha}}{\left(v_{\alpha}^{*}\right)^{3}}\right]\right|_{t,\vec{v}} & =\\
\left[\frac{\left.\partial_{t}\hat{f}_{\alpha}\right|_{x,\vec{v}}}{\left(v_{\alpha}^{*}\right)^{3}}+\hat{f}_{\alpha}\partial_{t}\left.(v_{\alpha}^{*})^{-3}\right|_{x}\right]+v_{\parallel}\left[\frac{\left.\partial_{x}\widehat{f}_{\alpha}\right|_{t,\vec{v}}}{\left(v_{\alpha}^{*}\right)^{3}}+\widehat{f}_{\alpha}\partial_{x}\left.\left(v_{\alpha}^{*}\right)^{-3}\right|_{t}\right] & =\\
\frac{1}{\left(v_{\alpha}^{*}\right)^{3}}\left[\partial_{t}\left.\widehat{f}_{\alpha}\right|_{x,\vec{\widehat{v}}}-\frac{\left(\partial_{t}v_{\alpha}^{*}+v_{\parallel}\partial_{x}v_{\alpha}^{*}\right)}{v_{\alpha}^{*}}\vec{\widehat{v}}\cdot\widehat{\nabla}_{v}\widehat{f}_{\alpha}-\frac{3\widehat{f}_{\alpha}}{v_{\alpha}^{*}}\left(\partial_{t}v_{\alpha}^{*}+v_{\parallel}\partial_{x}v_{\alpha}^{*}\right)+\hat{v}_{\parallel}v_{\alpha}^{*}\left.\partial_{x}\widehat{f}_{\alpha}\right|_{t,\vec{\widehat{v}}}\right] & =\\
\frac{1}{\left(v_{\alpha}^{*}\right)^{3}}\left[\partial_{t}\left.\widehat{f}_{\alpha}\right|_{x,\vec{\widehat{v}}}-\frac{3\widehat{f}_{\alpha}+\vec{\widehat{v}}\cdot\widehat{\nabla}_{v}\widehat{f}_{\alpha}}{v_{\alpha}^{*}}\left(\partial_{t}v_{\alpha}^{*}+v_{\parallel}\partial_{x}v_{\alpha}^{*}\right)+\left(\left.\partial_{x}(v_{\alpha}^{*}\hat{v}_{\parallel}\widehat{f}_{\alpha})\right|_{t,\vec{\widehat{v}}}-\hat{v}_{\parallel}\widehat{f}_{\alpha}\partial_{x}v_{\alpha}^{*}\right)\right] & .
\end{eqnarray*}
Using $\hat{\nabla}_{\widehat{v}}\cdot\hat{\vec{v}}=3$ and $\widehat{\nabla}_{\widehat{v}}\cdot\left(\widehat{v}_{||}\vec{\widehat{v}}\right)=4\widehat{v}_{||}$,
we find}

\begin{equation}
\left.\partial_{t}f_{\alpha}\right|_{x,\vec{v}}+\partial_{x}\left.\left(v_{||}f_{\alpha}\right)\right|_{\vec{v},t}=\frac{1}{\left(v_{\alpha}^{*}\right)^{3}}\left[\left.\partial_{t}\hat{f}_{\alpha}\right|_{x,\hat{\vec{v}}}-\frac{1}{v_{\alpha}^{*}}\frac{\partial v_{\alpha}^{*}}{\partial t}\widehat{\nabla}_{\widehat{v}}\cdot(\vec{\widehat{v}}\hat{f}_{\alpha})-\partial_{x}v_{\alpha}^{*}\widehat{\nabla}_{\widehat{v}}\cdot\left(\vec{\widehat{v}}\widehat{v}_{||}\widehat{f}_{\alpha}\right)+\partial_{x}\left.\left(v_{\alpha}^{*}\widehat{v}_{||}\widehat{f}_{\alpha}\right)\right|_{t,\vec{\widehat{v}}}\right].
\end{equation}
Finally, the electrostatic-acceleration term is written as

\begin{equation}
\frac{q_{\alpha}}{m_{\alpha}}E_{||}\partial_{v_{||}}f_{\alpha}=\frac{q_{\alpha}}{m_{\alpha}}\frac{E_{||}}{\left(v_{\alpha}^{*}\right)^{4}}\partial_{\widehat{v}_{||}}\widehat{f}_{\alpha}.
\end{equation}

Substituting these results into the original Vlasov equation, Eq.
(\ref{eq:1d_vfp_eqn}), there results the following transformed equation
for $\widehat{f}_{\alpha}\left(x,\vec{\widehat{v}},t\right)$
\begin{eqnarray}
\partial_{t}\widehat{f}_{\alpha}+\partial_{x}\left(v_{\alpha}^{*}\widehat{v}_{||}\widehat{f}_{\alpha}\right) & + & \left(\frac{q_{\alpha}}{m_{\alpha}}\frac{E_{||}}{v_{\alpha}^{*}}\right)\partial_{\widehat{v}_{||}}\widehat{f}_{\alpha}-\left(\frac{\partial_{t}v_{\alpha}^{*}}{v_{\alpha}^{*}}\right)\widehat{\nabla}_{v}\cdot\left(\vec{\widehat{v}}\widehat{f}_{\alpha}\right)-\left(\partial_{x}v_{\alpha}^{*}\right)\widehat{\nabla}_{v}\cdot\left(\vec{\widehat{v}}\widehat{v}_{||}\widehat{f}_{\alpha}\right)\nonumber \\
 & = & \left(v_{\alpha}^{*}\right)^{3}\left(\sum_{\beta}C_{\alpha\beta}+C_{\alpha e}\right)=\sum_{\beta}\widehat{C}_{\alpha\beta}+\widehat{C}_{\alpha e},\label{eq:vfp_transformed_continuum-1}
\end{eqnarray}
where we have used the definition of the normalized collision operator,
$\hat{C}_{\alpha\beta}=\left(v_{\alpha}^{*}\right)^{3}C_{\alpha\beta}$
\citep{Taitano_2016_rfp_0d2v_implicit}.

\section{Details on discrete conservation strategy for collisions between
kinetic ions and fluid electrons\label{app:details_conservation_strat_col_btw_kin_ion_fluid_e}}

In Sec. \ref{subsec:simul_disc_mass_mom_ener_cons_for_kin_ion_fe_sys},
we introduced the following nonlinear conservation constraints 
\begin{equation}
\gamma_{G,\alpha e}=1+\sum_{l=0}^{N_{B}}C_{l}^{G,\alpha e}P_{l}^{G,\alpha e},\;\gamma_{H,\alpha e,u}=1+\sum_{l=0}^{N_{B}}C_{l}^{H,\alpha e,u}P_{l}^{H,\alpha e,u},\;\gamma_{H,\alpha e,F}=1+\sum_{l=0}^{N_{B}}C_{l}^{H,\alpha e,F}P_{l}^{H,\alpha e,F}
\end{equation}
to discretely ensure the conservation symmetries for the ion-electron
collision operator {[}Eqs. (\ref{eq:ie_col_mom_cons_diff}), (\ref{eq:ie_col_mom_cons_fric}),
(\ref{eq:ie_col_mom_cons_fric_F}), (\ref{eq:ie_col_ener_cons_diff}),
(\ref{eq:ie_col_ener_cons_fric}), (\ref{eq:ie_col_ener_cons_fric-F}){]}.
Here, $C_{l}$ and $P_{l}$ are the coefficients and corresponding
basis functions that will be used to ensure the conservation symmetries.
In this study, we use the Fourier basis in both $v_{||}$ and $v_{\perp}$
directions,

\begin{equation}
\sum_{l=0}^{N_{b}}P_{l}\equiv\sum_{l_{||}=0}^{N_{B,||}}P_{l_{||}}\left(v_{||}\right)\sum_{l_{\perp}=0}^{N_{B,\perp}}P_{l_{\perp}}\left(v_{\perp}\right)\label{eq:basis_expansion}
\end{equation}
where

\begin{equation}
P_{l_{||}}=\begin{cases}
1 & \text{\textnormal{if}}\;l_{||}=0\\
\textnormal{sin}\left[l_{||}k_{||}v_{||}\right] & \textnormal{if}\;\textnormal{mod}\left(l_{||},2\right)=0\\
\textnormal{cos}\left[(l_{||}-1)k_{||}v_{||}\right] & \textnormal{if}\;\textnormal{mod}\left(l_{||},2\right)=1
\end{cases}\label{eq:parallel_basis_expansion}
\end{equation}

\begin{equation}
P_{l_{\perp}}=\begin{cases}
1 & \text{\textnormal{if}}\;l_{\perp}=0\\
\textnormal{sin}\left[l_{\perp}k_{\perp}v_{\perp}\right] & \textnormal{if}\;\textnormal{mod}\left(l_{\perp},2\right)=0\\
\textnormal{cos}\left[(l_{\perp}-1)k_{\perp}v_{\perp}\right] & \textnormal{if}\;\textnormal{mod}\left(l_{\perp},2\right)=1
\end{cases}\label{eq:perpendicular_basis_expansion}
\end{equation}
and $k_{||}=2\pi/L_{||},\;k_{\perp}=2\pi/L_{\perp}$ are the wave
vectors. We also choose $l_{||}=l_{\perp}=\left(0,1,2\right)$. The
coefficients are obtained by solving a constrained-minimization problem
for the following objective functions:

\begin{equation}
F\left(\vec{C}^{q},\vec{\lambda}_{q}\right)=\frac{1}{2}\sum_{l=0}^{N}C_{l}^{q}-\boldsymbol{\lambda}_{q}^{T}\cdot\vec{M}_{q},\textnormal{ where }q=G,\;u,\;F_{\alpha e},
\end{equation}
which satisfies the continuum symmetries in Eqs. (\ref{eq:ie_col_mom_cons_diff}),
(\ref{eq:ie_col_mom_cons_fric}), (\ref{eq:ie_col_mom_cons_fric_F}),
(\ref{eq:ie_col_ener_cons_diff}), (\ref{eq:ie_col_ener_cons_fric}),
(\ref{eq:ie_col_ener_cons_fric-F}). Here, $\vec{\lambda}$ is the
vector of Lagrange multipliers and $\vec{M}$ is the vector of vanishing
constraints that ensures the following discrete-momentum and energy-conservation
symmetries for the ion-electron-collision operator:

\begin{equation}
\vec{M}_{G}=\begin{bmatrix}m_{\alpha}\left\langle v_{||},\Gamma_{\alpha e}\nabla_{v}\cdot\left[\gamma_{G}\tensor D_{\alpha e}\cdot\nabla_{v}f_{\alpha}\right]\right\rangle _{\vec{v}}\\
\Gamma_{\alpha e}m_{\alpha}\left\langle \frac{v^{2}}{2},\nabla_{v}\cdot\left[\gamma_{G}\tensor D_{\alpha e}\cdot\nabla_{v}f_{\alpha}\right]\right\rangle _{\vec{v}}-3\nu_{e\alpha}\frac{m_{e}}{m_{\alpha}}n_{e}T_{e}
\end{bmatrix},
\end{equation}

\begin{equation}
\vec{M}_{u}=\begin{bmatrix}m_{\alpha}\Gamma_{\alpha e}\left\langle v_{||},\nabla_{v}\cdot\left[\frac{m_{\alpha}}{m_{e}}\gamma_{u}\vec{A}_{\alpha e,u}f_{\alpha}\right]\right\rangle _{\vec{v}}\\
m_{\alpha}\Gamma_{\alpha e}\left\langle \frac{v^{2}}{2},\nabla_{v}\cdot\left[\frac{m_{\alpha}}{m_{e}}\gamma_{u}\vec{A}_{\alpha e,u}f_{\alpha}\right]\right\rangle _{\vec{v}}-3\nu_{e\alpha}\frac{m_{e}}{m_{\alpha}}n_{e}T_{\alpha}
\end{bmatrix},
\end{equation}

\begin{equation}
\vec{M}_{F}=\begin{bmatrix}m_{\alpha}\Gamma_{\alpha e}\left\langle v_{||},\nabla_{v}\cdot\left[\frac{m_{\alpha}}{m_{e}}\gamma_{F}\vec{A}_{\alpha e,F}f_{\alpha}\right]\right\rangle _{\vec{v}}-F_{\alpha e,||}\\
m_{\alpha}\Gamma_{\alpha e}\left\langle \frac{v^{2}}{2},\nabla_{v}\cdot\left[\frac{m_{\alpha}}{m_{e}}\gamma_{F}\vec{A}_{\alpha e,F}f_{\alpha}\right]\right\rangle _{\vec{v}}+\vec{u}_{\alpha}\cdot\vec{F}_{\alpha e}
\end{bmatrix}.
\end{equation}
We solve the separate linear systems,

\begin{equation}
\begin{bmatrix}\partial_{\vec{C}^{G}}F\\
\partial_{\vec{\lambda}}F
\end{bmatrix}=\vec{0},\;\begin{bmatrix}\partial_{\vec{C}^{u}}F\\
\partial_{\vec{\lambda}}F
\end{bmatrix}=\vec{0},\;\begin{bmatrix}\partial_{\vec{C}^{F}}F\\
\partial_{\vec{\lambda}}F
\end{bmatrix}=\vec{0},
\end{equation}
for the coefficients. 

We chose the Fourier functions as the projection basis for robustness
and practical efficiency considerations. The natural basis in a cylindrical
geometry is given by the Bessel functions. However, they are very
costly to evaluate numerically. Since the conservation strategy is
based on projecting out discrete truncation errors in the conservation
error (integral measure) from parts of the flux, we find the choice
of Fourier basis to be both robust and efficient for practical applications.
Additionally, there are efficient libraries, such as MKL by Intel
(trademark), which supports fast evaluation of sine and cosine functions,
making the approach amenable to optimization.

\section{Robust generalization of discrete conservation scheme for collision
operator and temporal Vlasov inertial term\label{app:robust_generalizaiton_of_discrete_cons_scheme_for_col_op}}

In Ref. \citep{Taitano_2016_rfp_0d2v_implicit}, a discrete conservation
scheme for a spatially homogeneous system was developed for a multiple
velocity-space grid approach. In the reference, constraint coefficients
were introduced in various temporal terms to enforce targeted conservation
properties. These constraint coefficients were defined in terms of
moments in velocity space. In this study, to enhance computational
robustness, we have extended the constrained minimization approach
introduced in the previous section to these temporal terms, as well
as the collision operator itself. We provide some detail next.

We begin with the temporal terms in the Vlasov equation. Similarly
to the spatial terms, one can derive the following form for the temporal
piece of the Vlasov equation:

\begin{equation}
\partial_{t}\left[\left(v^{*}\right)^{2}\widehat{f}\right]-\partial_{t}\left(v^{*}\right)^{2}\left[\widehat{f}_{\alpha}+\frac{\widehat{\nabla}_{\widehat{v}}}{2}\cdot\left(\gamma_{t}\vec{\widehat{v}}\widehat{f}\right)\right]+\xi_{t}^{*}.
\end{equation}
Here,
\begin{eqnarray}
\xi_{t}^{*}=\left\{ v^{*}\partial_{t}\left(v^{*}\widehat{f}\right)-v^{*}\partial_{t}v^{*}\left[\widehat{f}+\widehat{\nabla}_{\widehat{v}}\cdot\left(\Upsilon_{t}\vec{\widehat{v}}\widehat{f}\right)\right]\right\} \nonumber \\
-\left\{ \partial_{t}\left[\left(v^{*}\right)^{2}\widehat{f}\right]-\partial_{t}\left(v^{*}\right)^{2}\left[\widehat{f}_{\alpha}+\frac{\widehat{\nabla}_{\widehat{v}}}{2}\cdot\left(\zeta_{t}\vec{\widehat{v}}\widehat{f}\right)\right]\right\} +\eta_{t}
\end{eqnarray}
and
\begin{eqnarray}
\eta_{t}=\left\{ \left(v^{*}\right)^{2}\partial_{t}\widehat{f}-\partial_{t}\left(v^{*}\right)^{2}\frac{\widehat{\nabla}_{\widehat{v}}}{2}\cdot\left(\vec{\widehat{v}}\widehat{f}\right)\right\} \nonumber \\
-\left\{ v^{*}\partial_{t}\left(v^{*}\widehat{f}\right)-v^{*}\partial_{t}v^{*}\widehat{f}-\partial_{t}\left(v^{*}\right)^{2}\frac{\widehat{\nabla}_{\widehat{v}}}{2}\cdot\left(\vec{\widehat{v}}\widehat{f}\right)\right\} .
\end{eqnarray}
We have introduced suitable constraint coefficients $\gamma_{t}$
and $\Upsilon_{t}$, defined as

\begin{equation}
\gamma_{t}=1+\sum_{l=0}^{N}C_{l}^{\gamma_{t}}P_{l}^{\gamma_{t}}\left(\widehat{v}_{||},\widehat{v}_{\perp}\right)
\end{equation}
and

\begin{equation}
\Upsilon_{t}=1+\sum_{l=0}^{N}C_{l}^{\Upsilon_{t}}P_{l}^{\Upsilon_{t}}\left(\widehat{v}_{||},\widehat{v}_{\perp}\right).
\end{equation}
The expansion coefficients, $C_{l}^{\gamma_{t}}$ and $C_{l}^{\Upsilon_{t}}$,
are determined by separately minimizing the following objective functions:
\begin{equation}
F_{\gamma_{t}}\left(\vec{C}^{\gamma_{t}},\lambda\right)=\frac{1}{2}\sum_{l=0}^{N}\left(C_{l}^{\gamma_{t}}\right)^{2}-\lambda\left[\left\langle \frac{\widehat{v}^{2}}{2},\left[\partial_{t}\left[\left(v^{*}\right)^{2}\widehat{f}\right]-\left(v^{*}\right)^{2}\partial_{t}\widehat{f}\right]\right\rangle _{\widehat{v}}-\partial_{t}\left(v^{*}\right)^{2}\left\langle \frac{\widehat{v}^{2}}{2},\frac{\widehat{\nabla}_{\widehat{v}}}{2}\cdot\left(\gamma_{t}\vec{\widehat{v}}\widehat{f}\right)\right\rangle _{\widehat{v}}\right]
\end{equation}
and

\begin{equation}
F_{\Upsilon_{t}}\left(\vec{C}^{\Upsilon_{t}},\lambda\right)=\frac{1}{2}\sum_{l=0}^{N}\left(C_{l}^{\Upsilon_{t}}\right)^{2}-\lambda\left[\left\langle \widehat{v}_{||},\left[v^{*}\partial_{t}\left[v^{*}\widehat{f}\right]-\left(v^{*}\right)^{2}\partial_{t}\widehat{f}\right]\right\rangle _{\widehat{v}}-v^{*}\partial_{t}v^{*}\left\langle \widehat{v}_{||},\widehat{\nabla}_{\widehat{v}}\cdot\left(\Upsilon_{t}\vec{\widehat{v}}\widehat{f}\right)\right\rangle _{\widehat{v}}\right].
\end{equation}
We have also introduced the coefficient $\zeta_{t}$,

\begin{equation}
\zeta_{t}=1+\sum_{l=0}^{N}C_{l}^{\zeta_{t}}P_{l}^{\zeta_{t}}\left(\widehat{v}_{||},\widehat{v}_{\perp}\right),
\end{equation}
where the coefficients $C_{l}^{\zeta}$ are determined by minimizing
the following function:

\begin{equation}
F_{\zeta_{t}}\left(\vec{C}^{\zeta_{t}},\vec{\lambda}\right)=\frac{1}{2}\sum_{l=0}^{N}\left(C_{l}^{\zeta_{t}}\right)^{2}-\boldsymbol{\lambda}^{T}\cdot\left(\vec{S}_{t}-\vec{M}_{t}\right)
\end{equation}
with

\begin{equation}
\vec{S}_{t}=\begin{bmatrix}\left\langle \widehat{v}_{||},\widehat{\nabla}_{\widehat{v}}\cdot\left(\gamma_{x}\vec{\widehat{v}}\widehat{f}\right)\right\rangle _{\widehat{v}}\\
2v^{*}\partial_{t}v^{*}\left\langle \frac{\widehat{v}^{2}}{2},\widehat{\nabla}_{\widehat{v}}\cdot\left(\Upsilon_{x}\vec{\widehat{v}}\widehat{f}\right)\right\rangle _{\widehat{v}}
\end{bmatrix},\;\vec{M}_{t}=\begin{bmatrix}\left\langle \widehat{v}_{||},\widehat{\nabla}_{\widehat{v}}\cdot\left(\zeta_{x}\vec{\widehat{v}}\widehat{f}\right)\right\rangle _{\widehat{v}}\\
\partial_{t}\left(v^{*}\right)^{2}\left\langle \frac{\widehat{v}^{2}}{2},\widehat{\nabla}_{\widehat{v}}\cdot\left(\zeta_{x}\vec{\widehat{v}}\widehat{f}\right)\right\rangle _{\widehat{v}}
\end{bmatrix}.
\end{equation}

For the collision operator, we begin my modifying the fast-on-slow
collision operator as:

\begin{equation}
C_{fs}=\Gamma_{fs}\nabla_{v}\cdot\left[\gamma_{fs,G}\vec{J}_{fs,G}-\frac{m_{f}}{m_{s}}\gamma_{fs,H}\vec{J}_{fs,H}\right]
\end{equation}
where $\gamma_{fs,G}=1+\sum_{l=0}^{N}C_{l}^{\gamma_{fs,G}}P_{l}^{*}\left(v_{||},v_{\perp}\right)$
and $\gamma_{fs,H}=1+\sum_{l=0}^{N}C_{l}^{\gamma_{fs,H}}P_{l}^{*}\left(v_{||},v_{\perp}\right)$,
$P_{l}^{*}\in\Omega_{\eta,fs}$ , $\Omega_{\eta,fs}$ is the domain
of overlap between the fast and slow species grid, and the coefficients
are determined by separately minimizing the following set of objective
functions:

\begin{equation}
F\left(\vec{C}^{\gamma_{fs,G}},\vec{\lambda}\right)=\frac{1}{2}\sum_{l=0}^{N}\left(C_{l}^{\gamma_{fs,G}}\right)^{2}-\boldsymbol{\lambda}^{T}\cdot\left[\vec{S}_{G}-\vec{M}_{G}\right]
\end{equation}
and

\begin{equation}
F\left(\vec{C}^{\gamma_{fs,H}},\vec{\lambda}\right)=\frac{1}{2}\sum_{l=0}^{N}\left(C_{l}^{\gamma_{fs,H}}\right)^{2}-\boldsymbol{\lambda}^{T}\cdot\left[\vec{S}_{H}-\vec{M}_{H}\right].
\end{equation}
Here:

\begin{equation}
\vec{S}_{G}=\begin{bmatrix}\left\langle 1,J_{||,fs,G}\right\rangle _{v}\\
\left\langle \vec{v},J_{||,fs,G}\right\rangle _{v}
\end{bmatrix},\;\vec{S}_{H}=\begin{bmatrix}\left\langle 1,J_{||,fs,H}\right\rangle _{v}\\
\left\langle \vec{v},J_{||,fs,H}\right\rangle _{v}
\end{bmatrix},
\end{equation}
and

\begin{equation}
\vec{M}_{G}=\begin{bmatrix}\left\langle 1,J_{||,sf,H}\right\rangle _{v}\\
\left\langle \vec{v},J_{||,sf,H}\right\rangle _{v}
\end{bmatrix},\;\vec{M}_{H}=\begin{bmatrix}\left\langle 1,J_{||,sf,G}\right\rangle _{v}\\
\left\langle \vec{v},J_{||,sf,G}\right\rangle _{v}
\end{bmatrix}.
\end{equation}

\section{Physics of a shock traveling through a density jump\label{app:physics_of_shock_traveling_through_a_density_jump}}

\begin{figure}[t]
\begin{centering}
\includegraphics[width=6cm]{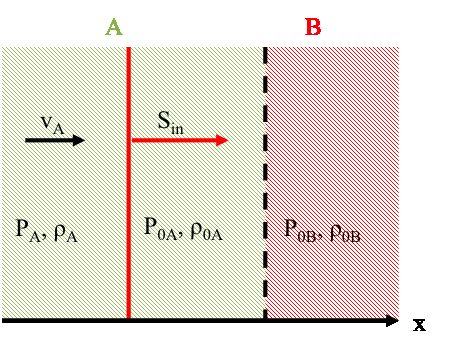}
\par\end{centering}
\caption{Shock propagates from left to right towards density discontinuity
at the boundary of materials A and B.\label{fig:app_illustration_shock_prop_through_dens_inc}}
\end{figure}

Assume we have semi-infinite materials A and B, with pre-shock pressures,
$P_{0A}=P_{0B}=P_{0}$, and densities, $\rho_{0A}$ and $\rho_{0B}$
(the system is assumed to be in a pressure equilibrium); refer to
Fig. \ref{fig:app_illustration_shock_prop_through_dens_inc}. 
The shock propagates from left to right with a speed $S_{in}$. After
it passes through the material A, the material acquires a velocity
$v_{A}$ from left to right. Its pressure and density become $P_{A}$
and $\rho_{A}$, respectively. The quantities $v_{A}$, $P_{A}$ and
$\rho_{A}$ can be found from

\begin{equation}
\frac{P_{A}}{P_{0}}=\frac{2\gamma M_{in}^{2}-\left(\gamma-1\right)}{\gamma+1},
\end{equation}

\begin{equation}
\frac{\rho_{A}}{\rho_{0A}}=\frac{S_{in}}{S_{in}-v_{A}}=\frac{M_{in}^{2}\left(\gamma+1\right)}{M_{in}^{2}\left(\gamma-1\right)+2},
\end{equation}
so that 

\begin{equation}
v_{A}=\frac{2c_{0A}}{\gamma+1}\frac{M_{in}^{2}-1}{M_{in}}
\end{equation}
where $\gamma=5/3$, $c_{0A}\equiv\sqrt{\gamma P_{0}/\rho_{0A}}$
is the upstream sound speed in the material A, and $P_{0}$, $\rho_{0A}$,
and $M_{in}\equiv S_{in}/c_{0A}$ are the input parameters.

\begin{figure}[t]
\begin{centering}
\includegraphics[width=6cm]{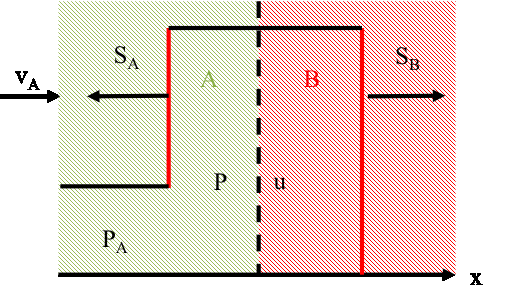}
\par\end{centering}
\caption{After encountering the material interface, the shock $S_{in}$ splits
into a transmitted shock $S_{B}$, propagating through the material
B from left to right; and a reflected shock (or a rarefaction wave)
$S_{A}$, propagating through the material A from right to left.\label{fig:app_illustration_shock_prop_through_dens_inc_post}}
\end{figure}

Eventually the shock $S_{in}$ arrives at the interface between the
materials A and B and splits into a transmitted shock $S_{B}$, propagating
through the material B from left to right; and a reflected shock (or
a rarefaction wave) $S_{A}$, propagating through the material A from
right to left, see Fig. \ref{fig:app_illustration_shock_prop_through_dens_inc_post}.
 $S_{A}$ is a shock when $\rho_{0B}>\rho_{0A}$, the case considered
herein; and a rarefaction wave otherwise. The material between $S_{A}$
and $S_{B}$, within the so-called \emph{contact discontinuity region},
has the common pressure $P$ and flow velocity $u$ from left to right.
These quantities can be evaluated by demanding downstream pressures
and flow velocities for $S_{A}$ and $S_{B}$ to be equal. This is
shown schematically in Fig. \ref{fig:app_illustration_shock_prop_through_dens_inc_post}.

\begin{figure}[h]
\begin{centering}
\includegraphics[height=6cm]{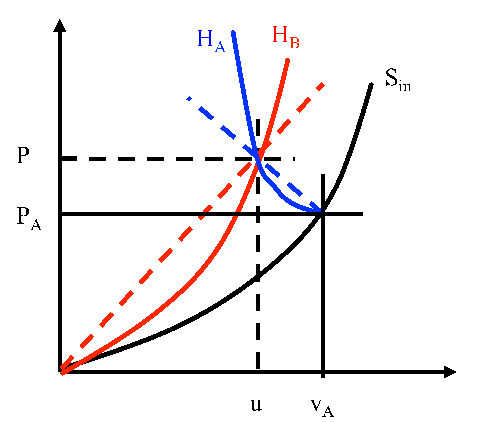}
\par\end{centering}
\caption{Black, red, and blue lines show Hugoniot curves (in the $P-v$ variables)
for shocks $S_{in}$, $S_{A}$, and $S_{B}$. The figure demonstrates
graphically the process of finding material pressure and flow velocity
in the contact discontinuity region. The situation depicted is only
possible when the red curve is steeper than the black curve, which
occurs for $\rho_{0B}>\rho_{0A}$. Then, the blue curve is as shown,
corresponding to a reflected shock. When the black curve is steeper
than the red one, which occurs for $\rho_{0B}<\rho_{0A}$, the reflected
wave is a rarefaction wave. }
\end{figure}
The downstream flow velocity for the shock $S_{B}$ is given as:

\begin{equation}
(S_{B}-u)^{2}=\frac{c_{0B}^{2}}{2\gamma}\frac{\sb{(\gamma+1)+(\gamma-1)P/P_{0}}^{2}}{(\gamma-1)+(\gamma+1)P/P_{0}}
\end{equation}
with $c_{0B}\equiv\sqrt{\gamma P_{0}/\rho_{0B}}$. The shock velocity,
$S_{B}$, is obtained from 

\begin{equation}
\frac{S_{B}}{S_{B}-u}=\frac{(\gamma-1)P_{0}+(\gamma+1)P}{(\gamma+1)P_{0}+(\gamma-1)P},
\end{equation}
resulting in 

\begin{equation}
\frac{S_{B}}{u}=\frac{(\gamma-1)+(\gamma+1)P/P_{0}}{2(P/P_{0}-1)}.\label{SB_vel}
\end{equation}
Combining these two results gives 

\begin{equation}
u^{2}=\frac{2c_{0B}^{2}}{\gamma}\frac{(P/P_{0}-1)^{2}}{(\gamma-1)+(\gamma+1)P/P_{0}}.\label{SA_ex}
\end{equation}
In the limit of a strong shock $S_{B}$, $P/P_{0}\gg1$, this becomes 

\begin{equation}
u^{2}\approx\frac{2P}{(\gamma+1)\rho_{0B}},\,\,\,\frac{S_{B}}{u}\approx\frac{\gamma+1}{2}.\label{SA}
\end{equation}

Downstream flow velocity for the shock $S_{A}$ is found similarly: 

\begin{equation}
(S_{A}+u)^{2}=\frac{c_{A}^{2}}{2\gamma}\frac{\sb{(\gamma+1)+(\gamma-1)P/P_{A}}^{2}}{(\gamma-1)+(\gamma+1)P/P_{A}}
\end{equation}
with $c_{A}\equiv\sqrt{\gamma P_{A}/\rho_{A}}$. The shock velocity
$S_{A}$ is obtained from 

\begin{equation}
\frac{S_{A}+v_{A}}{S_{A}+u}=\frac{(\gamma-1)P_{A}+(\gamma+1)P}{(\gamma+1)P_{A}+(\gamma-1)P},
\end{equation}
resulting in 

\begin{equation}
S_{A}=\frac{(\gamma-1)(Pv_{A}-P_{A}u)+(\gamma+1)(P_{A}v_{A}-Pu)}{2(P-P_{A})}.\label{rsh}
\end{equation}
Combining these two results gives 

\begin{equation}
(v_{A}-u)^{2}=\frac{2c_{A}^{2}}{\gamma}\frac{(P/P_{A}-1)^{2}}{(\gamma-1)+(\gamma+1)P/P_{A}}.\label{SB}
\end{equation}
While the transmitted shock can be strong, $P/P_{0}\text{\ensuremath{\gg}1}$,
the reflected one does not have to be so since $P/P_{A}=(P/P_{0})(P_{0}/P_{A})\sim M_{in}^{-2}(P/P_{0})$.
Thus, we should not expand Eq. (\ref{SB}) in $P/P_{A}\gg1$. 

\begin{figure}
\begin{centering}
\includegraphics[height=5cm]{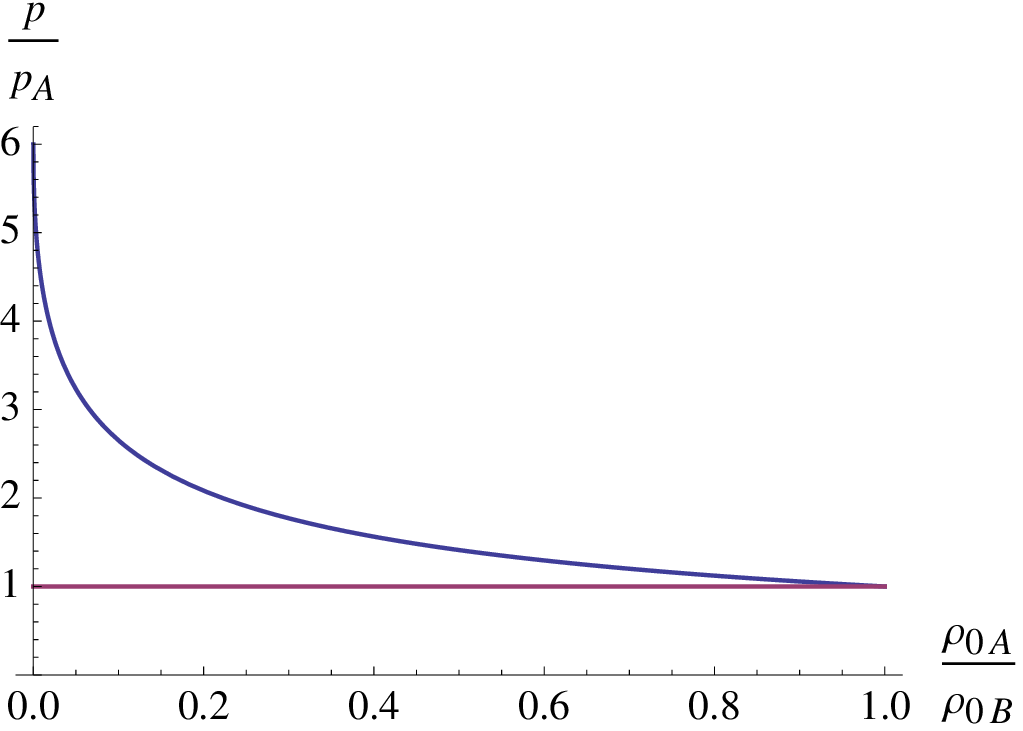}
\par\end{centering}
\caption{Solution of Eq. (\ref{disol}) for $\gamma=5/3$.\label{fig:solution_of_disol}}
\end{figure}

Equating $u$ from Eqs. (\ref{SA}) and (\ref{SB}) and assuming $M_{in}\gg1$
gives the following equation for $x\equiv\sqrt{P/P_{A}}$: 

\begin{equation}
\sqrt{\frac{\rho_{0A}}{\rho_{0B}}}x+\frac{x^{2}-1}{\sqrt{1+\frac{\gamma+1}{\gamma-1}x^{2}}}=1.\label{disol}
\end{equation}
The solution for $\gamma=5/3$ is shown in Fig. \ref{fig:solution_of_disol}
and confirms that $P/P_{A}\sim\text{O}\left(1\right)$. In general,
we have to solve a combination of Eqs. (\ref{SA_ex}) and (\ref{SB})
numerically without assuming $P/P_{0}\gg1$, $M_{in}\gg1$. Once $P/P_{A}$
is evaluated, all the other quantities can be obtained from the preceding
equations. E.g., 

\begin{equation}
\frac{u}{c_{0B}}\approx\sqrt{\frac{2}{\gamma(\gamma+1)}\frac{P}{P_{0}}}\approx\frac{2}{\gamma+1}M_{in}\sqrt{\frac{P}{P_{A}}}.\label{u_sol}
\end{equation}

\begin{figure}
\begin{centering}
\includegraphics[height=5cm]{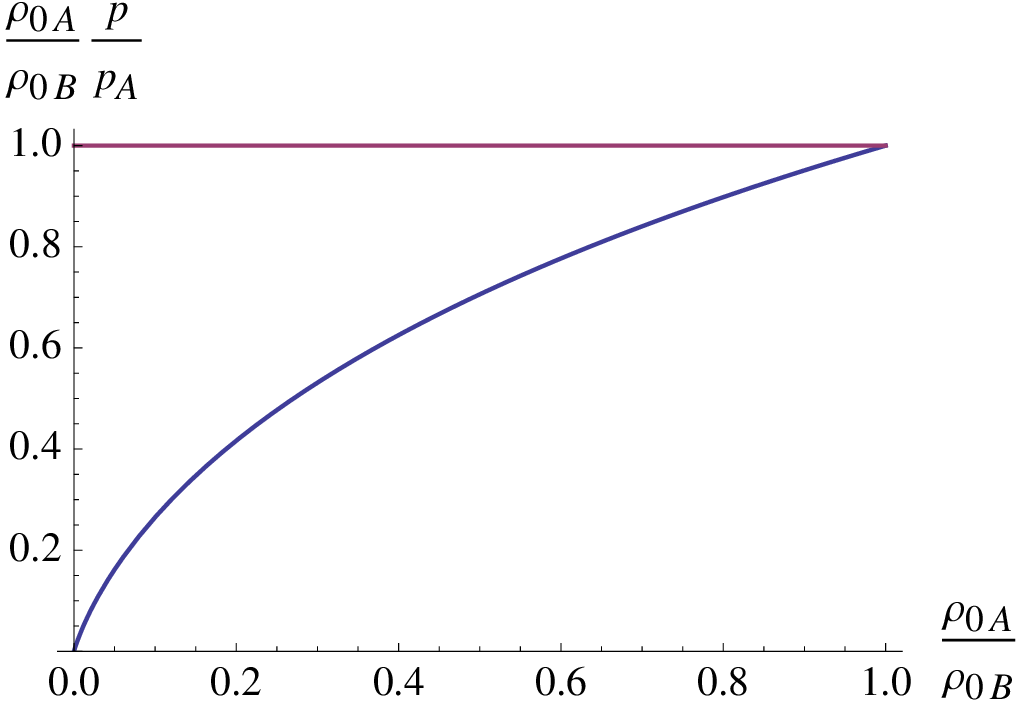}
\par\end{centering}
\caption{The factor $(p/p_{A})(\rho_{0A}/\rho_{0B})$ for $\gamma=5/3$.\label{fig:prhou_dens_jump}}
\end{figure}

The transmitted shock has a Mach number

\[
M_{B}\equiv\frac{S_{B}}{c_{0B}}\approx\sqrt{\frac{\gamma+1}{2\gamma}}\sqrt{\frac{P}{P_{0}}}\approx M_{in}\sqrt{\frac{P}{P_{A}}}>M_{in}.
\]
At the same time, when the shock velocity is normalized to $c_{0A}$
rather than $c_{0B}$, we can see that the transmitted shock is weaker
than the initial shock (see Fig. \ref{fig:prhou_dens_jump}): 
\[
M_{B}^{\prime}\equiv\frac{S_{B}}{c_{0A}}\approx M_{in}\sqrt{\frac{P}{P_{A}}\frac{\rho_{0A}}{\rho_{0B}}}<M_{in}.
\]
Finally, the reflected shock speed is given by (\ref{rsh}), 

\[
\frac{S_{A}}{v_{A}}=\frac{(\gamma-1)\rb{\frac{P}{P_{A}}-\frac{u}{v_{A}}}+(\gamma+1)\rb{1-\frac{P}{P_{A}}\frac{u}{v_{A}}}}{2\rb{\frac{P}{P_{A}}-1}}
\]
with 

\begin{equation}
\frac{u}{v_{A}}\approx\sqrt{\frac{\rho_{0A}}{\rho_{0B}}\frac{P}{P_{A}}}<1\label{uvA}
\end{equation}
or, alternatively, 

\[
M_{A}\equiv\frac{S_{A}}{c_{A}}\approx\sqrt{\frac{2}{\gamma(\gamma-1)}}\frac{S_{A}}{v_{A}}.
\]

The shocked material densities are evaluated in the usual fashion: 

\[
\frac{\rho_{B}}{\rho_{0B}}=\frac{S_{B}}{S_{B}-u}\approx\frac{\gamma+1}{\gamma-1},\frac{\rho_{A}^{\prime}}{\rho_{A}}=\frac{S_{A}+v_{A}}{S_{A}+u}=\frac{(\gamma+1)\frac{P}{P_{A}}+(\gamma-1)}{(\gamma-1)\frac{P}{P_{A}}+(\gamma+1)}>1,
\]
where $\rho_{A}^{\prime}$ is the material A density after passage
of the reflected shock $S_{A}$.

\section{Discrete momentum conservation proof for the Vlasov equation\label{app:Discrete-mass-and-momentum-conservation-proof}}

Expanding Eqs. (\ref{eq:discrete_mass_mom_cons_term_a}), (\ref{eq:discrete_mass_mom_cons_term_b}),
(\ref{eq:discrete_mass_mom_cons_term_c}), and discretely summing
over configuration space, we obtain:
\begin{eqnarray}
\int_{L_{min}}^{L_{max}}dx\left\{ v^{*}\partial_{x}\left(v^{*}\widehat{S}_{2,||}\right)\right\} -\left\{ \partial_{x}\left[\left(v^{*}\right)^{2}\widehat{S}_{2,||}\right]-\partial_{x}v^{*}\left[v^{*}\widehat{S}_{2,||}\right]\right\} \approx\nonumber \\
\sum_{i}^{N_{x}}v_{i}^{*}\left(v_{i+1/2}^{*}\widehat{S}_{2,||,i+1/2}-v_{i-1/2}^{*}\widehat{S}_{2,||,i-1/2}\right)-\sum_{i}^{N_{x}}\left(\left(v_{i+1/2}^{*}\right)^{2}\widehat{S}_{2,||,i+1/2}-\left(v_{i-1/2}^{*}\right)^{2}\widehat{S}_{2,||,i-1/2}\right)+\nonumber \\
\frac{1}{2}\sum_{i}^{N_{x}}\left\{ \left(v_{i+1}^{*}-v_{i}^{*}\right)v_{i+1/2}^{*}\widehat{S}_{2,||,i+1/2}+\left(v_{i}^{*}-v_{i-1}^{*}\right)v_{i-1/2}^{*}\widehat{S}_{2,||,i-1/2}\right\} .
\end{eqnarray}
With periodic boundary conditions, the second summation vanishes.
Gathering terms, we find: 
\begin{eqnarray}
\sum_{i}^{N_{x}}v_{i}^{*}\left(v_{i+1/2}^{*}\widehat{S}_{2,||,i+1/2}-v_{i-1/2}^{*}\widehat{S}_{2,||,i-1/2}\right)\nonumber \\
+\frac{1}{2}\sum_{i}^{N_{x}}\left\{ \left(v_{i+1}^{*}-v_{i}^{*}\right)v_{i+1/2}^{*}\widehat{S}_{2,||,i+1/2}+\left(v_{i}^{*}-v_{i-1}^{*}\right)v_{i-1/2}^{*}\widehat{S}_{2,||,i-1/2}\right\}  & =\\
\sum_{i}^{N_{x}}\left(v_{i+1/2}^{*}\right)^{2}\widehat{S}_{2,||,i+1/2}-\sum_{i}^{N_{x}}\left(v_{i-1/2}^{*}\right)^{2}\widehat{S}_{2,||,i-1/2} & = & 0.
\end{eqnarray}

\section{Discrete energy conservation proof for the Vlasov equation\label{app:Discrete-mass-and-energy-conservation-proof}}

Expanding Eqs. (\ref{eq:discrete_mass_ener_cons_term_a}), (\ref{eq:discrete_mass_ener_cons_term_b}),
(\ref{eq:discrete_mass_ener_cons_term_c}), and discretely summing
over configuration space, we obtain:
\begin{eqnarray}
\int_{L_{min}}^{L_{max}}dx\left\{ \left(v^{*}\right)^{2}\partial_{x}\left(v^{*}\widehat{S}_{3,||}\right)\right\} -\left\{ \partial_{x}\left[\left(v^{*}\right)^{3}\widehat{S}_{3,||}\right]-\partial_{x}\left(v^{*}\right)^{2}\left[v^{*}\widehat{S}_{3,||}\right]\right\} \approx\nonumber \\
\sum_{i}^{N_{x}}\left(v_{i}^{*}\right)^{2}\left(v_{i+1/2}^{*}\widehat{S}_{3,||,i+1/2}-v_{i-1/2}^{*}\widehat{S}_{3,||,i-1/2}\right)-\sum_{i}^{N_{x}}\left(\left(v_{i+1/2}^{*}\right)^{3}\widehat{S}_{3,||,i+1/2}-\left(v_{i-1/2}^{*}\right)^{3}\widehat{S}_{3,||,i-1/2}\right)+\nonumber \\
\frac{1}{2}\sum_{i}^{N_{x}}\left\{ \left(\left[\left(v_{i+1}^{*}\right)^{2}-\left(v_{i}^{*}\right)^{2}\right]\right)v_{i+1/2}^{*}\widehat{S}_{3,||,i+1/2}+\left[\left(v_{i}^{*}\right)^{2}-\left(v_{i-1}^{*}\right)^{2}\right]v_{i-1/2}^{*}\widehat{S}_{3,||,i-1/2}\right\} .
\end{eqnarray}
With periodic boundary conditions, the second summation vanishes.
Gathering terms, we find:
\begin{eqnarray}
\sum_{i}^{N_{x}}\left(v_{i}^{*}\right)^{2}\left(v_{i+1/2}^{*}\widehat{S}_{3,||,i+1/2}-v_{i-1/2}^{*}\widehat{S}_{3,||,i-1/2}\right)\nonumber \\
+\frac{1}{2}\sum_{i}^{N_{x}}\left\{ \left[\left(v_{i+1}^{*}\right)^{2}-\left(v_{i}^{*}\right)^{2}\right]v_{i+1/2}^{*}\widehat{S}_{3,||,i+1/2}+\left[\left(v_{i}^{*}\right)^{2}-\left(v_{i-1}^{*}\right)^{2}\right]v_{i-1/2}^{*}\widehat{S}_{3,||,i-1/2}\right\}  & =\\
\sum_{i}^{N_{x}}\left[\left(v_{i}^{*}\right)^{2}+\left(v_{i+1}^{*}\right)^{2}\right]v_{i+1/2}^{*}\widehat{S}_{3,||,i+1/2}-\sum_{i}\left[\left(v_{i-1}^{*}\right)^{2}+\left(v_{i}^{*}\right)^{2}\right]v_{i-1/2}^{*}\widehat{S}_{3,||,i-1/2} & = & 0.\label{eq:vlasov_inertial_discrete_energy_conservation_symmetry_spatial_interpolation}
\end{eqnarray}

\end{document}